\documentclass[twocolumn]{aastex631}
\newcommand{\paperone}{ \cite{dogruel2023} }

\usepackage{graphicx, xcolor}	
\usepackage{amsmath}
\usepackage{amssymb}	
\usepackage{bm, booktabs}
\usepackage{multirow, enumerate}
\usepackage[inline]{enumitem}

\shorttitle{GAMA Peculiar Velocities}

\begin{document}

\title{Galaxy And Mass Assembly (GAMA): Stellar-to-Dynamical Mass Relation II. Peculiar Velocities}

\correspondingauthor{M. Burak Dogruel}
\email{bdogruel@swin.edu.au}

\author[0000-0002-8688-4331]{M. Burak Dogruel}
\affiliation{Centre for Astrophysics and Supercomputing, Swinburne University of Technology, Hawthorn, VIC 3122, Australia}

\author[0000-0002-5522-9107]{Edward N. Taylor}
\affiliation{Centre for Astrophysics and Supercomputing, Swinburne University of Technology, Hawthorn, VIC 3122, Australia}

\author[0000-0002-9871-6490]{Michelle Cluver}
\affiliation{Centre for Astrophysics and Supercomputing, Swinburne University of Technology, Hawthorn, VIC 3122, Australia}

\author[0000-0001-9552-8075]{Matthew Colless}
\affiliation{Research School of Astronomy and Astrophysics, Australian National University, Canberra, ACT 2611, Australia}
\affiliation{ARC Centre of Excellence for All Sky Astrophysics in 3 Dimensions (ASTRO 3D), Canberra, ACT 261, Australia}

\author[0000-0002-2380-9801]{Anna de Graaff}
\affiliation{Max-Planck-Institut f\"ur Astronomie, K\"onigstuhl 17, D-69117, Heidelberg, Germany}

\author[0000-0002-6061-5977]{Alessandro Sonnenfeld}
\affiliation{Department of Astronomy, School of Physics and Astronomy, Shanghai Jiao Tong University, Shanghai 200240, China}

\author[0000-0002-9748-961X]{John R. Lucey}
\affiliation{Centre for Extragalactic Astronomy, Durham University, Durham DH1 3LE, United Kingdom}

\author[0000-0003-2388-8172]{Francesco D’Eugenio}
\affiliation{Kavli Institute for Cosmology, University of Cambridge, Madingley
Road, Cambridge, CB3 0HA, United Kingdom}
\affiliation{Cavendish Laboratory - Astrophysics Group, University of Cambridge,
19 JJ Thomson Avenue, Cambridge, CB3 0HE, United Kingdom}

\author[0000-0002-1081-9410]{Cullan Howlett}
\affiliation{School of Mathematics and Physics, The University of Queensland, Brisbane, QLD 4072, Australia}

\author[0000-0002-1809-6325]{Khaled Said}
\affiliation{School of Mathematics and Physics, The University of Queensland, Brisbane, QLD 4072, Australia}



\begin{abstract}

Empirical correlations connecting starlight to galaxy dynamics (e.g., the fundamental plane (FP) of elliptical/quiescent galaxies and the Tully--Fisher relation of spiral/star-forming galaxies) provide cosmology-independent distance estimation and are central to local Universe cosmology.
In this work, we introduce the mass hyperplane (MH), which is the stellar-to-dynamical mass relation $(M_\star/M_\mathrm{dyn})$ recast as a linear distance indicator. 
Building on recent FP studies, we show that both star-forming and quiescent galaxies follow the same empirical MH, then use this to measure the peculiar velocities (PVs) for a sample of 2496 galaxies at $z<0.12$ from GAMA.
The limiting precision of MH-derived distance/PV estimates is set by the intrinsic scatter in size, which we find to be $\approx$0.1~dex for both quiescent and star-forming galaxies (when modeled independently) and $\approx$0.11~dex when all galaxies are modeled together; showing that the MH is as good as the FP.
To empirically validate our framework and distance/PV estimates, we compare the inferred distances to groups as derived using either quiescent or star-forming galaxies. A good agreement is obtained with no discernible bias or offset, having a scatter of $\approx$0.05~dex $\approx$12\% in distance.
Further, we compare our PV measurements for the quiescent galaxies to the previous PV measurements of the galaxies in common between GAMA and the Sloan Digital Sky Survey (SDSS), which shows similarly good agreement. 
Finally, we provide comparisons of PV measurements made with the FP and the MH, then discuss possible improvements in the context of upcoming surveys such as the 4MOST Hemisphere Survey (4HS).

\end{abstract}



\section{Introduction} \label{sec:intro}

\subsection{Peculiar Velocities}

Redshift-independent distances inferred via empirical, statistical correlations are  central to cosmology. In the local Universe ($z \lesssim 0.1$), galaxy scaling relations are particularly useful as a tool for measuring  deviations of galaxy velocities from the Hubble-Lema\^itre law for the expansion of the Universe (Hubble flow). To first order, cosmological expansion is seen locally as apparent recession velocities proportional to their comoving distance: $V\equiv cz \approx H_0 D$ \citep{hubble1929}, where $c$ is the speed of light, $z$ is the redshift, $D$ is the comoving distance (i.e., the radial distance) and $H_0$ is the current expansion rate of the Universe; i.e., the Hubble parameter. Deviations from the cosmic expansion are caused by the inhomogeneities in the matter distribution and they give rise to the peculiar velocities (PVs, $V_\text{pec}$) of galaxies: $1 + z_\mathrm{obs} = (1 + z_H)(1 + z_\mathrm{pec})$ \citep{harrison1974} where $z_\mathrm{obs}$ is the observed redshift, $z_H$ is the cosmological/comoving redshift due to the Hubble flow, $z_\mathrm{pec}$ is the peculiar redshift with $V_\mathrm{pec} = cz_\mathrm{pec}$ and $z_H\approx H_0 D(z_H)/c$.
In this picture, $z_\text{obs}$ is the only observable, thus known, quantity while the true comoving distance at the cosmological redshift, $D(z_H)$, and $V_\text{pec}$ are unknown but desired quantities.

The crucial importance of PVs in cosmological studies is better understood by tracing the origins of PV. The small anisotropy in the cosmic microwave background (CMB) already shows the existence of density inhomogeneities. According to the large scale structure formation and growth theory, these tiny density perturbations grow over time because of their self-gravity and generate local gravitational fields due to local density fluctuations (i.e., the differences between overdense and underdense regions creating density contrasts). These gravitational fields drive the matter flow and generate PVs. Thus, especially at low redshifts, PVs enable cosmography \citep[e.g.,][]{springob2014, tully2014, graziani2019}, bulk flow measurements \citep[e.g.,][]{qin2021, howlett2022}, fitting cosmological parameters such as the growth rate of structure, mass fluctuation amplitude within $8h^{-1}$Mpc-radius spheres $(\sigma_8)$ \citep[e.g.,][]{turnbull2012, carrick2015, khaled2020} and even tests of the standard $\Lambda$CDM model and General Relativity \citep[e.g.,][]{adams2017, howlett2017, khaled2020}. 

On the other hand, direct calculation of PVs requires measurements of both redshifts and redshift-independent distances. Some of the well-known distance indicators that are widely used in PV studies are the period--luminosity relation of Cepheid variables \citep[][]{leavitt1912}, Type Ia supernovae (SNe Ia) \citep{phillips1993}, the Tully--Fisher relation for disk galaxies \citep[TFR,][]{tullyfisher1977} and the fundamental plane \citep[FP,][]{djor1987, dressler1987} for elliptical galaxies and spheroids. Even though Cepheids provide the most accurate and precise extragalactic distances \citep[e.g.,][]{riess2016, riess2021}, they are only useful to distances short enough ($\lesssim 25$ Mpc) to be able to discern them within a galaxy in the first place. While SNe Ia provide precise distances \citep[][and references therein]{howlett2017, koda2014, scolnic2018} even for $z > 1$, their rarity in the local Universe and observational challenges limit SNe Ia distances to relatively small numbers of galaxies ($\sim$1000 galaxies).

\subsection{Dynamical Scaling Relations}

Empirical galaxy scaling relations that relate distance dependent properties (e.g., physical size, luminosity) to distance independent intrinsic properties (e.g., stellar kinematics, surface brightness, color), are much easier to measure and galaxies that obey these relations are quite abundant. For these reasons, 
TFR and FP have been the pillar of PV studies \citep[cf., the statistical measurement of PVs via redshift space distortions (RSD) for a sample of galaxies;][]{kaiser1987} thanks to their availability and sample sizes.
Even though both relations are seen to be remarkably tight, with modern data the intrinsic scatters around these relations are the limiting factor in the precision of distance measurements/estimates: $\sim 0.1$ dex or 20--25 \%.
In fact, these large errors in PVs have become the characteristic feature of PV surveys \citep{watkins2015}, and these errors grow linearly with increasing distance. This is another fundamental reason why PVs are most useful at low redshifts $(z\sim 0.1)$.

The traditional fundamental plane of elliptical/early-type galaxies is expressed in terms of luminosity with surface brightness, $\langle I_e \rangle \equiv L/2\pi R_e^2$;
\begin{equation}
    \log R_e = a \log \sigma_0 + b \log \langle I_e \rangle + c
    \label{eq:lfp}
\end{equation}
and can be referred to as the luminosity fundamental plane (LFP). Here, $R_e$ is the physical effective radius (within which half of the luminosity is emitted) in units of $h^{-1}$ kpc, where $h$ is the scaled Hubble constant as $H_0 = 100h$ km/s/Mpc, $\sigma_0$ is the central velocity dispersion in km/s, $\langle I_e \rangle$ is the surface brightness in $L_\odot/\text{pc}^2$ and the subscript $e$ denotes the measurements made within the effective radius. In terms of directly observed properties -- angular effective radius $(\theta_e)$, apparent magnitude $(m_\lambda)$ and redshift -- the LFP parameters for a band $\lambda$ are, 
\begin{align}
    R_e &= \theta_e D_A, \nonumber \\
    \log\langle I_e \rangle_\lambda &= 0.4(M_\odot^\lambda - \mu_e^\text{cor}) + 8.629, \nonumber \\
    \mu_e^\text{cor} &= m_\lambda - A_\lambda - k_\lambda (z_\mathrm{obs}) + 2.5\log (2\pi\theta_e^2) \nonumber \\ &- 2.5\log(1+z_\text{obs})^4 .
    \label{eq:fppars}
\end{align}
Here, $D_A$ is the angular diameter distance, $M_\odot^\lambda$ is the absolute magnitude of the Sun, $\mu_e^\text{cor}$ is the surface brightness in magnitudes per square-arcseconds corrected for galactic extinction $(A_\lambda)$, bandpass stretching ($k_\lambda(z_\mathrm{obs})$; correcting magnitudes to rest-frame) and surface brightness dimming $(\log(1+z_\text{obs})^4)$.
The form in equation~(\ref{eq:lfp}) separates the distance dependent quantity, $\log R_e$, from the distance independent quantities $\sigma_0$ and $\langle I_e \rangle$ \footnote{Even though $\langle I_e \rangle$ being in units of $L_\odot/\text{pc}^2$ might create a confusion as to its distance independence, as seen from equation~(\ref{eq:fppars}), it scales with the surface brightness dimming, $(1 + z_\text{obs})^4$, and requires $k_\lambda(z_\text{obs})$, however, $z_\text{obs}$ is directly observable.}. Multiplying $\langle I_e \rangle$ by the stellar-mass-to-light ratio $(M_\star/L)$ estimated from spectral energy distribution (SED) modeling gives the surface stellar-mass density within $R_e$, denoted with $\Sigma_\star$: $(M_\star/L)\langle I_e \rangle = M_\star/(2\pi R_e^2) \equiv \Sigma_\star$. Thus, when $\Sigma_\star$ replaces $\langle I_e\rangle$ in the LFP (equation \ref{eq:lfp}), we obtain what might be called the stellar-mass fundamental plane (SMFP):
\begin{equation}
    \log R_e = \alpha \log \sigma_e + \beta \log \Sigma_\star + \gamma.
    \label{eq:mfp}
\end{equation}

\subsection{Unification and Generalization of Dynamical Scaling Relations}

There have been some attempts at finding a universal FP, in the sense that it is applicable to all types of galaxies. \cite{zaritsky2008} and \cite{ortiz2020} have shown that it is possible to define a universal fundamental plane (UFP) at low redshifts, such that both elliptical and spiral galaxies reside on the same plane, with:
\begin{equation}
    \log \left( \frac{M_\text{dyn}}{L}\right)_e \equiv \log \Upsilon_e = \log (S_{0.5}^2) - \log\langle I_e \rangle - \log R_e + C .
    \label{eq:ufp}
\end{equation}
In this relation, $\Upsilon_e$ is the dynamical-mass-to-light ratio within $R_e$ and $S_K^2 = KV_{R_e}^2 + \sigma_e^2$ is the total velocity parameter \citep{weiner2006} where $K$ is assumed to be constant and is usually taken to be 0.5 \citep{cortese2014}. $S_{0.5}$ encompasses rotational velocity $(V_{R_e})$ and velocity dispersion within $R_e$ $(\sigma_e)$. These definitions make it clear that a UFP is possible when we account for both mass-to-light ratio and rotational velocity.

\cite{bezanson2015} have shown that star-forming (SF) and quiescent galaxies (Q) lie on the same SMFP for $z\sim 0$ galaxies from SDSS, albeit in different regions of that plane. 
Furthermore, using the data from LEGA-C, \cite{degraaff2020} have definitively shown that the SMFP relation is the same up to $z\sim 1$. 
Even more interestingly, again using LEGA-C data, \cite{degraaff2021} have demonstrated that both star-forming and quiescent galaxies not only share the same SMFP, but they also share the same LFP, just with different zero-points and so again in different regions of the plane. In both of these studies, authors have found larger intrinsic scatter for SF galaxies.

These results are in fact not shocking discoveries. Velocity dispersion when measured within the effective radius, $\sigma_e$, is a good approximation to the luminosity weighted root-mean-square of the line-of-sight velocity inside $R_e$, thus, includes both rotation and dispersion \citep[\cite{cappellari2006, cappellari2013} or see review,][]{courteau2014}. Therefore, using $\sigma_e$ instead of $\sigma_0$ in LFP and SMFP (equations \ref{eq:lfp} and \ref{eq:mfp}) will suffice to approximately account for rotational velocity, as required in the UFP of \cite{zaritsky2008} and \cite{ortiz2020}.

Furthermore, these results raise a rather obvious question: why should we discard spiral/star-forming/late-type/blue galaxies from a PV study carried out with the FP? Or, what really happens if we include those galaxies in such a study? In this paper, we are addressing these naturally arising questions in the wake of recently reached conclusions from \cite{bezanson2015} and \cite{degraaff2021} that we summarized above.

This work is structured as follows: We present our sample selection from GAMA and a description of our methodology in Section \ref{sec:data_method}. We give the fitting results of the FP and the MH, along with the investigation of possible systematics in Section \ref{sec:results_fp_mp}. We then present the main focus of this paper; measuring redshift-independent distances (and thus PVs) for GAMA in Section \ref{sec:vpec}. Finally, we summarize our results and discuss future work in Section \ref{sec:conclusions}.

Throughout this work, we assume a flat $\Lambda$CDM cosmology with $\Omega_m=0.3, \Omega_\Lambda=0.7$ and $H_0=100h$ km/s/Mpc.
Unless otherwise stated, redshifts should be understood as `flow-corrected' following \cite{baldry2012}; i.e., using the \cite{tonry2000} model and tapering to a CMB-centric frame for $z > 0.03$.

\section{Data and Method}\label{sec:data_method}

\subsection{GAMA Sample Selection}\label{sec:gama_sample}

\begin{figure*}
    \centering
    \includegraphics[width=\textwidth]{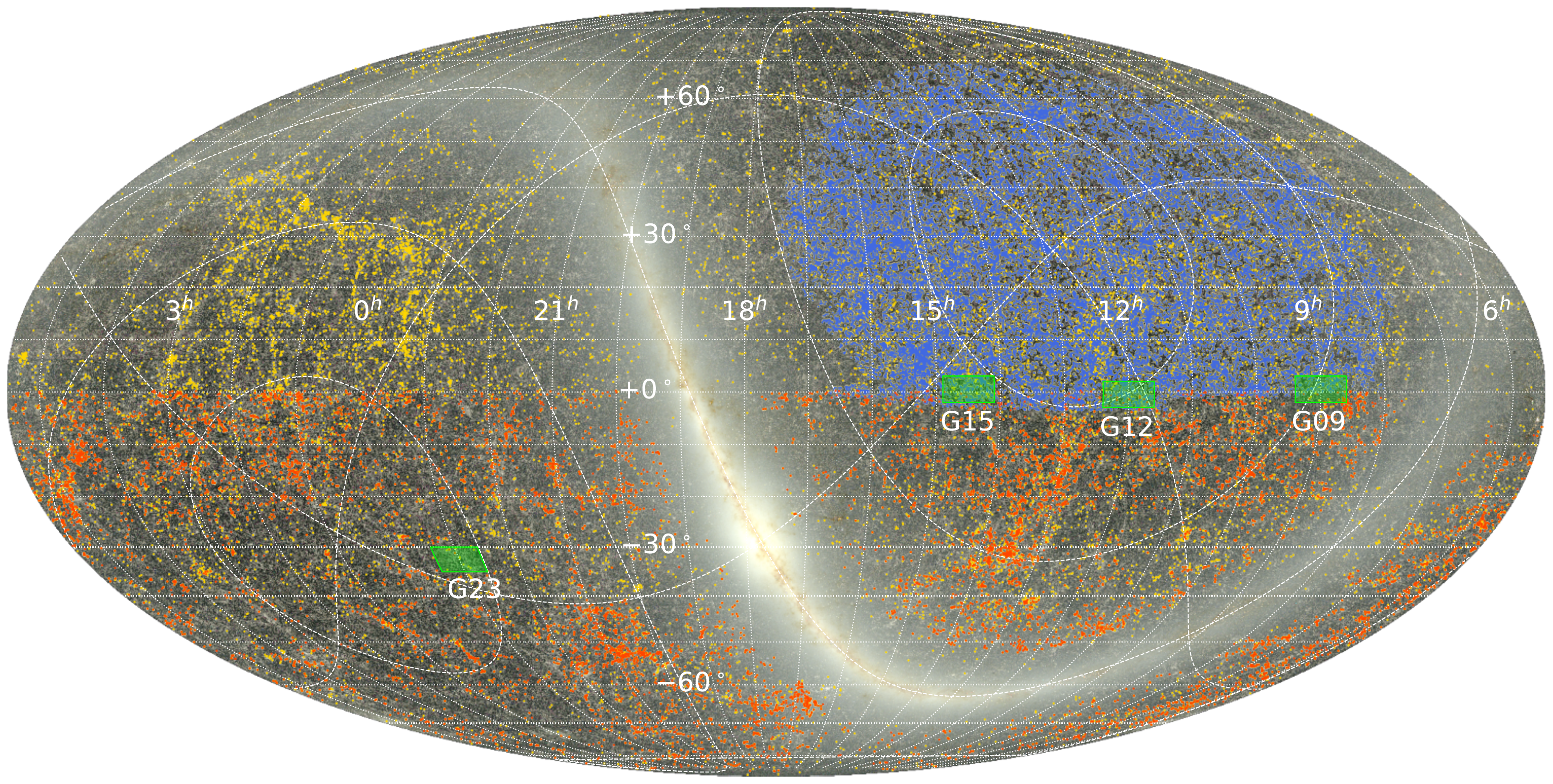}
    \caption{Distribution of the GAMA sample (green) in equatorial coordinates in comparison to 6dFGS (orange), CF4 (yellow) and SDSS (blue) shown in Mollweide projection, underlain by the 2MASS image of the Milky Way. The final GAMA PV sample consists of regions G09, G12 and G15, including 1489 star-forming (SF) and 1046 quiescent (Q) galaxies (a total of 2535 galaxies).}
    \label{fig:skycoverage}
\end{figure*}

The sky coverage of our GAMA PV sample in equatorial coordinates is presented in Figure \ref{fig:skycoverage} in comparison to some of the previous large scale PV studies carried out in the last decade: 6-degree Field Galaxy Survey \citep[6dFGS, using FP; ][]{springob2014}, CosmicFlows4 \citep[CF4, using TFR;][]{kourkchi2020} and Sloan Digital Sky Survey \citep[SDSS, using FP;][]{howlett2022}.

We use the same mass-limited sample drawn from GAMA as detailed in \paperone. In summary, the adopted selection criteria are:
\begin{enumerate}
    \item spectroscopic redshift quality flag $nQ \geqslant 3$ in the range $z < 0.12$, \footnote{The stellar-mass selection results in galaxies with $z>0.01$ being selected.}
    \item stellar-masses, $\log\,M_\star/M_\odot > 10.3$, and
    \item velocity dispersions, $60 < \sigma_e [\text{km s}^{-1}] < 450$ with uncertainties $\varepsilon_\sigma < 0.25\sigma_e+25$.
\end{enumerate}
In addition to these selection criteria, we also apply a cut on the projected axis ratio $(q\equiv b/a)$ and, as in \cite{howlett2022}, we select galaxies with $q > 0.3$ to limit our sample to exclude galaxies close to edge-on. We further elaborate this choice of 0.3 in section \ref{sec:systematics}.
For now, we note that there is the potential for complex systematics, especially for disk galaxies, arising from this selection, but that none of our results or conclusions change significantly if change our selection to, for example, $q > 0.5$  (see also Figure \ref{fig:kqcorrection}).
Thus it does not affect our main model (section \ref{sec:bayesian_summary}), though it causes a sizable reduction in the number of galaxies in our sample (from 2850 galaxies to 2535).

Before moving on, it is imperative to explain our approach to aperture corrections for the velocity dispersions. As stated in \paperone, GAMA velocity dispersion measurements have been calibrated to match those of SDSS, in which spectra have been taken using fibers with an aperture radius of $\theta_\mathrm{ap}=1.5''$. For $\sim84$\% of galaxies in our sample, the $\theta''_e$ exceed this value, which means that for these galaxies, the velocity dispersions measured through fibers might not reflect its value within the effective radius. In order to calculate the velocity dispersions within the effective radii, $\sigma_e$, we use the aperture correction in the form derived by \cite{jorgensen1995} and \cite{cappellari2006}: $\sigma_\mathrm{ap} / \sigma_e = (\theta_\mathrm{ap}/\theta_e)^\alpha$. Adopting the value $\alpha=-0.033\pm 0.003$ derived by \cite{degraaff2021} for SDSS, we find that the mean of this correction to $\sigma_e$ is $\sim 2\%$, corresponding to the corrections for $\log\,\sigma_e$ with a mean of 0.008 dex and a standard deviation of 0.008 dex. Therefore, as in \cite{taylor2010}, it is safe to say this correction does not play an important role in our results.

We make the distinction between quiescent and star-forming galaxies based on the equivalent width of H$\alpha$ emission lines, which is taken to be W$_{\text{H}\alpha} < 1$\AA~for quiescent (Q) galaxies \citep[][]{howlett2022} and for quiescent (Q) galaxies W$_{\text{H}\alpha} \geqslant 1$\AA, which naturally divides the two populations seen in the $M_*$--W$_{\text{H}\alpha}$ diagram \citep[see][]{dogruel2023}.

A common practice in PV studies conducted with TFR and FP is to use the group redshifts $(z_\text{group})$, instead of individual galaxy redshifts, to calculate $R_e$ for galaxies that are members of galaxy groups/clusters. $z_\text{group}$ is usually taken to be the median redshift of individual group/cluster members. This practice is adopted for partially reducing the effects of nonlinear motion stemming from intra-group/cluster PVs \citep[e.g.,][]{hong2014, springob2014, howlett2017}. In this work though, we still use the individual CMB-frame redshifts for all galaxies in our sample so that we can then calculate the group/cluster averaged distances. It should be pointed out here that, in the light of \cite{calcino2017}, we use heliocentric\footnote{Strictly, one should use the redshifts as observed; i.e., geocentric not heliocentric. The difference depends on declination, latitude, and time of observation, but is at most 30 km/s; i.e., small.} redshifts for conversions from comoving distance $(D_C)$ to angular diameter $(D_A)$ and luminosity distances $(D_L)$. We obtain the group information from the data management units (DMUs) named \verb|G3CGalv10| and \verb|G3CFoFGroupv10| \citep[][]{robotham2011}, and we find that 1794 of the 2535 galaxies in our GAMA sample reside in 180 unique groups.

\subsection{Method}\label{sec:method}

\subsubsection{Bayesian framework}\label{sec:bayesian_summary}

An extensive analytical description of the model is given in \paperone. Briefly, we define an 8-dimensional space of galaxy properties, namely, $r\equiv\log R_e~[h^{-1}\text{kpc}]$, $s\equiv\log\sigma_e~\mathrm{[km/s]}$, $i\equiv \log\langle I_e\rangle[L_\odot/\text{pc}^2]$, $m^*\equiv\log M_\star[M_\odot]$, $\ell\equiv\log L[L_\odot]$, $c=(g-i)_\text{rest}$ rest-frame color, $\nu\equiv\log\,n$ and $m^d\equiv\log M_\text{dyn}[M_\odot]$. Then, in a Bayesian framework, we model the distribution of galaxies in this 8D parameter space, $\mathbf{\hat{y}}=(\bm{r}, \bm{s}, \bm{i}, \bm{m^*}, \bm{\ell}, \bm{c}, \bm{\nu}, \bm{m^d})$, as a Gaussian mixture model with two components: (i) the core model for the underlying distribution as an 8D Gaussian with mean $\bm{\bar{y}}$ and covariance matrix $\bm{\Sigma}$, convolved with observational errors: $\mathbf{\hat{y}_j} \sim \mathcal{N}(\bm{\bar{y}}, \bm{\Sigma} + \mathbf{E_j})$, (ii) the outlier model, in which outliers are parameterized as a fraction $(f_\text{bad} = 1 - f_\text{good})$ of $N$ data points emerging from a bad distribution with the same mean but with a different covariance matrix $(\bm{\Sigma}_\text{bad})$. This forms our parent model with the posterior,
\begin{align}
    \ln p(\bm{\bar{y}}, \bm{\Sigma}, f_\text{good} | \mathbf{\hat{y}}) \propto \nonumber \\
    \sum_{j=1}^N \text{log-sum-exp}&\biggl( \ln f_\text{good} \nonumber \\ 
    & + w_j [\ln p(\mathbf{\hat{y}_j}|\bm{\bar{y}}, \bm{\Sigma}+\mathbf{E_j}) - \ln \hat{f}_j], \nonumber \\ 
    & \ln(1-f_\text{good}) + \ln p(\mathbf{\hat{y}_j}|\bm{\bar{y}}, \bm{\Sigma}_\text{bad}) \biggr).
    \label{eq:wbc_logsumexp}
\end{align}
Here, $j$ subscripts denote each galaxy, $\mathbf{E_j}$ is the observational error matrix, $\hat{f}_j$ is the normalization factor that accounts for the selection cuts in $\bm{s}$ and $\bm{m^*}$. Following the definition of the selection function, $S_j$, from \cite{magoulas2012}, $w_j\equiv 1/S_j$ is the inverse weighting similar to $1/V_\text{max}$ \citep{schmidt1968} that is commonly used to account for the magnitude and redshift limit of the selected sample.
Strictly, the $1/V_\mathrm{max}$ formalism depends on the non-trivial assumptions of uniform distribution in space, no evolution across the redshift interval, and no subsets/outliers missed entirely. We note, however, that our GAMA sample is very nearly mass-limited, and so our $1/V_\mathrm{max}$ corrections are small: only 2\% of our sample have $w>1$, with a maximum value of 2.5.
The expression log-sum-exp refers to the log-sum of exponentials, which is $\ln(a+b)=\ln[\exp{(\ln a)} + \exp{(\ln b)}] = \text{log-sum-exp}(\ln a, \ln b)$, providing a computational convenience for sampling. Note that $\bm{\bar{y}}, \bm{\Sigma}, f_\text{good}$ and $\bm{\Sigma}_\text{bad}$ are declared as free parameters, with $\bm{\Sigma}_\text{bad}$ assumed to be diagonal. We use the software \textit{PySTAN}, the Python interface of \textit{STAN} \citep[][]{carpenter2017}, to perform MCMC sampling: samples are drawn from the posterior (equation~\ref{eq:wbc_logsumexp}) in 4 chains with each chain consisting of 1000 draws 500 of which are warm up, summing up to 2000 draws after discarding the warm ups. We apply this model to the separate samples of quiescent and star-forming galaxies independently (but see sections \ref{sec:traditionalfp} and \ref{sec:mp_for_all} where we model both samples combined as a single population).

\subsubsection{Validation and verification}

As described in Appendix \ref{sec:mocks}, we have validated our model by verifying our ability to robustly recover the known input parameters for many mock samples. The results of this exercise are summarized in Figure \ref{fig:1000mocks}: each panel shows the histogram of the fitted parameter from 1000 mock samples, with black continuous curve showing the best-fitting Gaussian to the histogram. Additionally, at the top of each panel, we give the input value and the fitted mean value along with the standard deviation which provides the error estimate of the relevant parameter.

Figure \ref{fig:1000mocks} shows that our 8D Gaussian model is statistically rigorous and provides a good description of the 8D-parameter space, recovering all 20 input parameters without major biases. We note that for cosmological applications, these simulations could be used to calibrate and correct for such biases, as in \citet{magoulas2012}, but we have not done so here.

\subsubsection{Constructing linear distance predictors}

In \paperone, an SED-independent stellar-mass proxy has been calibrated, which is denoted with $\hat{M}_\star$, using the dynamical mass estimator $(\sigma_e^2 R_e)$, S\'ersic index $(n)$ and rest-frame color $(g-i)_\text{rest}$ as a hyperplane in the form,
\begin{align}
    \log M_\star = \alpha_0\log(\sigma_e^2 R_e) + \alpha_1 &\log\sigma_e + \alpha_2\log n \nonumber \\ &+ \alpha_3(g-i)_\text{rest} + \alpha_4.
    \label{eq:mstar_hyp}
\end{align}
This can be readjusted using,
\begin{equation}
    \frac{M_\star}{L} L \propto k(n)\sigma_e^2 R_e~,
    \label{eq:mstar_virial}
\end{equation}
where $k(n)$ is the structure correction factor as a function of S\'ersic index \citep[][]{bertin2002, cappellari2006}. Dividing both sides of equation (\ref{eq:mstar_virial}) by $R_e^2$ gives
\begin{align}
    R_e \propto k(n)\sigma_e^2 \left(\frac{M_\star}{L}\right)^{-1} \left( \frac{L}{R_e^2} \right)^{-1}.
    \label{eq:mdyn_re}
\end{align}
Since $\log M_\star/L_i \propto (g-i)$ \citep[e.g.,][]{bell2003, zibetti2009, taylor2011}, $L/2\pi R_e^2 = \langle I_e \rangle$ and especially for $n \gtrsim 2$ as a first order approximation, $\log k(n) \propto -\log n$, equation~(\ref{eq:mdyn_re}) can be recast in log-space as,
\begin{equation}
    \log R_e = \beta_0 \log \sigma_e + \beta_1 \log\langle I_e \rangle + \beta_2 \log n + \beta_3 (g-i) + \beta_4
    \label{eq:mdyn_distance}
\end{equation}
which can be regarded as an extension and/or generalization of the FP, now tracking the model dependence of $k(n)$ and $M_\star/L$ through $\log n$ and the $(g-i)$ color, respectively. Henceforth, we will refer to equation~(\ref{eq:mdyn_distance}) as the mass hyperplane (MH)\footnote{Not to be confused with the stellar-mass fundamental plane (SMFP) in equation (\ref{eq:mfp}) or the formulation considered by \cite{cappellari2006}, \cite{degraaff2021} etc.}.

Following the same procedure in \paperone, we define a subspace $\bm{Y}=(\bm{r}, \bm{s}, \bm{i}, \bm{\nu}, \bm{c})$, which is obtained from our original 8D-parameter space, $\mathbf{\hat{y}}$, via the transformation $\bm{Y}=\bm{A} \mathbf{\hat{y}} + \bm{B}$. The best-fitting coefficients $\beta_i$ of equation~(\ref{eq:mdyn_distance}) can then be calculated from the mean vector and covariance matrix $(\bm{\bar{Y}}, \bm{\Sigma_Y})$ extracted from the parent model with $\bm{\Sigma_Y}=\bm{A}\bm{\Sigma}\bm{A}^\intercal$ and $\bm{\bar{Y}}=\bm{A}\bm{\bar{y}} + \bm{B}$. Note that deriving the coefficients from $\bm{\bar{Y}}$ and $\bm{\Sigma_Y}$ is equivalent to using conditional distributions $\bm{Y_a}\,|\,\bm{Y_b}$ obtained by partitioning $\bm{Y}$ as $(\bm{Y_a}, \bm{Y_b})$. In this case, the slopes, $\beta_j = \partial \bm{r} / \partial \bm{x}$ for each $\bm{x}\in \bm{Y_b}=(\bm{s}, \bm{i}, \bm{\nu}, \bm{c})$, are calculated from the covariance matrices of conditional distributions $(\bm{r}, \bm{x}\,|\,\bm{Y_b \setminus \{x\}})$. In \paperone, the slopes derived from such conditional distributions have been referred to as isolated trends and these characterize the variation in one parameter that can be uniquely tied to another parameter. The detailed calculations are given in section 4.2 of \paperone.

As seen in this framework, we parameterize our models by their mean vector and covariance matrices, from which we can then derive any other linear correlation as a form of post-processing. This is not just convenient but also efficient considering that, in MCMC applications, covariance matrices have a better sampling behavior \citep[e.g.,][]{dam2020} than a model parameterized with a 3D Gaussian but expressed via the plane coefficients, mean vector and scatters \citep[as done in][]{magoulas2012}, particularly when the zero-point is correlated with the slopes, e.g., $c=\Bar{r} - a\bar{s} - b\bar{i}$ for the FP.

\subsubsection{Best-fitting planes and ensuing errors in distances}\label{sec:fit_planes}

Redshift-independent distances and thus, peculiar velocities are calculated from the offset along the $r-$direction, when using the FP. \cite{magoulas2012} have shown that the distribution of galaxies about the FP in $r-$direction is not symmetrical when the underlying distribution is modeled with a 3D Gaussian, even though it provides a perfect empirical match. Their results imply that the plane which maximizes $p(r | s, i)$, that is the probability density distribution of $r$ at fixed observables $s$ and $i$, does not align with the principal axes of the 3D Gaussian. In this case, the coefficients that are derived by maximizing $p(r | s, i)$ give us the direct coefficients that minimize the residuals in $r-$direction, which are ideal for distance estimations as pointed out by \cite{bernardi2003c} (see, appendix \ref{sec:direct_vs_orth_planes}).

It is crucial to point out here that the direct coefficients are systematically and inescapably different (especially $a$) from the orthogonal coefficients that minimize the residuals perpendicular to the plane and that are derived from the eigenvector corresponding to the smallest eigenvalue of the covariance matrix, $\bm{\Sigma_\text{fp}}$.
This is because the two sets of coefficients answer two different questions.
Where the orthogonal coefficients give the best description of the `true', underlying relation within the data, the ordinary least squares (OLS) coefficients are the ones that give the best prediction (formally, the best linear unbiased estimator, or BLUE) for the `true' size of any given galaxy, given the data.
The same argument applies to the MH in equation~(\ref{eq:mdyn_distance}). Within our framework, it is surely possible to fit the 5D parameter space with a hyperplane using orthogonal distance regression, though it is not straightforward to visualize, to describe the underlying distribution of the galaxy population in parameter space. Notwithstanding, direct coefficients obtained via OLS provide the answer that we want for distance estimation, by providing the BLUE values for the size, $r$. Therefore, we work with the direct coefficients for both the traditional FP and the mass hyperplane.

Now the issue is to calculate the intrinsic scatter in $r-$direction $(\sigma_{r,\text{int}})$, which will propagate through the distances and peculiar velocities. As \cite{magoulas2012} and \cite{khaled2020} have shown, orthogonal coefficients $(a_\perp, b_\perp)$ cannot yield the actual scatter in distances. In fact, for the 3D Gaussian model, when the intrinsic orthogonal scatter about the plane ($\sigma_1=\sqrt{\lambda_1}$ where $\lambda_1$ is the smallest eigenvalue of $\bm{\Sigma_\text{fp}}$) is projected onto $r-$direction using $\sigma_{r,\text{int}} = \sigma_1 \sqrt{1 + a_\perp^2 + b_\perp^2}$, we heavily overestimate the scatter in distance. Throughout this work, we adopt the same approach as \paperone, for calculating the errors in distances. We define a projection vector using the direct coefficients as $\bm{P} = (1, -\beta_0, -\beta_1, -\beta_2, -\beta_3)$, then the observed (total) scatter can be calculated from,
\begin{align}
    \sigma_{r,\mathrm{tot}} &= \sqrt{\text{rms}_j \left(\bm{P (\Sigma_Y + \mathbf{E_{j,Y}}) P}^\intercal \right)} \nonumber \\ 
    &= \sqrt{\sigma_{r,\mathrm{int}}^2 + \sigma_\mathrm{err}^2} ,
    \label{eq:scatter}
\end{align}
where $\mathbf{E_{j,Y}}$ is the error matrix of the quantities in $\bm{Y}$ for the $j-$th galaxy and $\sigma_\mathrm{err}$ is the scatter due to the measurement uncertainties. Equation (\ref{eq:scatter}) means that $\sigma_{r,\mathrm{int}} = \sqrt{ \bm{P\Sigma_Y P}^\intercal }$ and $\sigma_\mathrm{err} = \sqrt{ \mathrm{rms}_j (\sqrt{ \bm{P} \mathbf{E_{j,Y}} \bm{P}^\intercal }) }$.

\section{Stellar-to-dynamical mass relation as a distance indicator}\label{sec:results_fp_mp}

In this section, we present the properties of the MH in comparison to the FP and discuss the scatter that will propagate through the distances derived from these planes. We also demonstrate what (dis)advantages arise under different treatments of galaxy samples (combining or separating Q and SF populations). Finally, we investigate the possible sources of systematics/biases for both the FP and the MH, by studying their residual trends with several other galaxy properties.

In finding the best-fitting planar relations, we pursue an approach slightly different from what has been done before. For instance, \cite{degraaff2021} have used the FP slopes from \cite{hyde2009} that minimize the orthogonal residuals, then they have shown that star-forming galaxies lie on the same FP as the quiescent galaxies, but with a different zero-point and a larger scatter around the plane. In other words, the FP slopes, naturally, have not been fitted by also including the SF galaxies in the sample.
In this work, however, we calibrate the slopes of both the FP and the MH, first from the models independently applied to the separated Q and SF samples, then we repeat the procedure using the model applied to the entire galaxy sample (i.e., the combined sample of SFs and Qs, with $N=2535$). A point worth noting here is that the astrophysical implications of this treatment for galaxy formation and evolution are beyond the scope of this paper, but clearly worth further study in a future work.

\subsection{The traditional FP}\label{sec:traditionalfp}

As summarized in section \ref{sec:bayesian_summary}, we can easily acquire the FP fit from our parent model via a linear transformation, which results in the model of the FP space, $\bm{x}\equiv(r, s, i)$, as a 3D Gaussian, i.e., $\bm{x} \sim \mathcal{N}(\bm{\bar{x}}, \bm{\Sigma}_\mathrm{fp})$.

Following \cite{magoulas2012} and \cite{khaled2020}, a rough estimate for the total scatter around the FP in $r-$direction can be obtained from the orthogonal coefficients through,
\begin{equation}
    \sigma_{r,\mathrm{tot}}^\dagger = \left[\epsilon_r^2 + (a_\perp \epsilon_s)^2 + (b_\perp \epsilon_i)^2 + (\sigma_{r,\mathrm{int}}^\dagger)^2\right]^{1/2}~,
    \label{eq:wrong_scatter}
\end{equation}
where $\epsilon_{r,s,i}$ are the rms of the uncertainties in $r, s$ and $i$ respectively, and $\sigma_{r,\mathrm{int}}^\dagger$ is the intrinsic scatter in the $r-$direction projected from the intrinsic orthogonal scatter about the plane, as discussed in section \ref{sec:fit_planes}.
We note that this estimate is conservative, in that it is assumes no unmodelled sources of error.

\begin{figure*}
    \gridline{ \fig{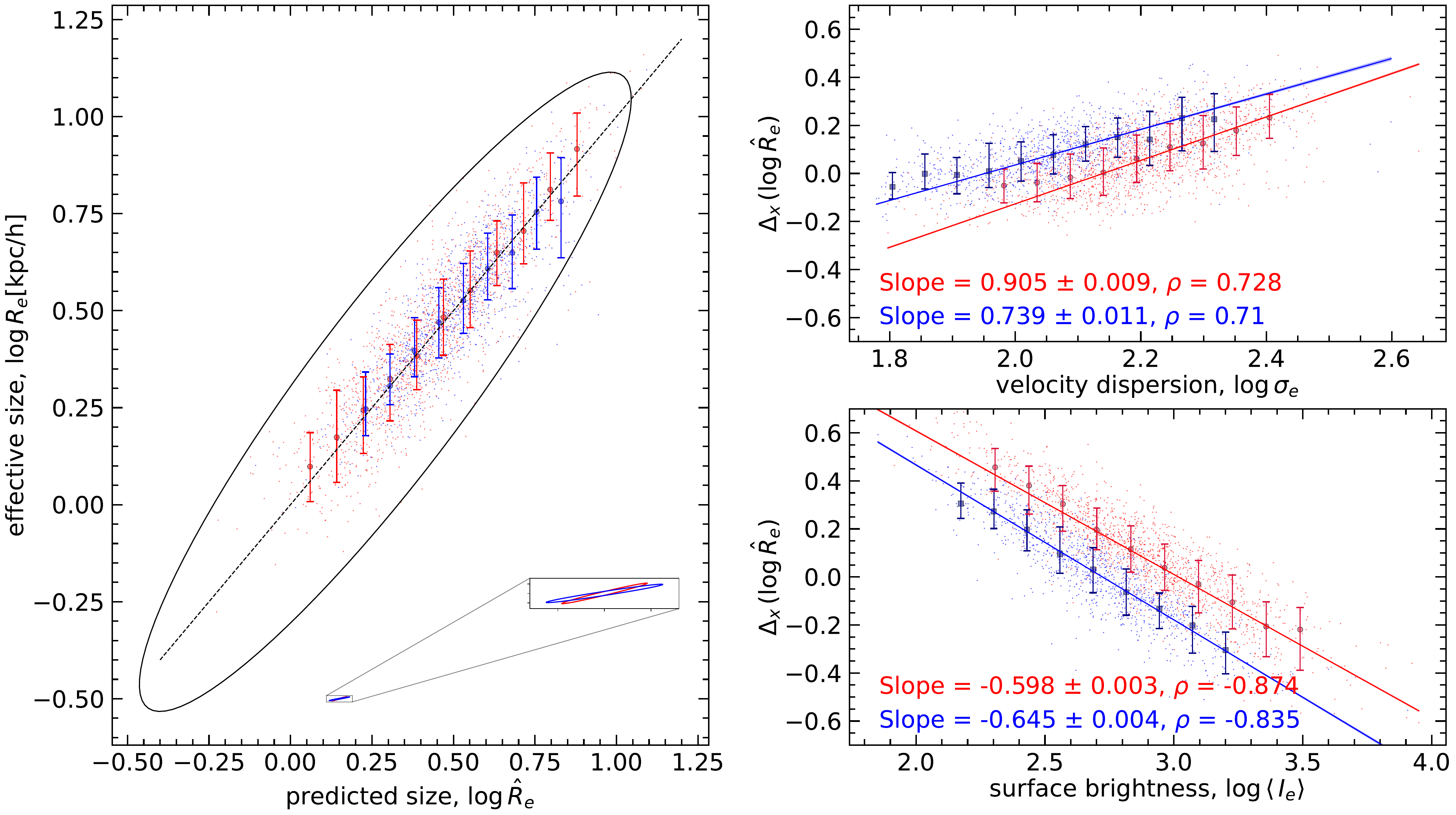}
                {0.78\textwidth}{(a) Separate and independent modeling of Q and SF galaxy samples} 
            }
    \gridline{\fig{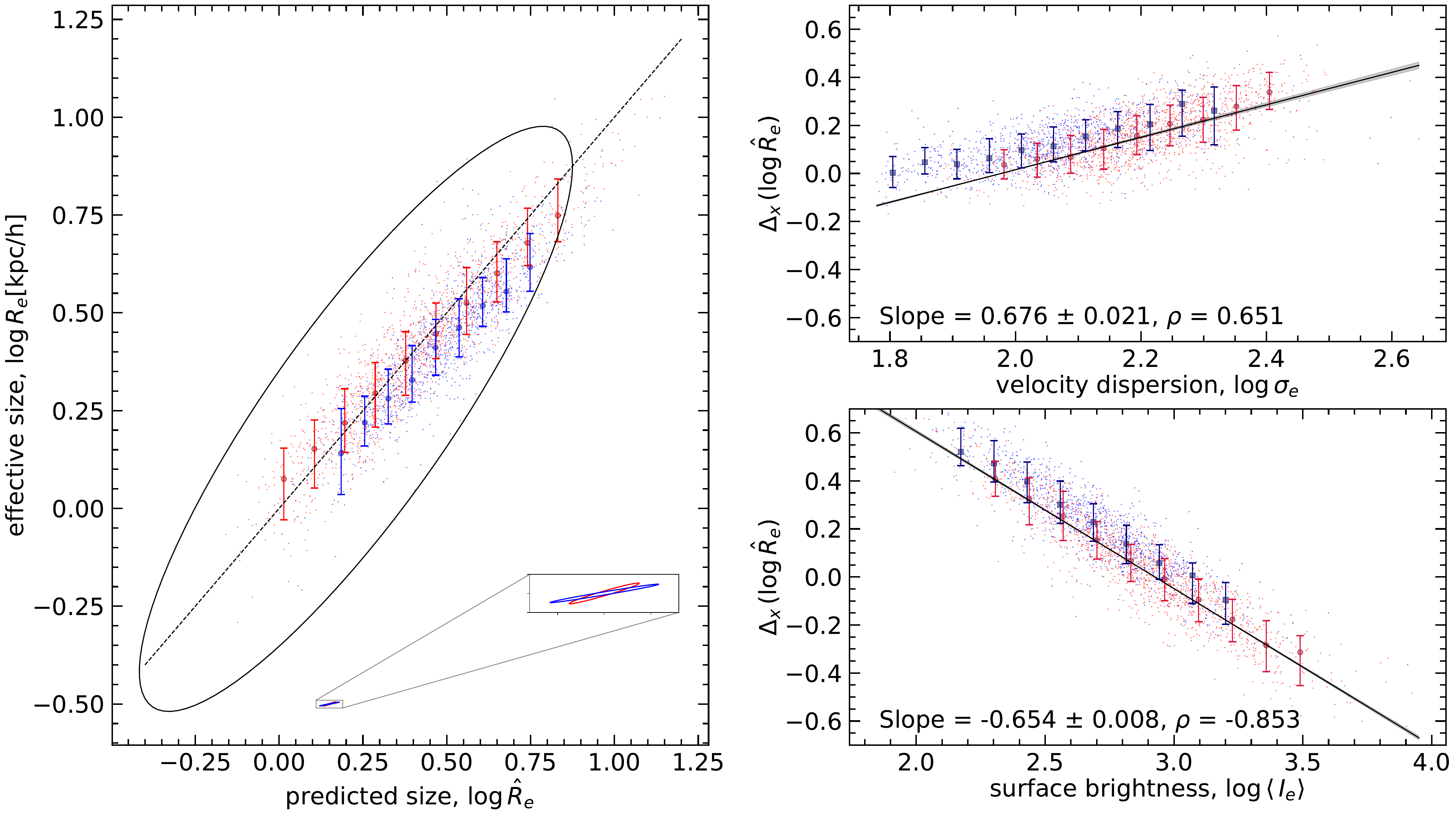}
                {0.78\textwidth}{(b) Modeling the combined sample of Q and SF galaxies}
            }
    \caption{Fit to the traditional FP of GAMA galaxies in Z-band, minimizing the residuals in $r-$direction. Blue and red colors represent the star-forming and quiescent galaxies respectively. Points show the data while the large blue squares and red circles show the median, and the error bars show the 16/84 percentiles of the $y$-axis parameter in bins of the $x$-axis parameter. The fits are derived from our parent model applied (a) to the separate samples of SF and Q galaxies, and (b) to the combined sample. \textit{Left-hand panels}: Comparison of the observed effective radii to the ones predicted from the FP, along with the underlying 3$\sigma$ Gaussian distribution shown with the ellipse. Zoomed region shows the median error ellipses. Dashed line is the one-to-one relation. \textit{Right-hand panels}: Isolated trends showing the slopes of each component of the plane. Shaded regions around the lines show the $1\sigma$ uncertainty in the relevant slope. $\rho$ in these panels represent the true correlation between $R_e$ and the $x-$axis quantity, with every other quantity fixed in the parameter space. The results are normalized such that the mean value of the $\Delta$ for each sample is zero. Since the slopes are similar, the apparent offset between the two relations primarily reflects differences in the mean values of the $x$ quantity for the two samples.}
    \label{fig:lfp_trends}
\end{figure*}

In Figure \ref{fig:lfp_trends}, we present the FP of GAMA galaxies in the Z-band and the corresponding fit derived from the model in which the quiescent and star-forming galaxies are treated separately and independently. While the left-hand panel shows the observed effective radii plotted against the ones predicted from direct fits to the FP, the right-hand panels show the isolated trends of velocity dispersion and surface brightness. In the same way, Figure \ref{fig:lfp_trends} shows the fit results from the model obtained when galaxies are treated together.

A more detailed look into these fit results is provided in Table \ref{tab:lfp_fits}, in which we also include the orthogonal coefficients $(a_\perp, b_\perp, c_\perp)$, the conservative scatter estimates $(\sigma_r^\dagger)$, the intrinsic orthogonal scatter about the FP $(\sigma_1)$ and the rms of uncertainties $(\epsilon_r, \epsilon_s, \epsilon_i)$ for each FP observable, $r, s$ and $i$ respectively.

In Figure \ref{fig:lfp_trends}, we see that Q and SF galaxies in the local Universe are on the same FP, as expected from the results of \cite{bezanson2015} and \cite{degraaff2021}. Additionally, we further verify these results in a slightly different way and show in Figure \ref{fig:lfp_trends} that tighter FP relations can be achieved for Qs and SFs separately.

Following the arguments on the scatter of the FP by \cite{magoulas2012} and \cite{khaled2020}, the true distance error is proportional to $\sqrt{(\sigma_r^\dagger)^2 - \sigma_{r, \mathrm{int}}^2}$. Hence, Table \ref{tab:lfp_fits} suggests that, in principle, we might expect the PV/distance errors for our data set to be $\sim$0.05 dex, when the traditional FP is used. This value is similar to the estimate that can be derived from \cite{khaled2020}, where $\sigma_r^\dagger=0.099$ and $\sigma_{r, \mathrm{int}}=0.089$ dex, thus the relative distance/PV errors are expected to be $\sim$0.04 dex.

As seen in Table \ref{tab:lfp_fits}, when the Q and SF galaxies are treated together, scatters $\sigma_1, \sigma_{r,\mathrm{int}}, \sigma_r^\dagger$ are all larger than the case of when Q and SF are treated separately and individually. Furthermore, it is interesting to see that under separate and individual modeling, the FP has a $\sim 6\%$ smaller intrinsic scatter in the $r-$direction for the SF population, whereas the orthogonal intrinsic scatter about the plane seems to be the same for both populations at 0.069 dex. However, the total scatters for the SF are always larger than the ones for Q, due to SF galaxies having velocity dispersions with significantly larger uncertainties ($\epsilon_s = 0.043$ and 0.062 for Qs and SFs respectively), while the uncertainties in both size and surface brightness are similar for both populations. Nevertheless, both populations having the same $\sigma_1$ when they are modeled independently and separately, may seem to be contradictory to \cite{degraaff2021}, who have found that the larger observed scatter seen for the SFs implies a larger intrinsic scatter because SFs and Qs in their LEGA-C sample have similar uncertainties. Though, as we stated in the beginning of section \ref{sec:results_fp_mp}, it should be considered here that \cite{degraaff2021} have used the slopes from \cite{hyde2009} obtained through fitting exclusively quiescent galaxies, whereas we calibrated these slopes by also including star-forming galaxies, an approach which historically has not been adopted in FP studies.

\subsection{The mass hyperplane for quiescent and star-forming galaxies}\label{sec:mp_for_q_and_sf}

The mass hyperplane, MH, in equation (\ref{eq:mdyn_distance}) derived from the stellar-to-dynamical mass $(M_\star/M_\mathrm{dyn})$ relation can be regarded as an enhanced FP, now accounting for physical differences between galaxies via S\'ersic index $(n)$ and rest-frame color $(g-i)_\mathrm{rest}$; two of the commonly used properties that separate galaxy populations into two broad classes of early/late-types or ellipticals/spirals, etc. \citep[e.g.,][]{blanton2009, lange2015}. Natural inclusion of these common separators within the $M_\star/M_\mathrm{dyn}$ relation is in fact crucial. For example, \cite{degraaff2021} have stated that the physical differences between galaxies are likely to dominate the observed scatter of the FP when both Q and SF galaxies are considered. Thus, the MH here may be expected to alleviate the uncertainties attributed to these differences. Obviously the easiest way to test this expectation is to compare the scatters of the MH and FP. 

\begin{figure*}
    \gridline{ \fig{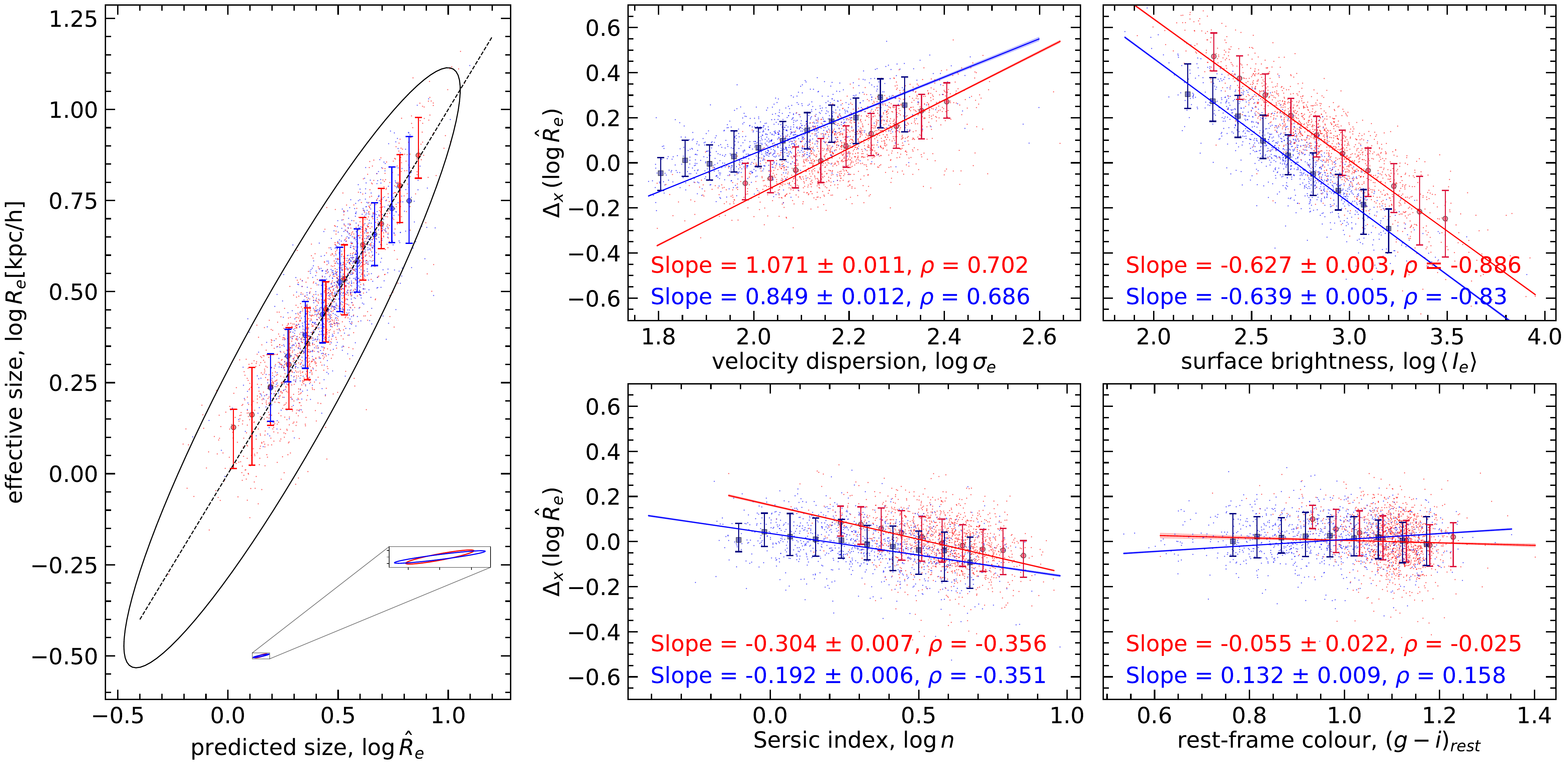}
                {0.98\textwidth}{(a) Separate and independent modeling of Q and SF galaxy samples} 
            }
    \gridline{\fig{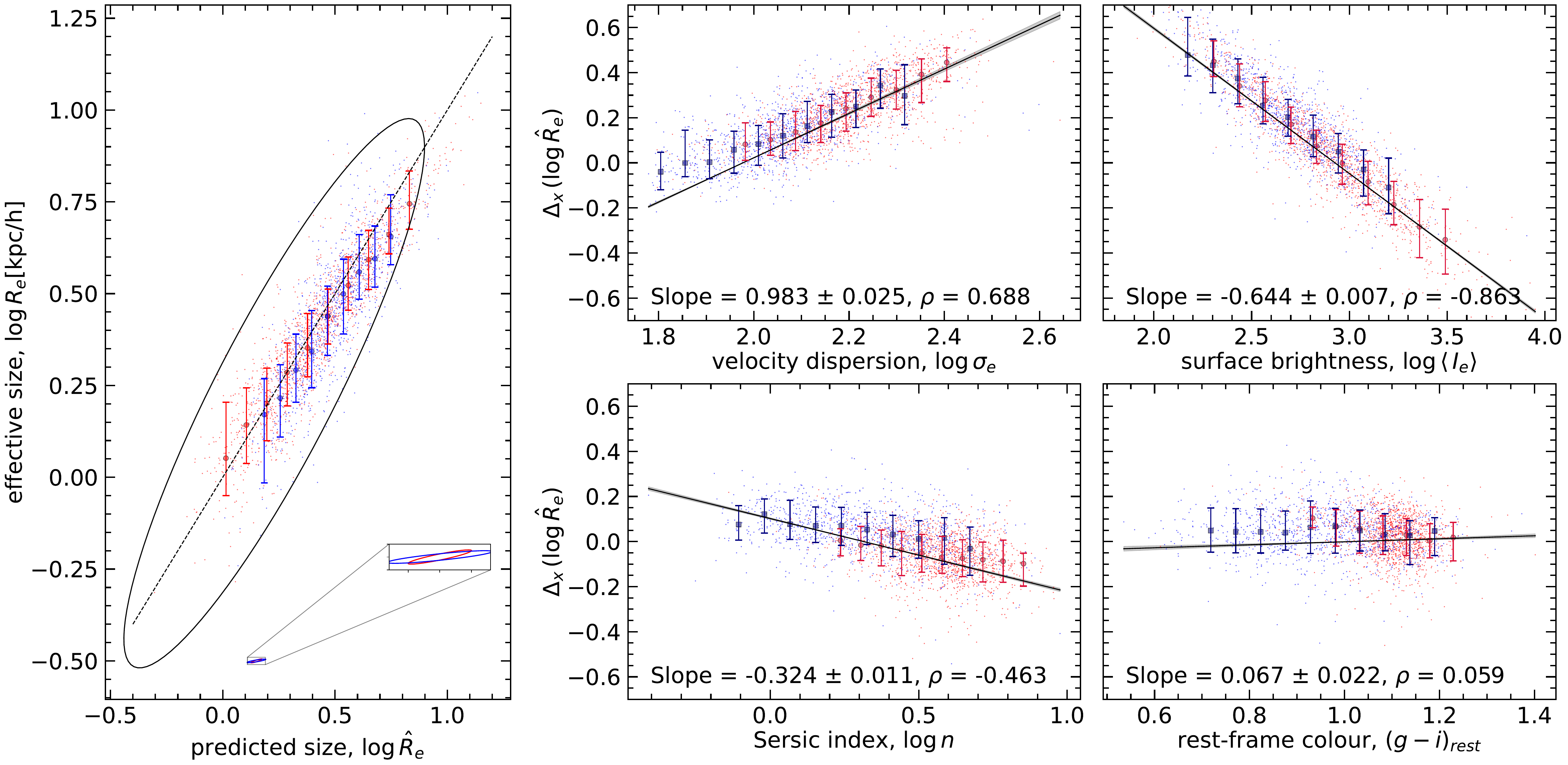}
                {0.98\textwidth}{(b) Modeling the combined sample of Q and SF galaxies}
            }
    \caption{Same as Figure \ref{fig:lfp_trends} but for the mass hyperplane.}
    \label{fig:lfpnc_trends}
\end{figure*}

In Figure \ref{fig:lfpnc_trends}, we show the MH of Q and SF galaxies when they are modeled separately and independently. As in Figure \ref{fig:lfp_trends}, the left hand panel shows the comparison between the observed and predicted $\log\,R_e$, while the right hand panels show the isolated trends of $\sigma_e, \langle I_e\rangle, n$ and $(g-i)_\mathrm{rest}$. We present the more detailed fitting results in Table \ref{tab:lfpnc_fits}.

A comparison between Figures \ref{fig:lfp_trends} and \ref{fig:lfpnc_trends} reveals that the correlations between the pairs $(\bm{r}, \bm{s})$ and $(\bm{r}, \bm{i})$ have significantly changed with the inclusion of $\bm{\nu}\equiv \log n$ and $\bm{c}\equiv(g-i)_\mathrm{rest}$. As explained in more detail in \paperone, this is because in the case of FP, these correlations include the contributions from the interrelations of galaxy properties e.g., $\bm{r}-\bm{\nu}$, $\bm{s}-\bm{c}$, $\bm{i}-\bm{\nu}$.

Unsurprisingly, S\'ersic index plays quite an important role for both populations, despite its shallower slope and lower correlation with $\bm{r}$, compared to the ones of $\bm{s}$ and $\bm{i}$ with $\bm{r}$. On the other hand, $(g-i)_\mathrm{rest}$ does not seem to have a noteworthy role even for the SF population which covers a much wider range in color than the Q population.
This suggests that the shift from luminosity and the FP to stellar mass and the MH captures a large part of the stellar population effects that are seen to be correlated with FP residuals \citep[see, e.g.,][but see also Figure \ref{fig:sps_residuals_fp}]{graves2010_3, springob2012}.

As for the scatter of the plane in the $r-$direction, which is the main parameter of interest concerning the distance errors, it can be seen by comparing Table \ref{tab:lfpnc_fits} to Table \ref{tab:lfp_fits} that the MH reduces $\sigma_{r,\mathrm{int}}$ from 0.109 dex to 0.102 dex ($\sim 6$\% decrease) for Qs and from 0.103 dex to 0.095 dex ($\sim 8$\% decrease) for SFs. This is promising since it implies that we should be able to see an improvement in the precision of the distances from the MH by $\sim 6$\% and $\sim 8$\% for Qs and SFs, respectively, compared to the FP, when galaxy populations are modeled separately and independently.

\subsection{The mass hyperplane for all galaxies}\label{sec:mp_for_all}

Following the seemingly promising improvement obtained with the MH in section \ref{sec:mp_for_q_and_sf}, we now turn our attention to fitting the MH when the combined sample of Q and SF galaxies is modeled. We present the resulting MH for this case in Figure \ref{fig:lfpnc_trends}, in the same fashion as Figure \ref{fig:lfp_trends}, and give further details of the MH parameters in Table \ref{tab:lfpnc_fits}. At first sight, these results show that the isolated trends are similar to the case of separate and individual treatment. However, as seen for the FP in section \ref{sec:traditionalfp}, when the combined galaxy sample is treated as one single population, both intrinsic and total $\sigma_r$ slightly increase while the correlation between observed and predicted $\bm{r}$ slightly decreases. This is expected because the inclusion of SF galaxies which have larger uncertainties in $\bm{s}$ (at least in our sample) will naturally introduce more scatter to the plane. 
Nevertheless, the value we get when fitting the MH for all galaxies is still just 0.11 dex, as compared to 0.124 dex for the FP for quiescent galaxies only.
That is, the limiting precision for MH-derived distances is $\sim$ 10\% better for MH relative to the traditional FP, but with the significant advantage of much larger sample sizes through the inclusion of star-forming galaxies (see Section \ref{sec:mp_vs_fp} for further discussion).

\begin{table*}[h!]
    \centering
    \caption{Properties of the FP for GAMA galaxies in Z-band. Listed uncertainties correspond to 1$\sigma$ confidence intervals derived from MCMC chains.}
   \begin{tabular}{lcccc}
    \cmidrule{2-5}     & \multicolumn{2}{c}{modeling the combined sample} & \multicolumn{2}{c}{modeling individual samples} \\
       \midrule
       Parameter & Quiescent & Star-forming & Quiescent & Star-forming \\
       \midrule
       \midrule
       $N$ & \multicolumn{2}{c}{2535} & 1489 & 1046 \\
       $a$ & \multicolumn{2}{c}{0.676$\pm$0.021} & 0.905$\pm$0.009 & 0.739$\pm$0.011 \\
       $b$ & \multicolumn{2}{c}{-0.654$\pm$0.008} & -0.598$\pm$0.003 & -0.645$\pm$0.004 \\
       $c$ & \multicolumn{2}{c}{0.805$\pm$0.042} & 0.158$\pm$0.02 & 0.73$\pm$0.02 \\
       $a_\perp$ & \multicolumn{2}{c}{0.984$\pm$0.028} & 1.257$\pm$0.012 & 1.02$\pm$0.015 \\
       $b_\perp$ & \multicolumn{2}{c}{-0.74$\pm$0.008} & -0.599$\pm$0.003 & -0.755$\pm$0.005 \\
       $c_\perp$ & \multicolumn{2}{c}{0.448$\pm$0.053} & -0.594$\pm$0.028 & 0.481$\pm$0.026 \\
       $\sigma_{r,\mathrm{int}}$ & \multicolumn{2}{c}{0.124$\pm$0.002} & 0.109$\pm$0.001 & 0.103$\pm$0.001 \\
       $\rho$ & \multicolumn{2}{c}{0.882$\pm$0.006} & 0.916$\pm$0.002 & 0.848$\pm$0.003 \\
       $\sigma_1$ & \multicolumn{2}{c}{0.085$\pm$0.001} & 0.069$\pm$0.0004 & 0.069$\pm$0.0005 \\
       $\sigma_{r,\mathrm{tot}}$ & 0.127$\pm$0.022 & 0.131$\pm$0.022 & 0.116$\pm$0.013 & 0.113$\pm$0.014 \\
       $\sigma_\mathrm{err}$ & 0.029 & 0.042 & 0.04 & 0.047 \\
       $\sigma_{r,\mathrm{int}}^\dagger$ & 0.134 & 0.134 & 0.118 & 0.111 \\
       $\sigma_{r,\mathrm{tot}}^\dagger$ & 0.141 & 0.148 & 0.13 & 0.128 \\
       $\epsilon_r$ & 0.008 & 0.006 & 0.008 & 0.006 \\
       $\epsilon_s$ & 0.043 & 0.062 & 0.043 & 0.062 \\
       $\epsilon_i$ & 0.015 & 0.013 & 0.015 & 0.013 \\
       \bottomrule
   \end{tabular}%
    \label{tab:lfp_fits}
\end{table*}

\begin{table*}[h!]
  \centering
  \caption{Properties of the mass hyperplane in the Z-band. Properties regarding the orthogonal fits (orthogonal coefficients, $\sigma_1, \sigma_{r, \mathrm{int}}^\dagger$ and $\sigma_{r, \mathrm{tot}}^\dagger)$ are omitted.}
   \begin{tabular}{lcccc}
    \cmidrule{2-5}     & \multicolumn{2}{c}{modeling the combined sample} & \multicolumn{2}{c}{modeling individual samples} \\
       \midrule
       Parameter & Quiescent & Star-forming & Quiescent & Star-forming \\
       \midrule
       \midrule
       $\beta_0$ & \multicolumn{2}{c}{0.983$\pm$0.025} & 1.071$\pm$0.011 & 0.849$\pm$0.012 \\
       $\beta_1$ & \multicolumn{2}{c}{-0.644$\pm$0.007} & -0.627$\pm$0.003 & -0.639$\pm$0.005 \\
       $\beta_2$ & \multicolumn{2}{c}{-0.324$\pm$0.011} & -0.304$\pm$0.007 & -0.192$\pm$0.006 \\
       $\beta_3$ & \multicolumn{2}{c}{0.067$\pm$0.022} & -0.055$\pm$0.022 & 0.132$\pm$0.009 \\
       $\beta_4$ & \multicolumn{2}{c}{0.205$\pm$0.046} & 0.113$\pm$0.022 & 0.411$\pm$0.023 \\
       $\rho$ & \multicolumn{2}{c}{0.895$\pm$0.004} & 0.927$\pm$0.002 & 0.873$\pm$0.003 \\
       $\sigma_{r,\mathrm{int}}$ & \multicolumn{2}{c}{0.11$\pm$0.002} & 0.102$\pm$0.001 & 0.095$\pm$0.001 \\
       $\sigma_{r,\mathrm{tot}}$ & 0.118$\pm$0.02 & 0.126$\pm$0.021 & 0.112$\pm$0.012 & 0.109$\pm$0.014 \\
       $\sigma_\mathrm{err}$ & 0.042 & 0.061 & 0.047 & 0.054 \\
       \bottomrule
   \end{tabular}
  \label{tab:lfpnc_fits}
\end{table*}

\subsection{Possible systematics}\label{sec:systematics}

In the preceding two sections, \ref{sec:mp_for_q_and_sf} and \ref{sec:mp_for_all}, we have shown quantitatively how the MH is a tighter linear relation that is potentially capable of providing redshift-independent distances with slightly improved precision compared to the FP. In this section, we investigate whether this improvement comes with any biases and/or systematic issues.

To do this, we adopt the same approach as \paperone ~and analyze the residuals of both the MH and the FP as a function of several galaxy properties: 
\begin{enumerate*}[label=(\roman*)]
    \item projected axis ratio, $q=b/a$,
    \item H$\alpha$ and H$\delta$ line equivalent widths,
    \item 4000\AA~break strength, $D_n 4000$,
    \item dust attenuation, $E(B-V)$,
    \item specific star formation rate, sSFR $=$ SFR$/M_\star$,
    \item luminosity-weighted mean stellar age, $\log \langle t_\star \rangle_\mathrm{lw}$, and,
    \item metallicity, $\log Z_\star/Z_\odot$.
\end{enumerate*}
Using \textit{STAN}, we model the residuals $(\Delta\bm{r}\equiv\log R_e/\hat{R}_e)$ as a function of each of these properties as a 2D Gaussian, then derive the best-fitting relation from this model in the form $y=Ax+B+\mathcal{N}(0,\sigma)$ where $\sigma$ is the Gaussian scatter.

\begin{figure}
    \centering
    \includegraphics[width=0.47\textwidth]{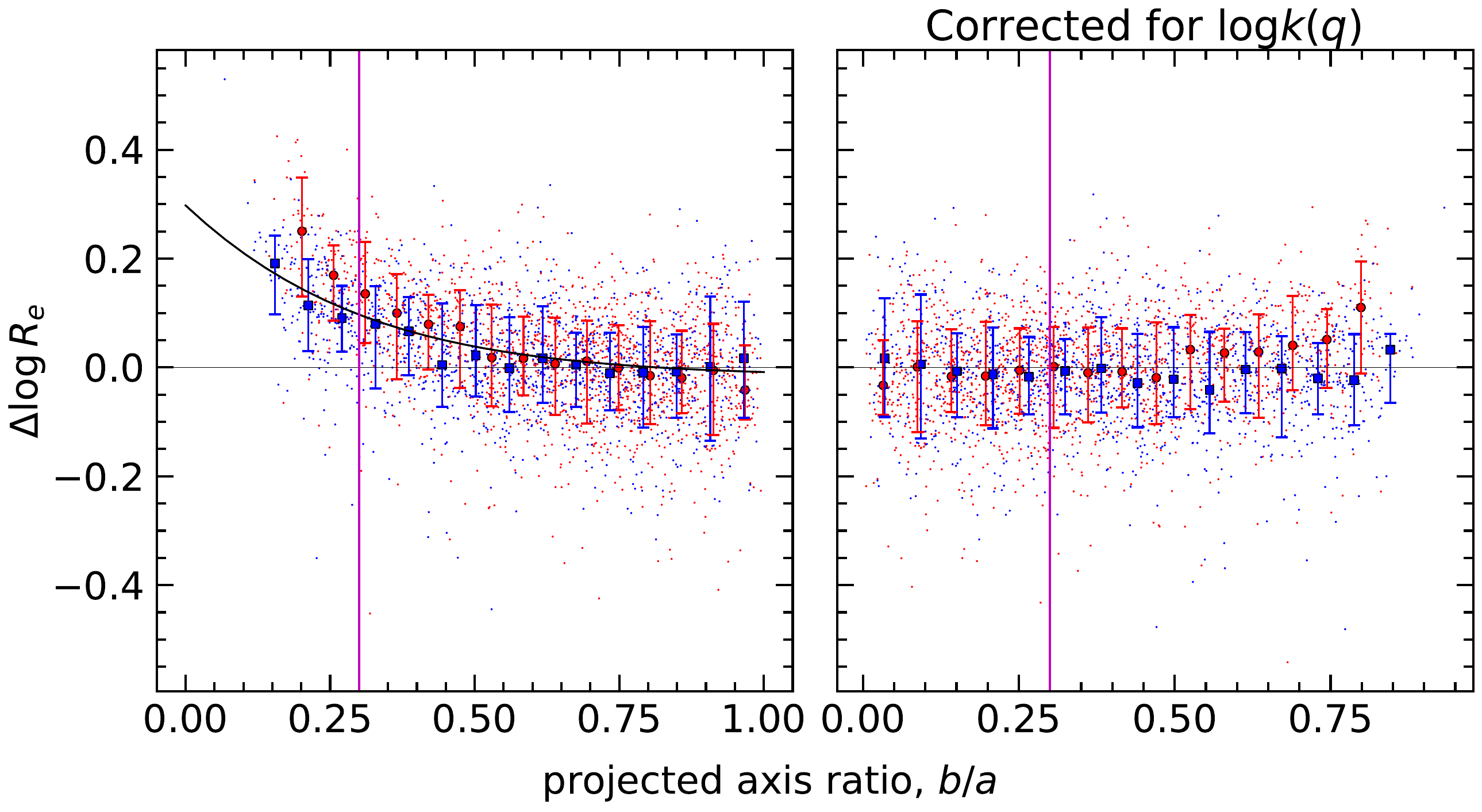}
    \caption{Residuals in the $r-$direction as a function of axis ratio with symbols and colors being the same as Figure \ref{fig:lfp_trends}. Left-hand panel shows the residuals before the $k(q)$ correction was applied to $R_e$. Here, the smooth black curve show the $\log k(q)$ from \cite{wel2022}. Right-hand panel shows the residuals corrected for $k(q)$. The vertical magenta line shows the $q=0.3$ lower limit adopted for our sample selection.}
    \label{fig:kqcorrection}
\end{figure}

Starting from the axis ratio, Figure \ref{fig:kqcorrection} shows that the residuals have a curved trend with $q$, which follows the $k(q)$ correction factor empirically calibrated by \cite{wel2022}, just as shown in Figure 7 of \paperone. This is particularly prevalent for $q<0.5$, which if adopted as the lower limit for $q$ will reduce the sample size by $\sim$600 galaxies. However, if we simply apply this correction as $\log R_{e,\mathrm{corr}} = \log [R_e/k(q)]$, the residuals flatten as shown in the right-hand panel of Figure \ref{fig:kqcorrection}. At face value, this suggests that with the inclusion of the \cite{wel2022} prescription as a $q$-dependent correction to account for the variations in the amount of rotation versus dispersion measured by $\sigma$, we could do away with any $q$ selection. However, to be conservative, we continue to adopt the lower limit of $q=0.3$ as in \cite{howlett2022}.

\begin{figure*}
    \gridline{ 
            \fig{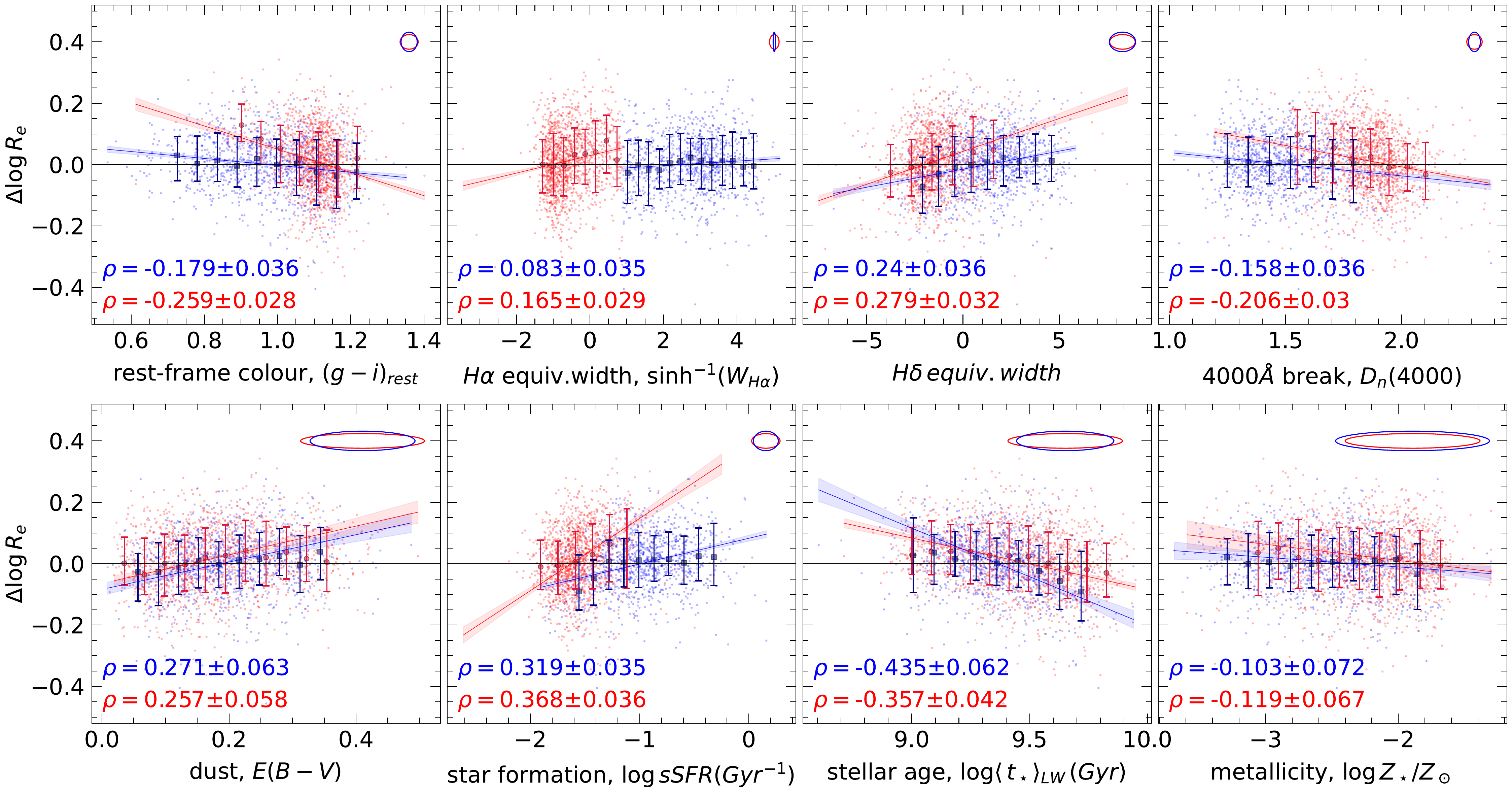}{0.95\textwidth}{(a) Separate and independent modeling of Q and SF populations} 
            }
    \gridline{ 
            \fig{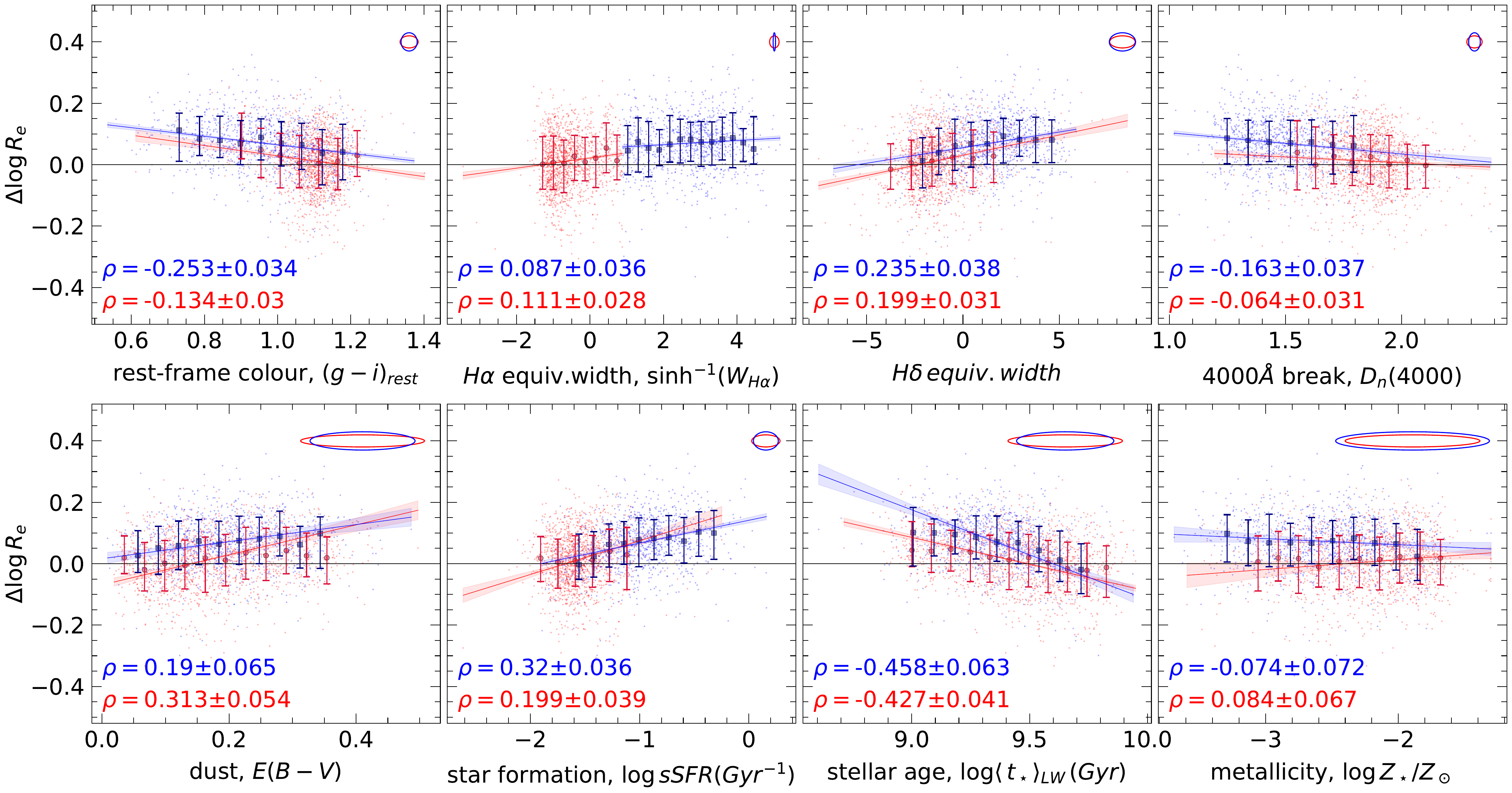}{0.95\textwidth}{(b) Modeling the combined sample of Q and SF populations} 
            }
    \caption{FP residuals in the $r-$direction as a function of rest-frame color, spectral indices (H$\alpha$, H$\delta$ and $D_n 4000$) and stellar population parameters $E(B-V), \log s\mathrm{SFR}, \log\langle t_\star\rangle_\mathrm{LW}$, and $\log Z_\star/Z_\odot$ for (a) when Q and SF populations are modeled separately and independently, and (b) when the combined sample of Q and SF is modeled. The first row of each figure contains direct observables and the second row contains ancillary SP parameters derived through SED-fitting. Symbols and colors are the same as Figure \ref{fig:lfp_trends}. In each panel, smooth lines show the best-fitting line derived from the 2D Gaussian model, shaded regions around the lines correspond to 1$\sigma$ uncertainty in the fit, ellipses on the upper right corner represent the median uncertainties of the data, and correlations are denoted with $\rho$.}
    \label{fig:sps_residuals_fp}
\end{figure*}

\begin{figure*}
    \gridline{ 
            \fig{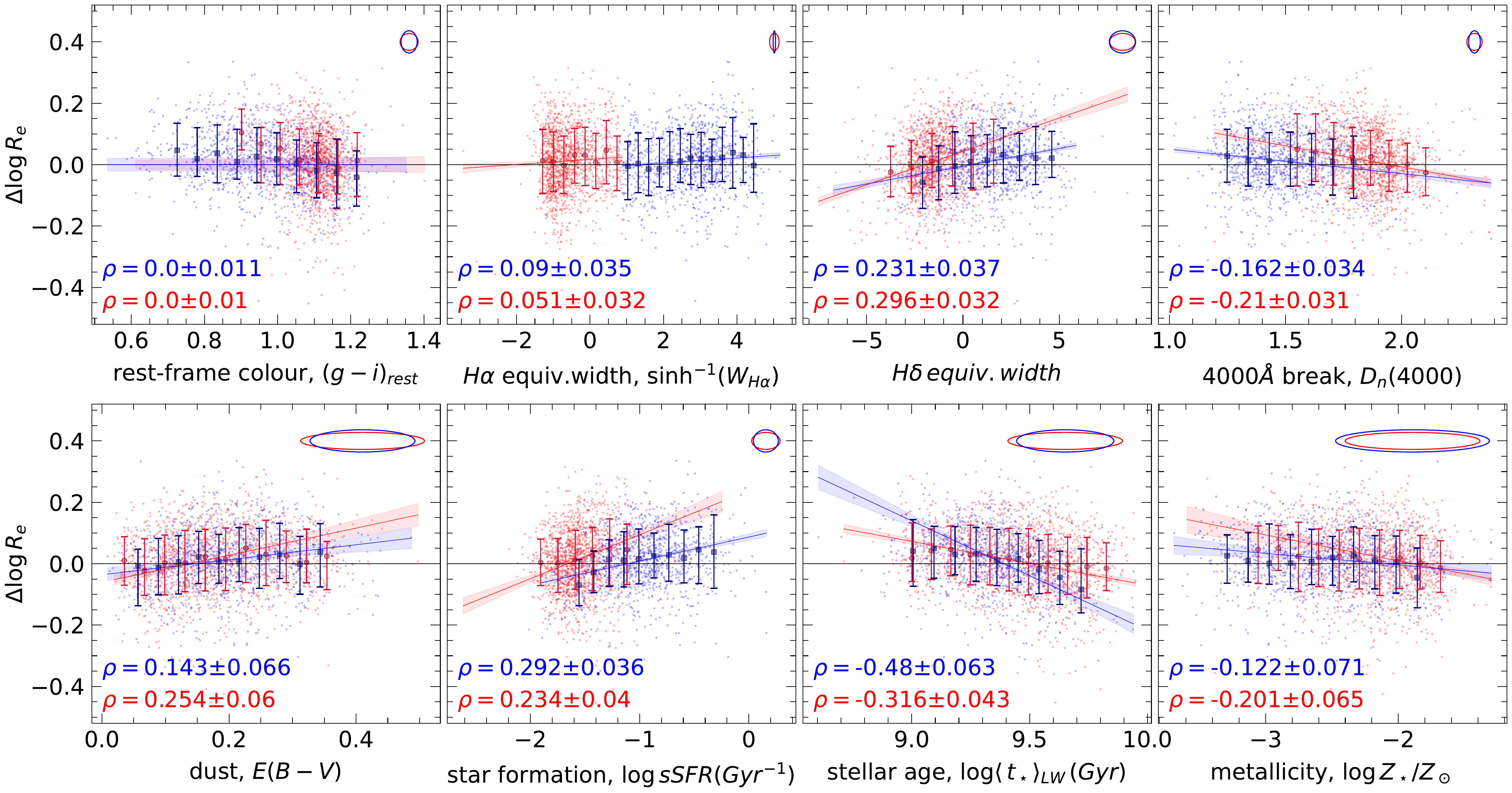}{0.98\textwidth}{(a) Separate and independent modeling of Q and SF populations} 
            }
    \gridline{ 
            \fig{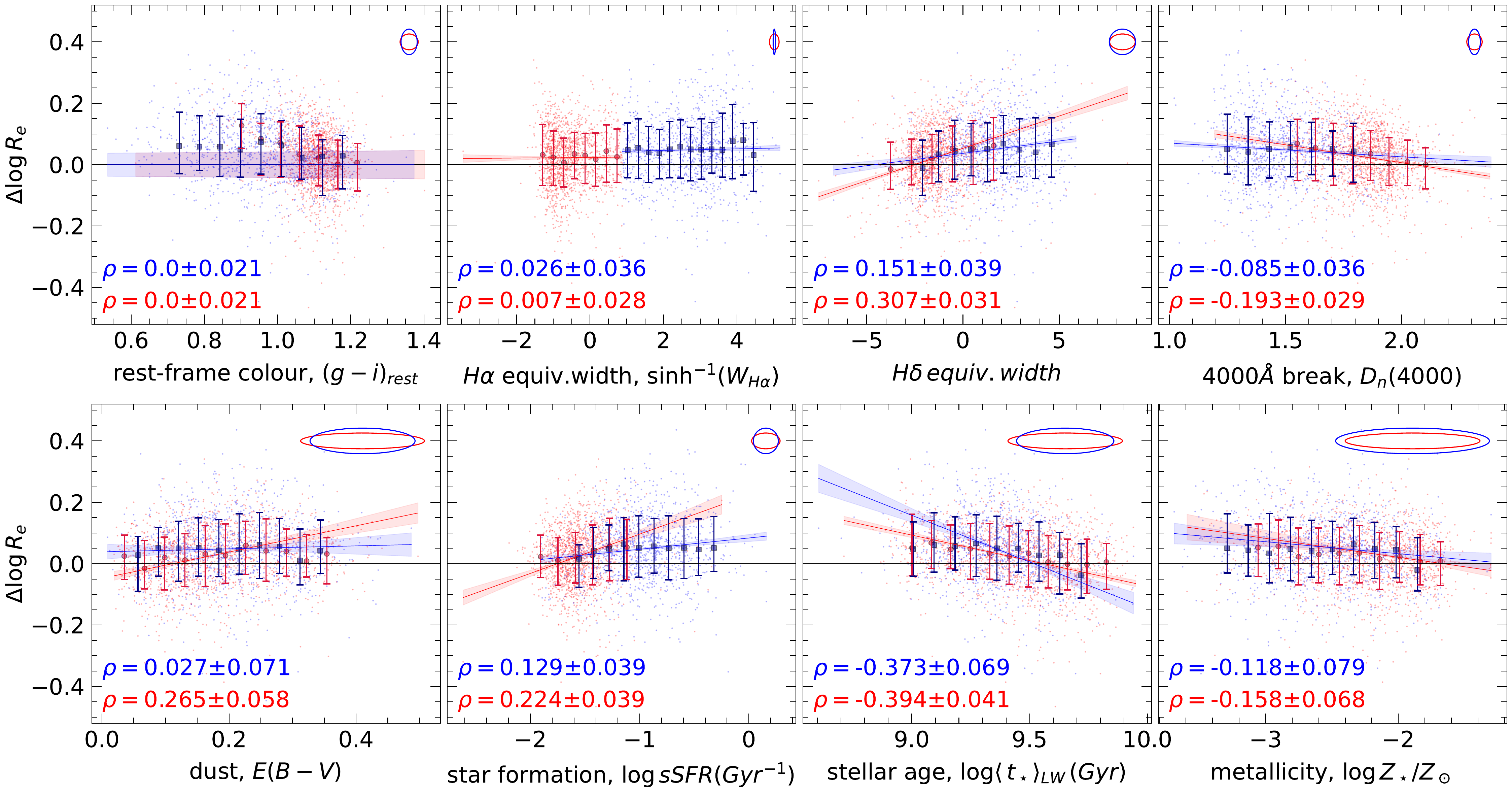}{0.98\textwidth}{(b) Modeling the combined sample of Q and SF populations} 
            }
    \caption{Same as Figure \ref{fig:sps_residuals_fp} but for MH residuals in the $r-$direction.}
    \label{fig:sps_residuals_mh}
\end{figure*}

We summarize the results for the modeling of the combined sample in Figures \ref{fig:sps_residuals_fp} and \ref{fig:sps_residuals_mh}, which show that both FP and MH have similar trends, indicating practically the same systematics with stellar population (SP) parameters. If anything, the MH provides slight reductions in most of these residual correlations compared to the FP.

\begin{figure}[htbp]
    \gridline{
        \fig{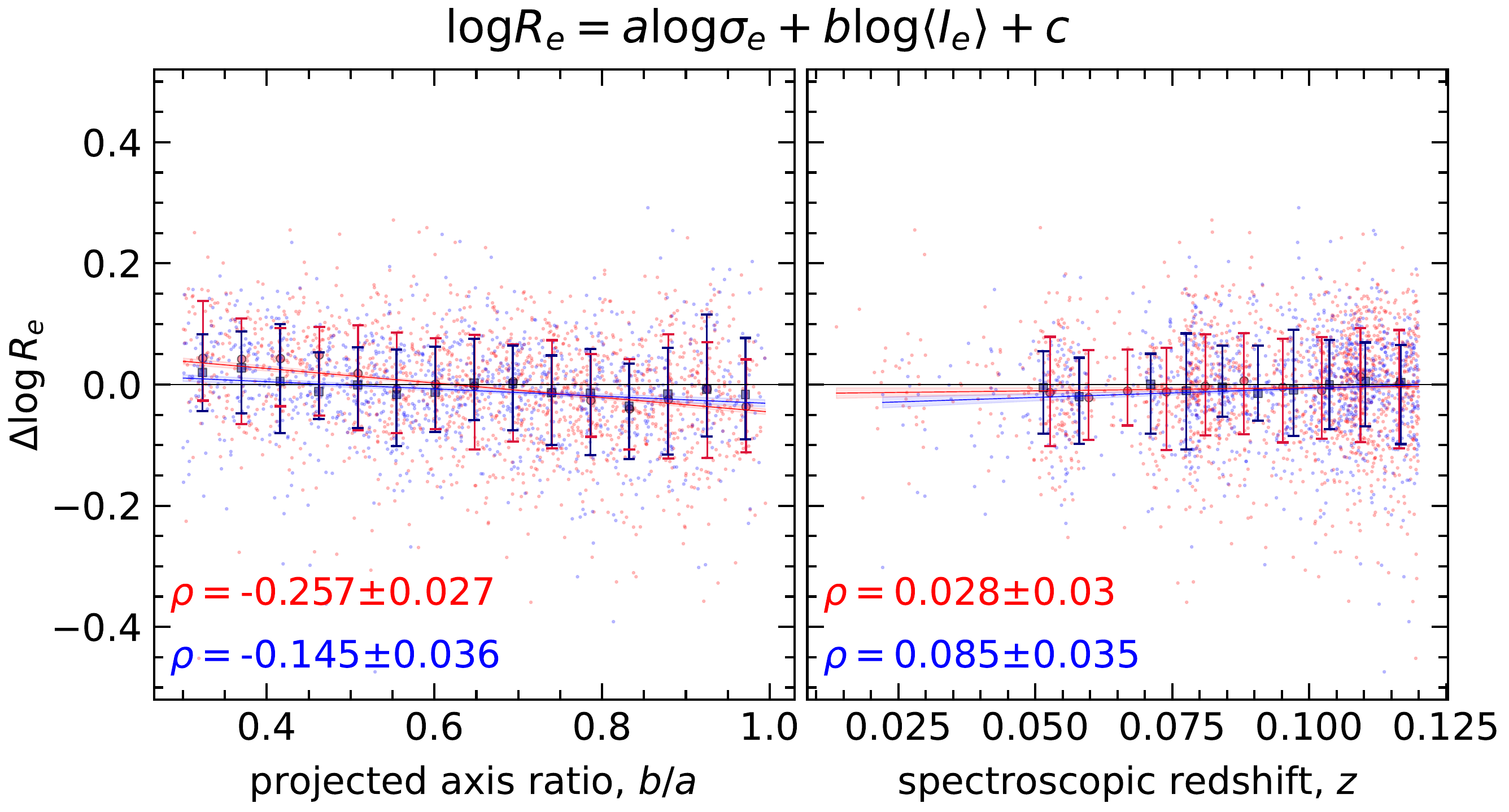}{0.46\textwidth}{(a) For the fundamental plane}
    }
    \gridline{
        \fig{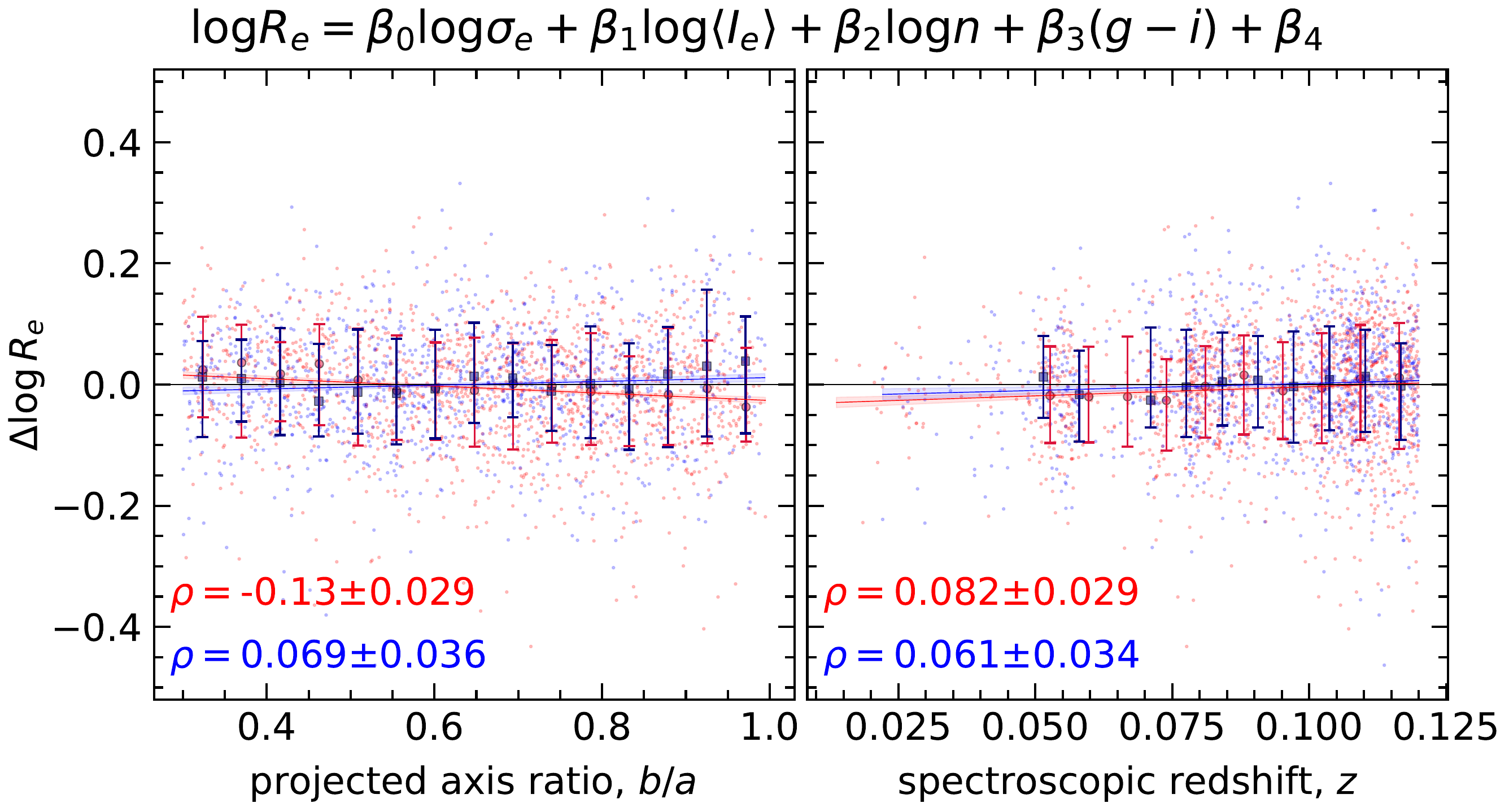}{0.46\textwidth}{(a) For the mass hyperplane}  
    }
    \caption{Same as Figure \ref{fig:sps_residuals_fp} but for the residuals in the $r-$direction as a function of axis ratio and spectroscopic redshift.}
    \label{fig:q_z_residuals}
\end{figure}

Furthermore, we present the results for the residuals as a function of axis ratio and observed redshift in Figure \ref{fig:q_z_residuals}, which shows that despite the $k(q)$-correction, the axis ratio has a noteworthy systematic effect on the FP, just as shown in \cite{bernardi2020}, whereas this is largely reduced in the MH. Critically, redshift is not a source of systematics for either the FP or the MH.

The results are almost identical under separate and independent modeling of Q and SF populations. We should note that this approach we have applied here is to statistically compare the residual trends of both the FP and the MH, notwithstanding the astrophysical interpretation of these trends is reserved for a future work.

\section{Peculiar Velocities}\label{sec:vpec}

\subsection{Distance ratios}\label{sec:distratios}

For calculating the peculiar velocities, we adopt the same approach as \cite{springob2014} 
and \cite{howlett2022}, where we calculate the posterior probability distributions for the distances of each galaxy. We convert the observed effective angular radii $(\theta_e)$ to physical effective radii $(R_e)$ with $R_e=D_A(z)\theta_e$, where $D_A(z)$ is the angular diameter distance at observed redshift $(z)$, which may not be the same as the unknown true cosmological redshift (i.e., Hubble redshift, $z_H$), due to the peculiar motions of galaxies. In this case, the observed effective radii, $r_z\equiv \log [D_A(z)\theta_e]$, will be different from the true intrinsic effective radii, $r_H\equiv \log [D_A(z_H)\theta_e]$. Therefore, the offset between them is the log-distance ratios: $r_z - r_H = \log[D_A(z)/D_A(z_H)] = \log[D_C(z)/D_C(z_H)] \equiv \eta$ \citep[see][footnote 4 for a detailed explanation]{howlett2022} from which we will derive the peculiar velocities. We use the estimator to convert $\eta$ to $V_\text{pec}$ given by \cite{watkins2015}, 
\begin{align}
    V_\text{pec} &\approx \frac{cz_\text{mod}}{1 + z_\text{mod}}\eta \ln(10)~,\\
    z_\text{mod} &= z[1 + 0.5(1-q_0)z - (1/6)(j_0 - q_0 - 3q_0^2 + 1)z^2],
    \label{eq:vpec_approx}
\end{align}
where $q_0$ is the deceleration parameter and $j_0$ is the jerk parameter; i.e., the first and second linear derivatives of the cosmic expansion history, $H(z)$. For $\Lambda$CDM cosmology, $q_0=-0.535$ and $j_0=1$. Finally, we can also calculate the cosmological distance modulus $(\mu_\mathrm{H})$ using the log-distance ratios as follows,
\begin{equation}
    \mu_\mathrm{H} = 5\log D_L(z) - 5\eta + 25 \equiv \mu_z - 5\eta,
    \label{eq:etadistmod}
\end{equation}
where $D_L(z)$ is the luminosity distance in units of $h^{-1}$Mpc, calculated from the observed redshift, thus $\mu_z = 5\log D_L(z) + 25$. For our calculations here, we use the redshift in the cosmic microwave background frame $(z_\text{CMB})$.

Our goal is to calculate the posterior probability distribution for the log-distance ratio $\eta$ for each galaxy, given the data and the best-fitting mass hyperplane (equation (\ref{eq:mdyn_distance})): $p\left(\eta_j\,|\,\bm{Y}_j, \bm{\overline{Y}}, \bm{C}^Y_j\right)$. Changing variables from $\bm{Y}$ to $\bm{Y_H}=(\bm{r_z}-\bm{\eta}, \bm{s}, \bm{i}, \bm{\nu}, \bm{c})^\intercal=(\bm{r_H}, \bm{s}, \bm{i}, \bm{\nu}, \bm{c})^\intercal$ results in $d{\bm{Y_H}}_j/d\bm{Y}_j=1$. Thus, using Bayes' theorem gives,
\begin{equation}
    p(\eta_j\,|\,\bm{Y}_j, \bm{\overline{Y}}, \bm{C}_j^Y) = \frac{p({\bm{Y_H}}_j\,|\, \eta_j, \bm{\overline{Y}}, \bm{C}_j^Y) p(\eta_j)}{p(\bm{Y}_j\,|\, \bm{\overline{Y}}, \bm{C}_j^Y)},
    \label{eq:p_eta}
\end{equation}
where $p(\eta_j)$ is the prior. Following \cite{howlett2022}, we assume a flat prior (the natural choice given the approximately normal PDFs for $r$ and $\log D$) and generate 1000 values for $\eta_j$, uniformly distributed in the range $[-1, 1]$ for each individual galaxy, $j$. This means that each galaxy will have a posterior distribution estimated at $N_\eta=1000$ different possible distances.

Since our model is a 5D Gaussian, we would expect the posterior for each galaxy in equation~(\ref{eq:p_eta}) to have a normal distribution as well. However, due to selection effects, the posterior will be a (slightly) skewed Gaussian rather than a perfect one \citep{springob2014, howlett2022}. Thus, before fitting the posteriors, the selection effects must be corrected by normalizing each posterior by,
\begin{equation}
    f_j = \int\limits_{Y_\text{cut}}^\infty p(\eta_j\,|\,\bm{Y}_j, \bm{\overline{Y}}, \bm{C}_j^Y) d^5 \bm{Y}_j,
    \label{eq:fj_eta}
\end{equation}
where ${Y_\text{cut}}$ represents the limits stemming from our sample selection criteria. $f_j$ will weight each galaxy to each possible distance that the galaxy could be at, so that it accounts for the galaxies missing from our sample due to our imposed selection criteria. We show how we tackle this issue in the next subsection, \ref{sec:malm_bias}.

\begin{figure}
    \centering
    \includegraphics[width=0.47\textwidth]{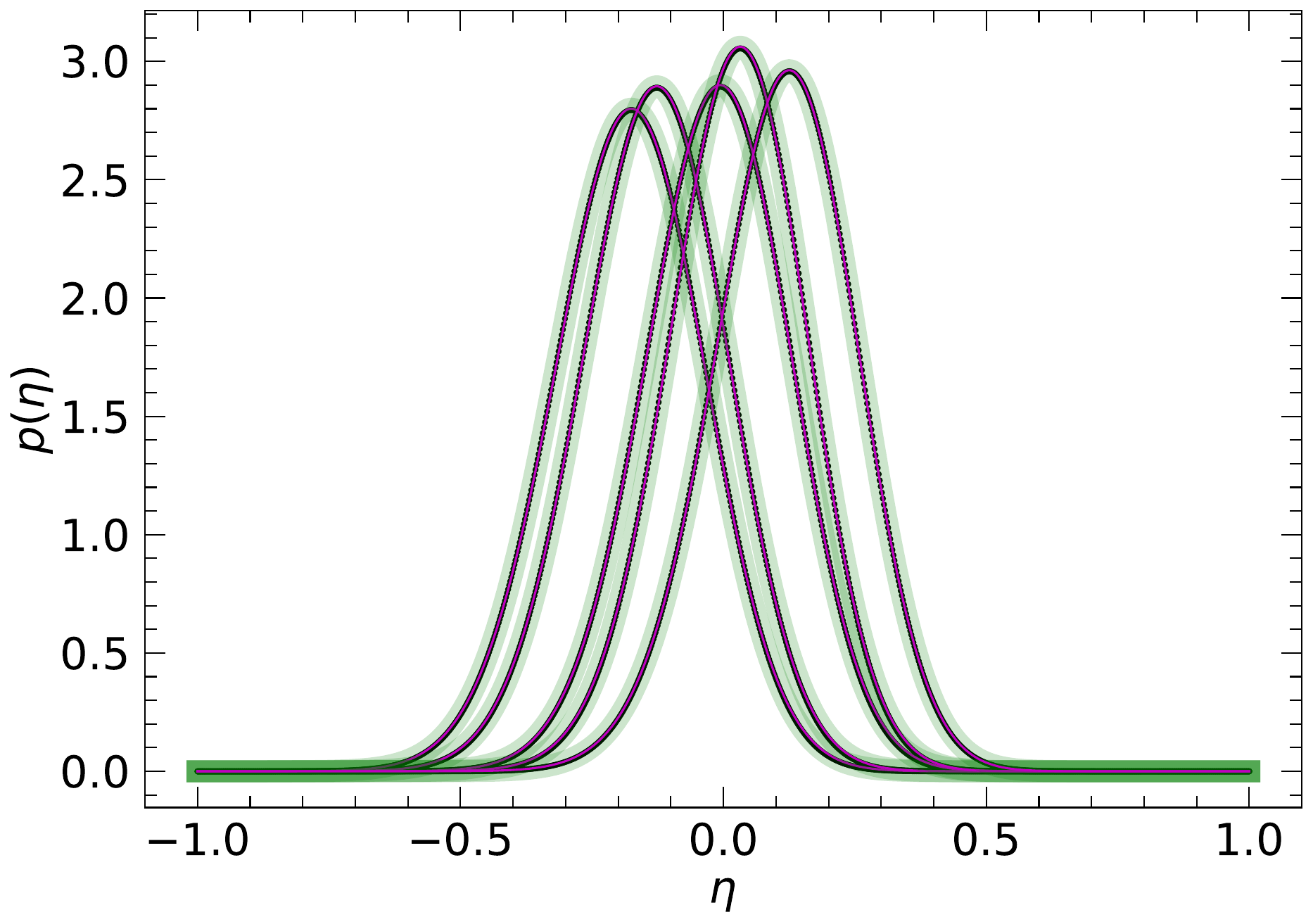}
    \caption{Normalized PDFs of log-distance ratios for 5 randomly chosen quiescent galaxies in our GAMA sample. Filled black circles show the probability distribution and the shaded green curves show the skew-normal fits.}
    \label{fig:p_eta_fitting}
\end{figure}

\begin{figure}
    \centering
    \includegraphics[width=0.47\textwidth]{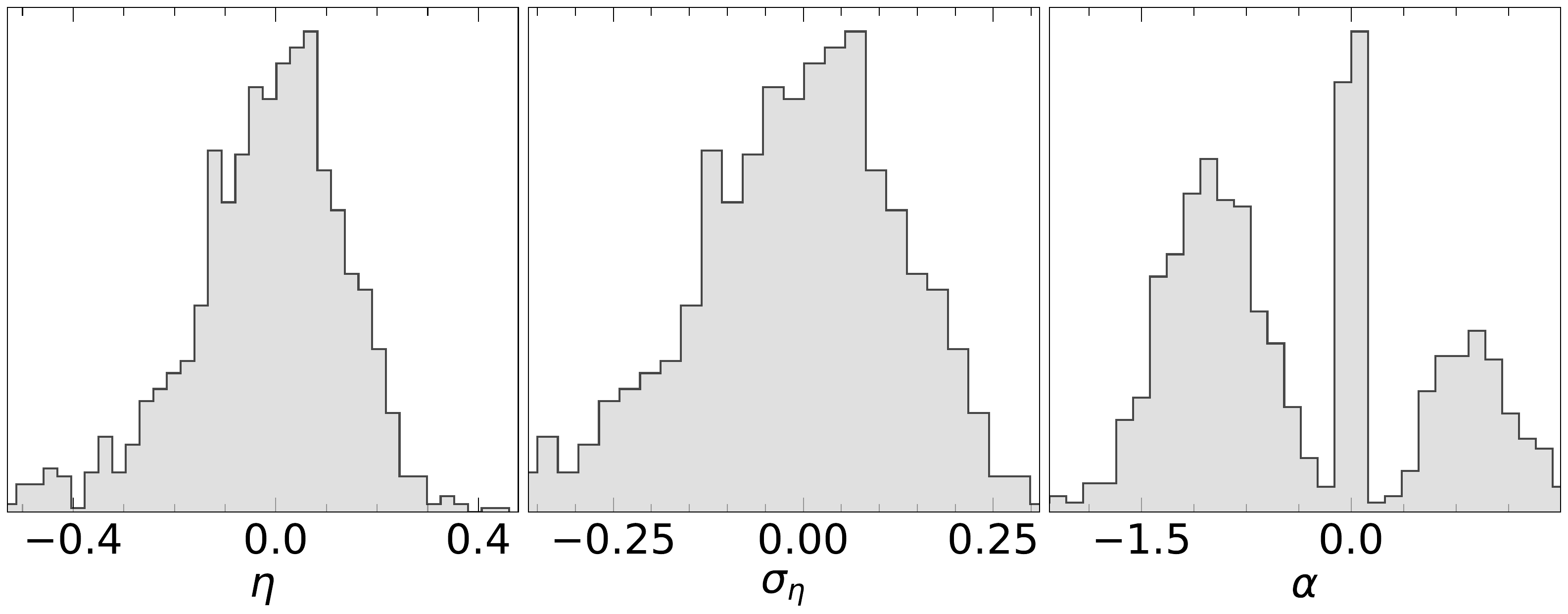}
    \caption{Distributions of the mean, standard deviation and shape values obtained from the skew-normal fits to the PDFs of log-distance ratios for quiescent galaxies.}
    \label{fig:p_eta_parameters}
\end{figure}

As in \cite{howlett2022} \citep[and similar to][]{springob2014}, we fit the normalized posteriors for each individual galaxy with a skewed normal distribution described by location $(\xi_j)$, scale $(\omega_j)$ and shape $(\alpha_j)$:
\begin{align}
    p(\eta_j\,|\,\bm{Y}_j, \bm{\overline{Y}}, \bm{C}_j^Y) = \frac{1}{\omega_j}\sqrt{\frac{2}{\pi}} &\exp{\left[ -\frac{(\eta_j - \xi_j)^2}{2\omega_j^2} \right]}\nonumber \\
    &\times\Phi\left(\alpha_j\frac{\eta_j - \xi_j}{w_j}\right)~,
    \label{eq:skewnormal}
\end{align}
where $\Phi$ is the cumulative distribution function (CDF) of the standard normal distribution. It can be seen from equation~(\ref{eq:skewnormal}) that $\alpha_j=0$ corresponds to normal distribution. In Figure \ref{fig:p_eta_fitting}, we show the example PDFs of 5 randomly chosen quiescent galaxies in our GAMA sample and their corresponding skew-normal fits. We calculate the parameters of interest, $\langle \eta_j \rangle$ and its standard deviation $\sigma_{\eta_j}$ from the fitted parameters $\xi_j, \omega_j$ and $\alpha_j$ using, 
\begin{align}
    &\langle\eta_j\rangle = \xi_j + \omega_j\delta_j\sqrt{\frac{2}{\pi}}~,~\sigma_{\eta_j} = \omega_j\sqrt{1 - \frac{2\delta_j^2}{\pi}}~,\nonumber \\
    &\mathrm{where}~\delta_j = \frac{\alpha_j}{\sqrt{1+\alpha_j^2}}~.
    \label{eq:skew_normal_pars}
\end{align}
We give the distributions of $\langle \eta_j\rangle, \sigma_{\eta_j}$ and $\alpha_j$ in Figure \ref{fig:p_eta_parameters}, which shows that the shape parameter $\alpha_j$ is consistently non-zero for most of the galaxies, albeit small. Lastly, we should point out that PV measurements of 2496 galaxies have been obtained with this procedure after discarding the fits that have not converged.

\subsection{Correction for Malmquist bias}\label{sec:malm_bias}

Due to the magnitude limit, more and more fainter galaxies will be excluded from a flux-limited sample with increasing distances. This is known as the (homogeneous, or the second type) Malmquist bias and it affects every galaxy regardless of their position on the sky \citep[e.g.,][]{springob2014}.

Even though we have already accounted for the magnitude and redshift limits on our sample using $w_j=1/S_j$ in our parent Gaussian mixture model given in equation~(\ref{eq:wbc_logsumexp}), each galaxy is now being inspected at varying distances which in turn causes varying completeness levels. Therefore, we must account for the selection effects for each individual galaxy again, but this time, we do not have to worry about $1/S_j$ weighting because, as stated in \cite{howlett2022}, it is fixed for each galaxy. The issue here is that our sample is stellar-mass-limited, not flux/magnitude-limited, which changes the aspect on how the Malmquist bias applies.

Some past studies have used samples like 6dFGS and SDSS which have been selected primarily by imposing an apparent magnitude limit. Therefore, they preferentially select intrinsically brighter galaxies at greater distances. Assuming all else being equal, this results in solutions with $\eta<0$ (i.e., distances less than implied by the observed redshift and thus $V_\mathrm{pec}<0$) being preferentially selected, which is the manifestation of the Malmquist bias in the PVs. The correction for the Malmquist bias for an apparent magnitude limited sample is obtained through averaging over all plausible values of the galaxy observables (given the data) that still satisfy the apparent magnitude selection. The usual way to do this is to determine the limiting absolute magnitude, as a function of distance, for which a given galaxy would still be included in the sample. This defines the limits of integration for the correction factor. 

However, because our sample is stellar-mass-limited, it is necessary to revise this correction scheme slightly, even though the basic idea remains the same. This time, the correction factor is obtained by averaging over all plausible values of galaxy observables which satisfy the mass selection condition.

\begin{figure*}[htbp]
    \centering
    \includegraphics[width=\textwidth]{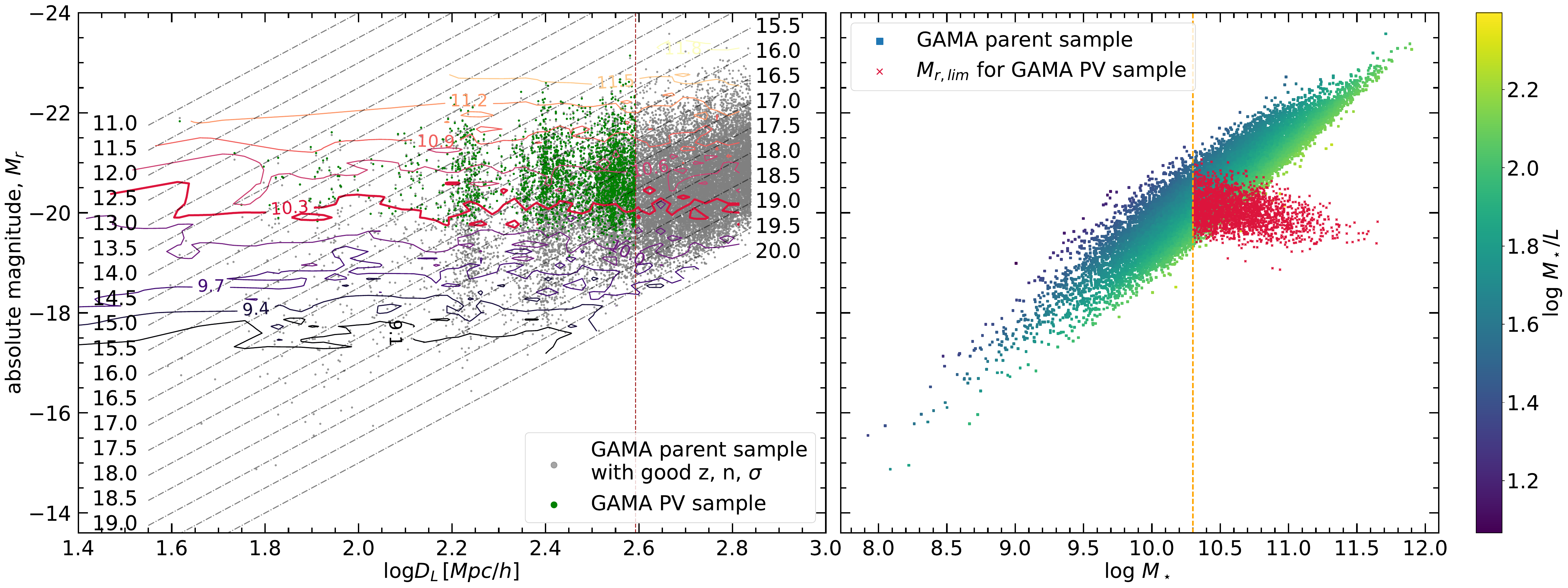}
    \caption{\textit{Left:} Demonstration of how Malmquist bias affects our sample in the $M_r$ vs $\log D_L$ plot. The gray dots are the entire GAMA sample with reliable redshift, S\'ersic index and velocity dispersion measurements, whereas the green dots are our adopted PV sample. Contours of $M_\star$ are also given, where the thick red contour represents $\log M_{\star, \mathrm{cut}}=10.3$, the lower mass limit. Dash-dotted gray lines are the Malmquist biases corresponding to the $r-$band limiting magnitudes given in the labels. Dashed vertical line is the redshift limit of our sample, $z=0.12$. \textit{Right:} $M_r$ plotted against the stellar-mass and color-coded with the stellar-mass-to-light ratio. Red crosses are the limiting absolute magnitudes, $M_{r,\mathrm{lim}}$, for each galaxy in our sample and orange dashed line is the lower mass limit. }
    \label{fig:absmag_lumdist}
\end{figure*}

Figure \ref{fig:absmag_lumdist} shows the distribution of GAMA galaxies in the $r-$band absolute magnitude, $M_r$, as a function of log-luminosity distance, $\log D_L$, in the left-hand panel and as a function of stellar-mass, $\log M_\star$, in the right-hand panel. Because mass and luminosity are directly correlated, it is natural to expect that a lower mass limit would cause a lower limit in absolute magnitude, $M_{r,\mathrm{lim}}$, which will be the integration limit.

The question is now to find to what distance a galaxy could still remain in the stellar-mass-limited sample if its distance were changed, which means that we need to find $M_{r, \mathrm{lim}}^j$ corresponding to $M_{\star, \mathrm{cut}}$. This can be done for each galaxy, $j$, via
\begin{equation}
    M_{r, \mathrm{lim}}^j = -2.5\log\left[ M_{\star,\mathrm{cut}} - \left(\frac{M_\star}{L_r}\right)_j \right] + M_{r, \odot}~,
    \label{eq:absmaglimit}
\end{equation}
where the stellar-mass-to-light ratio, $M_\star/L$, is measured from the SED and $M_{r,\odot}$ is the absolute magnitude of the Sun in $r-$band. The resulting $M_{r,\mathrm{lim}}$ are shown with red crosses on the right-hand panel of Figure \ref{fig:absmag_lumdist}.

However, $M_{\star, \mathrm{cut}}$ does not create a single, distance independent $M_{r,\mathrm{lim}}$ value (see, right panel of Figure \ref{fig:absmag_lumdist}), as opposed to what one might expect at first sight when looking at the left-hand panel of Figure \ref{fig:absmag_lumdist}. Instead, as equation \ref{eq:absmaglimit} shows, due to the $M_\star/L$ varying for each galaxy, the fixed stellar-mass cut produces a range in limiting absolute magnitude. Moreover, the trend in $M_\star/L$ with $M_\star$, as seen in the right-hand panel of Figure \ref{fig:absmag_lumdist}, leads to a corresponding trend in $M_{r,\mathrm{lim}}$ with $M_\star$ as shown in red crosses. Therefore, more massive galaxies have fainter $M_{r,\mathrm{lim}}$ due to their larger $M_\star/L$ ratios. Close to the $M_{\star, \mathrm{cut}}$, the $M_{r,\mathrm{lim}}$ are close to the actual $M_r$; this is not a coincidence, but rather a consequence of choosing the sample stellar-mass limit to correspond to the sample apparent magnitude limit at the sample's redshift limit, as seen in the left-hand panel of Figure \ref{fig:absmag_lumdist}.

We now turn our attention to finding the integration limits caused by the Malmquist bias. As in \cite{dam2020}, using $(L/L_\odot)_\lambda = 2\pi (R_{e,\lambda}^{\text{pc}})^2 \langle I_e \rangle_\lambda$, we find $i\equiv \log\langle I_e \rangle_\lambda = \log (L/L_\odot)_\lambda - 2(r+3) - \log 2\pi$ which leads to $i+2r=-0.4(M_\lambda - M_{\lambda, \odot} + 2.5\log 2\pi)-6$ where the absolute magnitudes in band $\lambda$ are $M_\lambda - M_{\lambda, \odot} = -2.5\log (L/L_\odot)_\lambda$. Considering that the distance modulus is $\mu_z \equiv m_\lambda - M_\lambda = 5\log D_L[\text{pc}]-5$, we can encapsulate the constants with $\mathcal{M}_0 = 2.5\log 2\pi - M_{\lambda, \odot}+15$ and we obtain an expression for the apparent magnitude,
\begin{equation}
    m_\lambda = -2.5(i+2r)+\mu_z-\mathcal{M}_0~.
    \label{eq:i2r}
\end{equation}
As stated in \cite{springob2014} and \cite{dam2020}, equation~(\ref{eq:i2r}) shows that a magnitude limit corresponds to a diagonal cut in $(\bm{r}, \bm{s}, \bm{i})$ space since magnitude is a function of both $i$ and $r$. On the other hand, equation (\ref{eq:i2r}) shows that a limit in absolute magnitude can also be expressed as a function of $i+2r$, with $M_{\lambda, \mathrm{lim}} = -2.5(i+2r) - \mathcal{M}_0.$ Therefore, $M_{\lambda, \mathrm{lim}}$ which is caused by the lower limit on $M_\star$, will also result in a cut in both $\bm{r}$ and $\bm{i}$. 

Even though both $r$ and $i$ are functions of S\'ersic index $n$, our sample still covers the full possible range of S\'ersic indices $(0.3\leqslant n \leqslant 10)$, which means that the absolute magnitude limit does not cause a cut in $n$. A similar result can be reached for $(g-i)_\text{rest}$. Therefore, we only need to account for the cut-offs in $\bm{r}, \bm{s}$ and $\bm{i}$ in the 5D parameter space of the mass hyperplane.

Following \cite{dam2020}, we change variables from $\bm{Y} = (\bm{r}, \bm{s}, \bm{i}, \bm{\nu}, \bm{c})^\intercal$ to $\bm{w} = (\bm{u}, \bm{s}, \bm{i}, \bm{\nu}, \bm{c})^\intercal = (\bm{i} + 2\bm{r}, \bm{s}, \bm{i}, \bm{\nu}, \bm{c})^\intercal$ by,
\begin{equation}
    \bm{W} = \bm{J} \bm{Y} = \begin{pmatrix} 
        2 & 0 & 1 & 0 & 0\\
        0 & 1 & 0 & 0 & 0\\
        0 & 0 & 1 & 0 & 0\\
        0 & 0 & 0 & 1 & 0\\
        0 & 0 & 0 & 0 & 1\\
        \end{pmatrix} \begin{pmatrix} \bm{r} \\ \bm{s} \\ \bm{i} \\ \bm{\nu} \\ \bm{c} \end{pmatrix} = \begin{pmatrix} 2\bm{r} + \bm{i} \\ \bm{s} \\ \bm{i} \\ \bm{\nu} \\ \bm{c} \end{pmatrix}
    \label{eq:wjy}
\end{equation}
where $\bm{J}$ is the Jacobian of the transformation $\bm{u} = \bm{i} + 2\bm{r}$. For a random variable having an $M$-dimensional normal distribution $Y\sim \mathcal{N}(\bm{\overline{Y}}, \bm{\Sigma_Y})$, the transformed variable $w = AY + B$, where $A \in \mathbb{R}^{M\times M}$ and $B \in \mathbb{R}^M$, also has an $M$-dimensional normal distribution with mean $\overline{w} = A\overline{Y} + B$ and covariance matrix $\bm{\Sigma_w}=A\bm{\Sigma_Y}A^\intercal$. Thus, equation~(\ref{eq:fj_eta}) becomes,
\begin{equation}
    f_j = \int\limits_{\bm{w}_{\text{cut}}}^\infty p(\bm{w_j} | \bm{\bar{w}}, \mathbf{C}^w_j) d^5 \bm{w_j}
    \label{eq:fj_w}
\end{equation}
where $\bm{\bar{w}} = \bm{J\overline{Y}} = (\bar{i} + 2\bar{r}, \bar{s}, \bar{i}, \bar{\nu}, \bar{c})^\intercal$ and $\mathbf{C}^w_j = \bm{J}\mathbf{C}^Y_j\bm{J}^\intercal$. With this transformation, apparent or absolute magnitude limit now corresponds to a cut in the new variable $\bm{u}$, leading to two orthogonal cuts in the $(\bm{u}, \bm{s}, \bm{i}, \bm{\nu}, \bm{c})^\intercal$ space: $\bm{w}_{\text{cut}} = (u_{\text{cut}}, s_{\text{cut}}, -\infty, -\infty, -\infty)$. From equation~(\ref{eq:i2r}), 
\begin{equation}
    u_{\text{cut}~j,k} = -0.4(m_{r, \text{lim}}^j - \mu_\mathrm{H}^{j,k} + \mathcal{M}_0)~,
    \label{eq:ucut}
\end{equation}
where the limiting apparent magnitude for each galaxy can be calculated from equation (\ref{eq:absmaglimit}) as $m_{r,\mathrm{lim}}^j = \mu_z - M_{r,\mathrm{lim}}^j$. Notice that the cut-off $u_\mathrm{cut}$ varies not only with each galaxy but also with the distance assumed for each galaxy (equation \ref{eq:etadistmod}), thus $j=1,2,\dots,N$ and $k=1,2,\dots, N_\eta$. 

We provide the details of how to reduce equation (\ref{eq:fj_w}) to a simpler form as a combination of PDF and CDF of normal distribution ($\phi$ and $\Phi$ respectively) in Appendix \ref{sec:fj_derivation}. Since these functions are built-in and optimized in \textit{PySTAN} (and some other computational packages), they are fast to compute. This is crucial considering that numerical evaluation of equation (\ref{eq:fj_w}) for large data sets is not feasible. We should also emphasize that this approach enables us to reach an exact solution instead of an approximation. 

\subsection{Zero-point calibration and comparison to the previous works}\label{sec:zpcal_and_comp}

The zero-points in the FP and the MH expressions (equations~\ref{eq:lfp} and \ref{eq:mdyn_distance}) are $c$ and $\beta_4$, respectively. These are obtained using the fitted slopes and means. Thus, determination of zero-points depends on the choice of which variable to minimize: residuals in $r-$direction or residuals in the direction perpendicular to the plane.

In the previous works of \cite{magoulas2012, springob2014, khaled2020} and \cite{howlett2022}, the FP has been fitted with a maximum likelihood Gaussian algorithm that models the underlying distribution as a 3D Gaussian and, thus, coefficients that minimize the orthogonal residuals have been derived. This has a key underlying assumption that the average radial peculiar velocity of the galaxies in the FP sample is zero. This is most likely not true in reality, therefore, it needs to be corrected. \cite{khaled2020} have fitted the FP and the velocity field simultaneously in a Bayesian framework, eliminating the possible need for this correction. However, as stated in the beginning of this section, we use the approach of deriving PVs after fitting the FP and the MH. Thus, we need to take the zero-point considerations into account.

We calibrate our zero-point by comparing the log-distance ratios that we calculated here (for the sample of quiescent galaxies) to the ones from SDSS \citep{howlett2022} that are also included in \textit{Cosmicflows-4} \citep[CF4;][]{kourkchi2020}. We found 268 galaxies in common after cross-matching and find the weighted mean difference in $\eta$ between SDSS and our sample to be $\langle \eta_\text{SDSS} - \eta_\text{GAMA} \rangle = 0.068\pm0.01$. This value includes the uncertainty in the zero-point of CF4 that \cite{howlett2022} found by cross-matching their SDSS sample with \textit{Cosmicflows-III} \citep[CF3;][]{tully2016} data (0.004 dex), their uncertainty in $\bar{r}$ (0.016 dex) and finally the uncertainty in $\bar{r}$ in our study, which is derived from our ensemble of mocks (Figure \ref{fig:1000mocks}) to be 0.014 dex. Quadratic subtraction of these three contributions gives us the uncertainty in our zero-point $(\beta_4)$ to be 0.064 dex. One final remark here is, \cite{howlett2022} noted that their overlapping sample was dominated by low redshift objects with $z<0.05$, whereas, our sample that overlaps with theirs is dominated with $z>0.05$.

\begin{figure*}
    \centering
    \includegraphics[width=\textwidth]{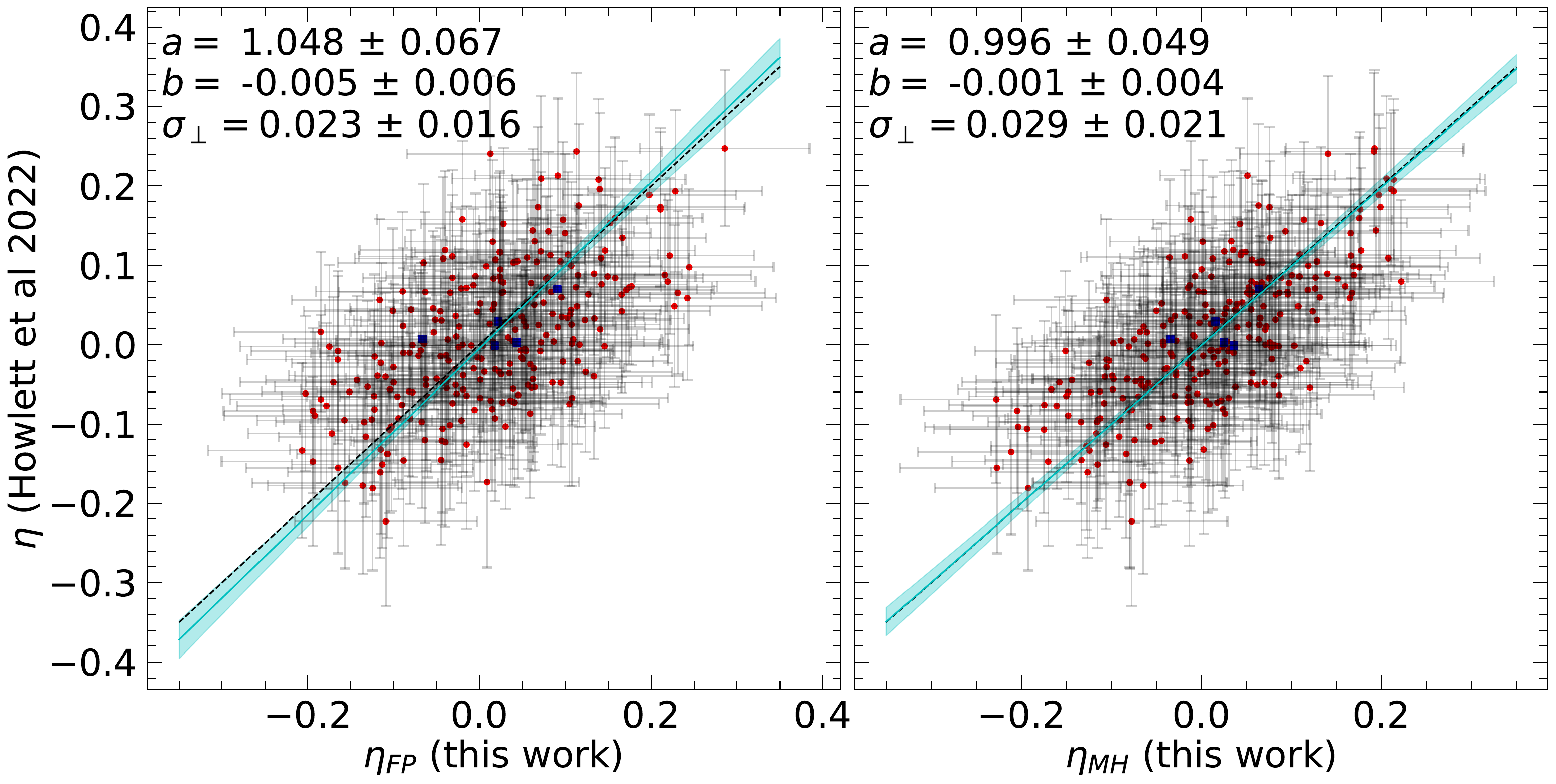}
    \caption{Comparison of our measurements of log-distance ratios, made from the fundamental plane (left) and from the mass hyperplane (right) of the combined sample of Qs and SFs, to the ones of \protect\cite{howlett2022}, for the overlapping sample of GAMA and SDSS which consists of 268 quiescent galaxies in common, shown in red circles. Blue squares show the five SF galaxies found in common with CF4. In each panel, the turquoise line shows the orthogonal distance regression fit to this sample, while the shaded region around it shows the $1\sigma$ standard deviation of the fit. Slope $a$, intercept $b$ and orthogonal scatter $\sigma_\perp$ from the ODR fit are given in the upper left corner. Black dashed line shows the one-to-one relation. Note that this fit is performed, excluding the three SF galaxies, after the zero-point calibration.}
    \label{fig:eta_gama_vs_sdss}
\end{figure*}

After the zero-point correction, we use ODR to fit the relation between $\eta_\text{SDSS}$ and $\eta_\text{GAMA}$, in the form $y=ax+b+\mathcal{N}(0,\sigma)$, accounting for measurement uncertainties in both\footnote{This fit was performed with the SciPy.ODR package of Python.}. Figure \ref{fig:eta_gama_vs_sdss} shows the comparison between these independent measurements for the overlapping sample of 268 quiescent galaxies. As seen here, the fit result is consistent with a one-to-one relation, verifying the apparent good agreement between these different data sets and methods. Additionally, this agreement is in fact so good that it does not require any correction to the slope. 

We then perform a $T-$test (including the uncertainties in variables) between our measurements, $\eta_\mathrm{gama, FP}$ and $\eta_\mathrm{gama, MH}$, and the ones from \cite{howlett2022}, $\eta_\mathrm{sdss}$, to further test the consistency of our measurements. We find that the $T-$statistic and the $p-$value to be $t=-0.007, p=0.994$ and $t=-0.005, p=0.996$ for the pairs of $\eta_\mathrm{SDSS}-\eta_\mathrm{GAMA, FP}$ and $\eta_\mathrm{SDSS}-\eta_\mathrm{GAMA, MH}$ respectively. This means that the difference between $\langle\eta_\mathrm{SDSS}\rangle$ and $\langle\eta_\mathrm{GAMA}\rangle$ (through both FP and MH) are not statistically significant.

It is straightforward to think that we can simply compare the distances for SF galaxies that we have computed in this work, to the ones from CF4 derived through TFR, and see whether this method really works for SFs. Unfortunately, the CF4 TFR sample and our GAMA sample have only 5 SF galaxies in common. Even relaxing our $\log\,M_\star/M_\odot > 10.3$ cut raises this number to only 12. 
Although Figure \ref{fig:eta_gama_vs_sdss} shows the consistency between the $\eta$ values of these 5 SF galaxies for CF4 and GAMA, it is obviously nowhere near being statistically sufficient to draw any conclusions on the relation between these different estimates. 

\begin{figure*}
    \includegraphics[width=\textwidth]{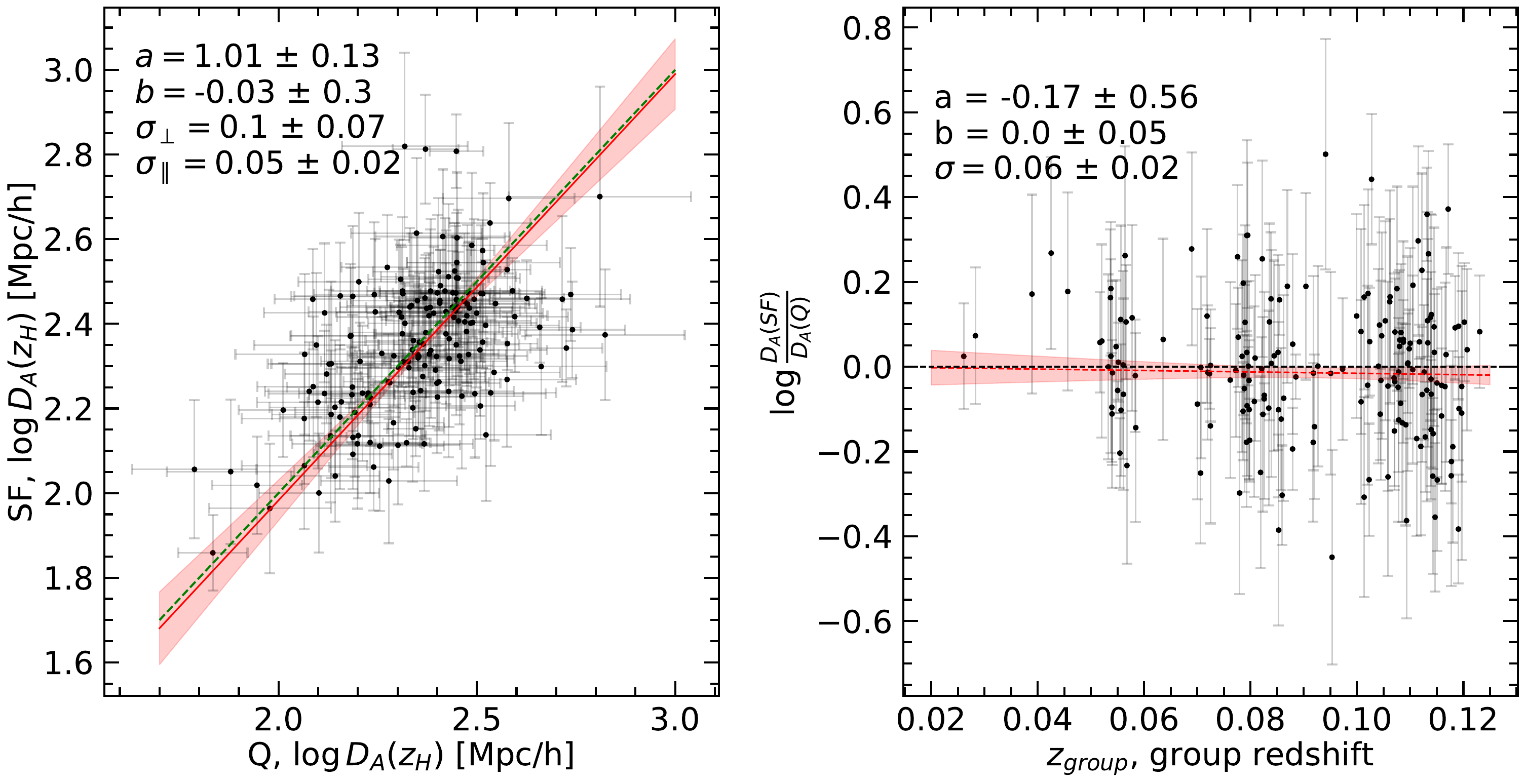}
    \caption{Comparison of the weighted means of the angular diameter distances derived from $\eta$ measured through the MH for SF and Q galaxies in the same groups. \textit{Left:} The dashed-green line shows the one-to-one relation, whereas the red line shows the mean relation between the Q and SF distances obtained by an ODR fit. The shaded red region around the mean line shows the $1\sigma$ uncertainty. Note that the implication is that the independent Q- and SF-derived distances for groups agree with a $\sigma_\parallel = \sigma_\perp / \sqrt{1 + a^2} \approx$ 0.17 dex scatter in each, and no appreciable bias. \textit{Right:} The difference between Q- and SF-derived group distances, plotted as a function of group redshift. Red line shows the fit obtained through OLS, which is consistent with zero. This lack of a significant trend with group redshifts means our PV measurements do not show biases with redshift, at least across the redshift range of our sample.}
    \label{fig:group_distances}
\end{figure*}

We can, however, use galaxy groups and the fact that galaxies in the same group should be at more or less the same distance. Thus, we can look at galaxy groups in our sample that have both SF and Q, then compare the distances of SF galaxies, $\log\,D(z_H)$, that we have derived from the log-distance ratios, to the ones of Q galaxies that are in the same galaxy group. Using the galaxy groups from \cite{robotham2011}, we find 180 groups in our sample that have at least one Q and one SF. For each group, we take the weighted mean of the $\log\,D(z_H)$ distances of Qs and SFs separately, and plot these in Figure \ref{fig:group_distances}. We then perform an ODR fit using \textit{STAN} and find that the relation between the $\log\,D(z_H)$ distances to Qs and SFs that are in the same galaxy groups is consistent with a one-to-one relation, as seen in the left-hand panel of Figure \ref{fig:group_distances}. Furthermore, in the right-hand panel of Figure \ref{fig:group_distances}, we plot the differences between the distances of Q and SF galaxies that are in the same group against the group redshift. An OLS fit to the distribution in this figure shows that the variation of distance differences with group redshift is consistent with zero, thus, showing no signs of biases.

The fact that the MH yields distance estimates with no appreciable bias for Q versus SF galaxies in the same group, and with $\approx$0.07 dex or 16\% scatter for each population, is truly remarkable.
This has the potential to significantly expand future PV cosmology samples through the inclusion of all (massive) galaxies.  At least as significantly, it also argues against any appreciable redshift-dependent bias that might creep in through the relative proportion of Q versus SF galaxies.

\subsection{Comparison to the FP: the value of the MH for PV cosmology}\label{sec:mp_vs_fp}

As discussed in more detail in section \ref{sec:results_fp_mp}, a comparison between Tables \ref{tab:lfp_fits} and \ref{tab:lfpnc_fits} shows that the intrinsic scatter, $\sigma_{r, \mathrm{int}}$, is smaller in the MH than in the FP, irrespective of the galaxy type and the sample used in fitting. This raises the question of whether this improvement in the tightness of the plane would manifest itself in the measurements of $\eta$. To address this question, we can simply repeat the measurements in sections \ref{sec:distratios} and \ref{sec:malm_bias} for the traditional FP and measure the rms scatter of the observed $\eta$ distributions for both FP and MH.

\begin{figure*}[htbp]
    \centering
    \includegraphics[width=0.8\textwidth]{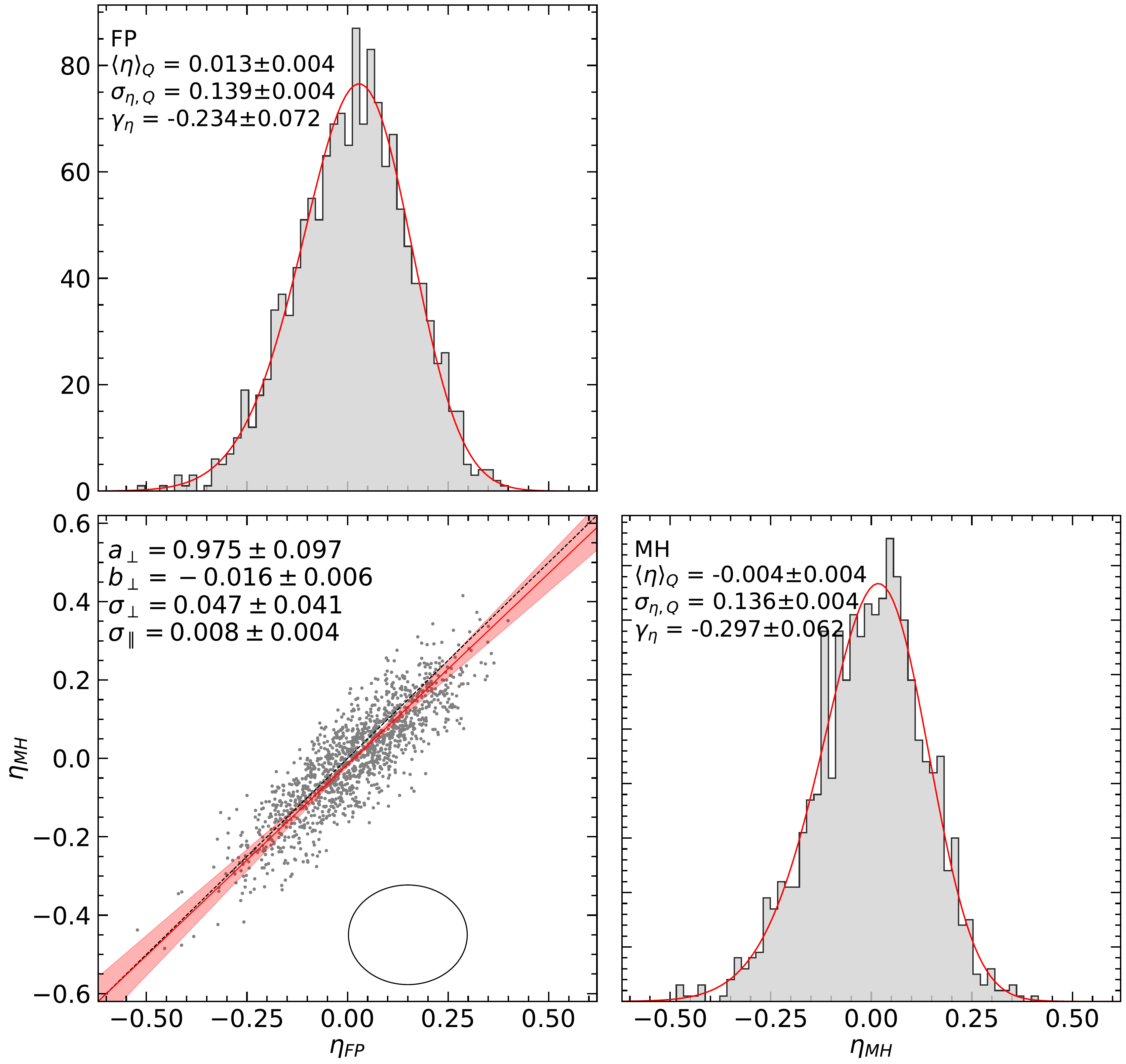}
    \caption{$\eta$ measurements carried out through the FP and the MH for quiescent galaxies. Diagonal panels show the histograms of $\eta_\mathrm{FP}$ and $\eta_\mathrm{MH}$. Smooth red curves show the best-fitting skew-normal distributions obtained via \textit{PySTAN}. The mean $(\langle \eta \rangle)$, standard deviation $(\sigma_\eta)$ and skewness $(\gamma_\eta)$ are given in the upper left corner in each diagonal panel. The lower left panel gives the comparison of these measurements plotted against each other. Dashed line shows the one-to-one relation, while the red line and shade show the best-fitting linear relation and the corresponding 1$\sigma$ region. Fit parameters are given in the upper left-hand corner. Finally, the mean error ellipse is shown in the lower right-hand corner.
    }
    \label{fig:fp_mh_comparison_q}
\end{figure*}

In Figure \ref{fig:fp_mh_comparison_q}, we show the distributions of $\eta$ for quiescent galaxies as measured from the FP and the MH. We fit these with skew-normal distributions using \textit{PySTAN}, then derive the mean $(\langle \eta \rangle)$ and standard deviation $(\sigma_\eta)$ values using equation (\ref{eq:skew_normal_pars}). 
While the standard deviation (which corresponds to the distance error in this case) is $0.139\pm0.004$ dex for the FP-derived distances, it is $0.136\pm0.004$ dex for the MH, which means that, in practice, the precision of the distances measured from the MH is not significantly better than with the FP. Repeating the same procedure to the star-forming sample, we see similar. We provide a similar discussion in appendix \ref{sec:pvs_separate} for the case of separate and independent modeling of galaxy populations.
This suggests that observational errors in the measurements of galaxy properties and/or in the derivation of Malmquist bias corrections may make a substantial contribution to the final error budget for $\eta$, at least for this sample.
We therefore conclude that while, in principle, the use of the MH can lead to a $\approx 10$\% improvement in the precision of distance estimations, in practice it may be hard to realize that potential.  
Even so, we have shown how our MH formalism and analysis can be applied not only to quiescent galaxies, but to the broader $\log M_* \gtrsim 10.3$ galaxy population, with an increase of $70$\% or greater in sample size and a commensurate increase in statistical power.

\begin{figure*}[htbp]
    \centering
    \includegraphics[width=0.8\textwidth]{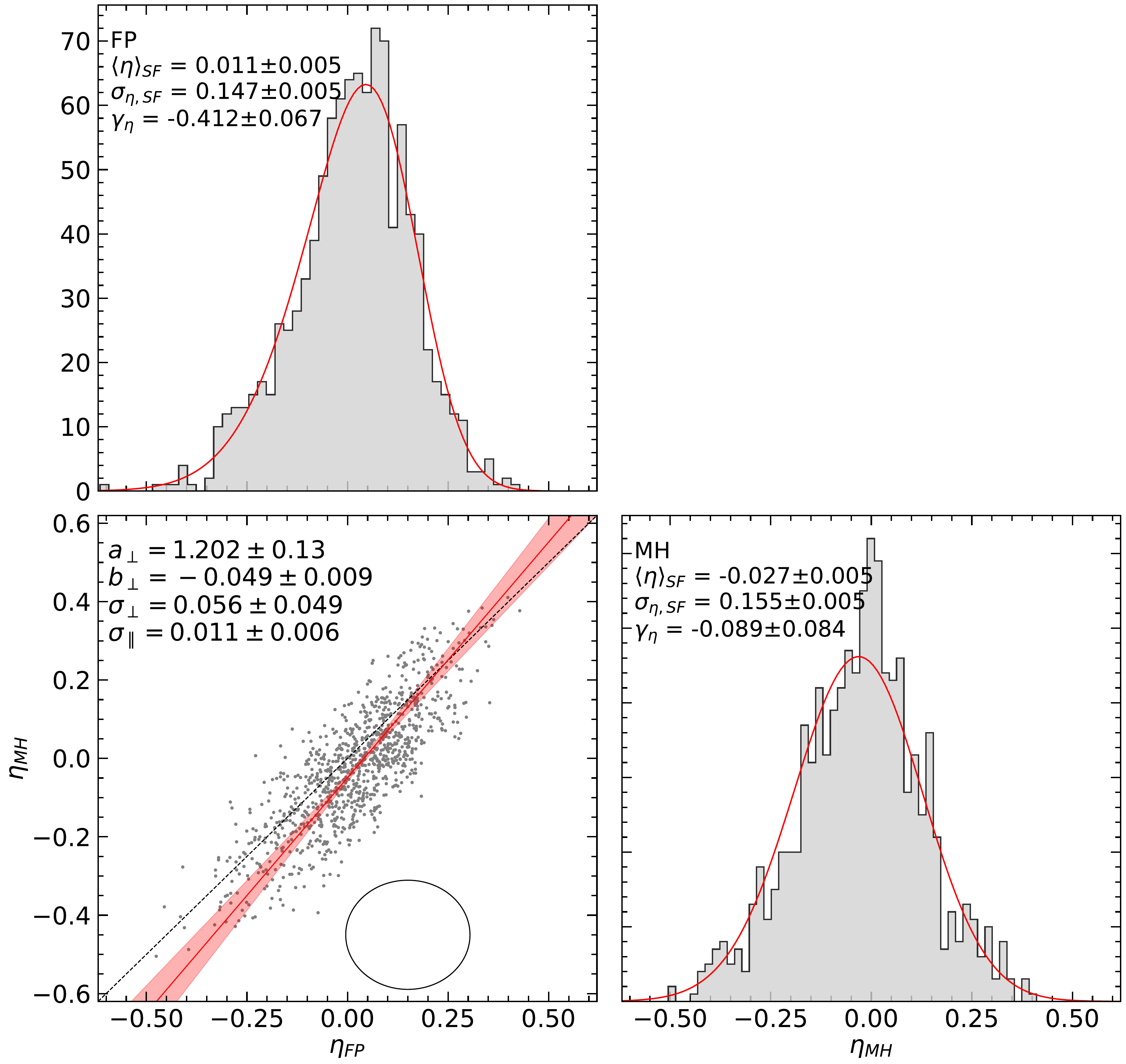}
    \caption{Same as Figure \ref{fig:fp_mh_comparison_q} but for the star-forming galaxies in our sample.}
    \label{fig:fp_mh_comparison_sf}
\end{figure*}

\section{Summary and Conclusions}\label{sec:conclusions}

In this work, we have introduced the mass hyperplane (MH) as a recasting of the $M_\star/M_\mathrm{dyn}$ relation, which provides an FP-like linear distance indicator. We have modeled both the FP and the MH for quiescent and star-forming galaxies separately, showing that both populations follow similar relations (Figures \ref{fig:lfp_trends} and \ref{fig:lfpnc_trends}). Further, we have demonstrated that both populations can be modeled as a single FP or MH relation with a slightly larger intrinsic scatter. That is, the FP and the MH are not specific to a certain class of galaxies, but apply to all (field) galaxies as a population. This result broadly agrees with the growing consensus, including authors like \cite{bezanson2015, degraaff2020} and \cite{degraaff2021}.

As well as showing the importance of accounting for projection effects (Figure \ref{fig:kqcorrection}), we have made a thorough comparison between the MH and the FP, and explored their possible systematics by studying their residuals as a function of stellar population parameters (Figure \ref{fig:sps_residuals_fp} and \ref{fig:sps_residuals_mh}), along with axis ratio and redshift (Figure \ref{fig:q_z_residuals}). Then, we have set out to determine the redshift-independent distances of both quiescent and star-forming galaxies of GAMA at $z<0.12$ simultaneously using the MH. We have tested the validity of our framework by comparing our distance/PV estimates made through both the FP and the MH to the previous measurements for quiescent galaxies from SDSS (Figure \ref{fig:eta_gama_vs_sdss}). To perform this test for the star-forming galaxies, we compare the distances to groups independently derived for quiescent and star-forming galaxies (Figure \ref{fig:group_distances}). Finally, we have compared distance/PV measurements from the FP to the ones from the MH (Figures \ref{fig:fp_mh_comparison_q} and \ref{fig:fp_mh_comparison_sf}). 

Our results are summarized as follows:
\begin{enumerate}
    \item Both the FP and the MH can be obtained by fitting either the individual quiescent and star-forming samples or the combined sample (Figures \ref{fig:lfp_trends} and \ref{fig:lfpnc_trends}), although separate and individual fitting results in smaller intrinsic scatters in the $r-$direction, i.e., tighter planes for distance estimation. However, in all cases of fitting, the MH slightly reduces the intrinsic scatter compared to the FP by at least $\sim$ 6\% (Tables \ref{tab:lfp_fits} and \ref{tab:lfpnc_fits}) which may slightly improve the precision of the redshift-independent distance estimates.
    \item The intrinsic scatter (in $r-$direction, $\sigma_{r,\mathrm{int}}$) of the MH is $\sim 10$\% smaller than for the FP. Therefore, the limiting precision for MH-derived distances is $\sim 5-10$\% better compared to the FP. While SF and Q galaxies can be described with a single FP or MH relation, obtaining separate/independent FP and MH relations result in $\sim 10$\% smaller $\sigma_{r,\mathrm{int}}$ relative to a single FP/MH fit. That is, using a Q- or SF-specific FP/MH relation gives the best precision in distance estimates (Tables \ref{tab:lfp_fits} and \ref{tab:lfpnc_fits}).
    \item Similar to \cite{graves2009_2} and \cite{springob2012}, we have found strong correlations between stellar population properties (such as age/$D_n 4000$, SFR) and $\Delta \log R_e$ residuals of either the FP or the MH relations (Figures \ref{fig:sps_residuals_fp} and \ref{fig:sps_residuals_mh}). Further, we see similar variations with spectral and SED-derived properties ($g-i$, $H_\alpha$, $H_\delta$, $D_n4000$, E(B-V), sSFR, $\log\langle t_\star\rangle_\mathrm{lw}$, $Z_\star$ for both the FP and MH, showing that the inclusion of the rest-frame color as a stellar population diagnostic, provides a modest benefit. The same can be said for the inclusion of S\'ersic index as a proxy for stellar/dynamical structure.
    \item The projection effects seen as a function of axis ratio $(q=b/a)$ are substantial. Although we have limited our analysis to $q>0.3$, we have also shown that the empirical description of \cite{wel2022} provides a good way to describe this effect (Figure \ref{fig:kqcorrection}). Future PV studies carried out via the FP/MH may either simply adopt the \cite{wel2022} prescription or may be able to incorporate a similar parameterization into the model to calibrate these effects, which can possibly eliminate the need for an axis ratio selection. 
    \item We have validated our MH framework as a tool for PV cosmology by directly comparing our MH-derived distances to the latest FP-derived distances of quiescent galaxies for SDSS from \cite{howlett2022}. We have found an excellent agreement with a random scatter of $\sim 0.1$ dex (Figure \ref{fig:eta_gama_vs_sdss}), i.e., one-to-one agreement within measurement errors and with no discernible bias.
    \item For validating our MH framework for star-forming galaxies, we compared the MH-derived group distances based on star-forming galaxies to the ones based on their quiescent counterparts in the same group. We have shown that both measurements are consistent with each other, with a random scatter of $\sim 0.1$ dex, while showing no discernible bias as a function of inferred distance or group redshift (Figure \ref{fig:group_distances}).
    \item The $\sim$10\% improvement in the intrinsic scatter in $r-$direction when the MH is used, does not seem to translate as an improvement in the precision of $z-$independent distance estimates, as seen in e.g., Figure \ref{fig:fp_mh_comparison_q}, which might be caused by the additional sources of observational uncertainties in the MH. This result is similar to \cite{springob2012, springob2014} who found that even though the FP residuals correlate strongly with stellar age, adding it to the FP does not provide improvement for distances. Furthermore, modeling the galaxy sample as one population may save time, however, may result in larger uncertainties. Meanwhile, using the same form of the planar relation, albeit with different coefficients acquired by modeling the populations separately and independently, most likely constitutes the best approach (Figures \ref{fig:fp_mh_comparison_q_separate} and \ref{fig:fp_mh_comparison_sf_separate}), as it reduces the scatter in distances compared to the combined treatment of galaxy populations.
\end{enumerate}

Overall, we have shown that the sample size for a PV study can be significantly increased (by $\gtrsim 70$\%) through the inclusion of star-forming galaxies in both the FP and the MH, which is particularly valuable when the data required for the Tully--Fisher relation are not available. For example, the 4MOST Hemisphere Survey (4HS) will measure the PVs for $\sim$650,000 early-type galaxies via the FP to map the cosmic velocity field and to measure the growth of structure in the local Universe, up to $z\sim0.12$ \citep{taylor2023}. In this context, the framework presented in this study can potentially improve the precision of the cosmological measurements as well as increasing the number of galaxies up to a million. 



\begin{acknowledgments}
    KS acknowledges support from the Australian Government through the Australian Research Council’s Laureate Fellowship funding scheme (project FL180100168)
\end{acknowledgments}

\bibliography{ref}{}

\begin{thebibliography}{}
\expandafter\ifx\csname natexlab\endcsname\relax\def\natexlab#1{#1}\fi
\providecommand{\url}[1]{\href{#1}{#1}}
\providecommand{\dodoi}[1]{doi:~\href{http://doi.org/#1}{\nolinkurl{#1}}}
\providecommand{\doeprint}[1]{\href{http://ascl.net/#1}{\nolinkurl{http://ascl.net/#1}}}
\providecommand{\doarXiv}[1]{\href{https://arxiv.org/abs/#1}{\nolinkurl{https://arxiv.org/abs/#1}}}

\bibitem[{{Abazajian} {et~al.}(2009){Abazajian}, {Adelman-McCarthy},
  {Ag{\"u}eros}, {Allam}, {Allende Prieto}, {An}, {Anderson}, {Anderson},
  {Annis}, {Bahcall}, {Bailer-Jones}, {Barentine}, {Bassett}, {Becker},
  {Beers}, {Bell}, {Belokurov}, {Berlind}, {Berman}, {Bernardi}, {Bickerton},
  {Bizyaev}, {Blakeslee}, {Blanton}, {Bochanski}, {Boroski}, {Brewington},
  {Brinchmann}, {Brinkmann}, {Brunner}, {Budav{\'a}ri}, {Carey}, {Carliles},
  {Carr}, {Castander}, {Cinabro}, {Connolly}, {Csabai}, {Cunha}, {Czarapata},
  {Davenport}, {de Haas}, {Dilday}, {Doi}, {Eisenstein}, {Evans}, {Evans},
  {Fan}, {Friedman}, {Frieman}, {Fukugita}, {G{\"a}nsicke}, {Gates},
  {Gillespie}, {Gilmore}, {Gonzalez}, {Gonzalez}, {Grebel}, {Gunn},
  {Gy{\"o}ry}, {Hall}, {Harding}, {Harris}, {Harvanek}, {Hawley}, {Hayes},
  {Heckman}, {Hendry}, {Hennessy}, {Hindsley}, {Hoblitt}, {Hogan}, {Hogg},
  {Holtzman}, {Hyde}, {Ichikawa}, {Ichikawa}, {Im}, {Ivezi{\'c}}, {Jester},
  {Jiang}, {Johnson}, {Jorgensen}, {Juri{\'c}}, {Kent}, {Kessler}, {Kleinman},
  {Knapp}, {Konishi}, {Kron}, {Krzesinski}, {Kuropatkin}, {Lampeitl},
  {Lebedeva}, {Lee}, {Lee}, {French Leger}, {L{\'e}pine}, {Li}, {Lima}, {Lin},
  {Long}, {Loomis}, {Loveday}, {Lupton}, {Magnier}, {Malanushenko},
  {Malanushenko}, {Mandelbaum}, {Margon}, {Marriner}, {Mart{\'\i}nez-Delgado},
  {Matsubara}, {McGehee}, {McKay}, {Meiksin}, {Morrison}, {Mullally}, {Munn},
  {Murphy}, {Nash}, {Nebot}, {Neilsen}, {Newberg}, {Newman}, {Nichol},
  {Nicinski}, {Nieto-Santisteban}, {Nitta}, {Okamura}, {Oravetz}, {Ostriker},
  {Owen}, {Padmanabhan}, {Pan}, {Park}, {Pauls}, {Peoples}, {Percival}, {Pier},
  {Pope}, {Pourbaix}, {Price}, {Purger}, {Quinn}, {Raddick}, {Re Fiorentin},
  {Richards}, {Richmond}, {Riess}, {Rix}, {Rockosi}, {Sako}, {Schlegel},
  {Schneider}, {Scholz}, {Schreiber}, {Schwope}, {Seljak}, {Sesar}, {Sheldon},
  {Shimasaku}, {Sibley}, {Simmons}, {Sivarani}, {Allyn Smith}, {Smith},
  {Smol{\v{c}}i{\'c}}, {Snedden}, {Stebbins}, {Steinmetz}, {Stoughton},
  {Strauss}, {SubbaRao}, {Suto}, {Szalay}, {Szapudi}, {Szkody}, {Tanaka},
  {Tegmark}, {Teodoro}, {Thakar}, {Tremonti}, {Tucker}, {Uomoto}, {Vanden
  Berk}, {Vandenberg}, {Vidrih}, {Vogeley}, {Voges}, {Vogt}, {Wadadekar},
  {Watters}, {Weinberg}, {West}, {White}, {Wilhite}, {Wonders}, {Yanny},
  {Yocum}, {York}, {Zehavi}, {Zibetti}, \& {Zucker}}]{abazajian2009}
{Abazajian}, K.~N., {Adelman-McCarthy}, J.~K., {Ag{\"u}eros}, M.~A., {et~al.}
  2009, \apjs, 182, 543, \dodoi{10.1088/0067-0049/182/2/543}

\bibitem[{{Adams} \& {Blake}(2017)}]{adams2017}
{Adams}, C., \& {Blake}, C. 2017, \mnras, 471, 839,
  \dodoi{10.1093/mnras/stx1529}

\bibitem[{{Aquino-Ort{\'\i}z} {et~al.}(2020){Aquino-Ort{\'\i}z}, {S{\'a}nchez},
  {Valenzuela}, {Hern{\'a}ndez-Toledo}, {Jin}, {Zhu}, {van de Ven},
  {Barrera-Ballesteros}, {Avila-Reese}, {Rodr{\'\i}guez-Puebla}, \&
  {Tissera}}]{ortiz2020}
{Aquino-Ort{\'\i}z}, E., {S{\'a}nchez}, S.~F., {Valenzuela}, O., {et~al.} 2020,
  \apj, 900, 109, \dodoi{10.3847/1538-4357/aba94e}

\bibitem[{{Baldry} {et~al.}(2012){Baldry}, {Driver}, {Loveday}, {Taylor},
  {Kelvin}, {Liske}, {Norberg}, {Robotham}, {Brough}, {Hopkins}, {Bamford},
  {Peacock}, {Bland-Hawthorn}, {Conselice}, {Croom}, {Jones}, {Parkinson},
  {Popescu}, {Prescott}, {Sharp}, \& {Tuffs}}]{baldry2012}
{Baldry}, I.~K., {Driver}, S.~P., {Loveday}, J., {et~al.} 2012, \mnras, 421,
  621, \dodoi{10.1111/j.1365-2966.2012.20340.x}

\bibitem[{{Bell} {et~al.}(2003){Bell}, {McIntosh}, {Katz}, \&
  {Weinberg}}]{bell2003}
{Bell}, E.~F., {McIntosh}, D.~H., {Katz}, N., \& {Weinberg}, M.~D. 2003, \apjs,
  149, 289, \dodoi{10.1086/378847}

\bibitem[{{Bernardi} {et~al.}(2020){Bernardi}, {Dom{\'\i}nguez S{\'a}nchez},
  {Margalef-Bentabol}, {Nikakhtar}, \& {Sheth}}]{bernardi2020}
{Bernardi}, M., {Dom{\'\i}nguez S{\'a}nchez}, H., {Margalef-Bentabol}, B.,
  {Nikakhtar}, F., \& {Sheth}, R.~K. 2020, \mnras, 494, 5148,
  \dodoi{10.1093/mnras/staa1064}

\bibitem[{{Bernardi} {et~al.}(2003){Bernardi}, {Sheth}, {Annis}, {Burles},
  {Eisenstein}, {Finkbeiner}, {Hogg}, {Lupton}, {Schlegel}, {SubbaRao},
  {Bahcall}, {Blakeslee}, {Brinkmann}, {Castander}, {Connolly}, {Csabai},
  {Doi}, {Fukugita}, {Frieman}, {Heckman}, {Hennessy}, {Ivezi{\'c}}, {Knapp},
  {Lamb}, {McKay}, {Munn}, {Nichol}, {Okamura}, {Schneider}, {Thakar}, \&
  {York}}]{bernardi2003c}
{Bernardi}, M., {Sheth}, R.~K., {Annis}, J., {et~al.} 2003, \aj, 125, 1866,
  \dodoi{10.1086/367794}

\bibitem[{{Bertin} {et~al.}(2002){Bertin}, {Ciotti}, \& {Del
  Principe}}]{bertin2002}
{Bertin}, G., {Ciotti}, L., \& {Del Principe}, M. 2002, \aap, 386, 149,
  \dodoi{10.1051/0004-6361:20020248}

\bibitem[{{Bezanson} {et~al.}(2015){Bezanson}, {Franx}, \& {van
  Dokkum}}]{bezanson2015}
{Bezanson}, R., {Franx}, M., \& {van Dokkum}, P.~G. 2015, \apj, 799, 148,
  \dodoi{10.1088/0004-637X/799/2/148}

\bibitem[{{Blanton} \& {Moustakas}(2009)}]{blanton2009}
{Blanton}, M.~R., \& {Moustakas}, J. 2009, \araa, 47, 159,
  \dodoi{10.1146/annurev-astro-082708-101734}

\bibitem[{Boys(1989)}]{richardj1989}
Boys, R.~J. 1989, Journal of the Royal Statistical Society. Series C (Applied
  Statistics), 38, 580.
\newblock \url{http://www.jstor.org/stable/2347755}

\bibitem[{{Calcino} \& {Davis}(2017)}]{calcino2017}
{Calcino}, J., \& {Davis}, T. 2017, \jcap, 2017, 038,
  \dodoi{10.1088/1475-7516/2017/01/038}

\bibitem[{{Caon} {et~al.}(1993){Caon}, {Capaccioli}, \& {D'Onofrio}}]{caon1993}
{Caon}, N., {Capaccioli}, M., \& {D'Onofrio}, M. 1993, \mnras, 265, 1013,
  \dodoi{10.1093/mnras/265.4.1013}

\bibitem[{{Cappellari}(2017)}]{cappellari2017}
{Cappellari}, M. 2017, \mnras, 466, 798, \dodoi{10.1093/mnras/stw3020}

\bibitem[{{Cappellari} {et~al.}(2006){Cappellari}, {Bacon}, {Bureau}, {Damen},
  {Davies}, {de Zeeuw}, {Emsellem}, {Falc{\'o}n-Barroso}, {Krajnovi{\'c}},
  {Kuntschner}, {McDermid}, {Peletier}, {Sarzi}, {van den Bosch}, \& {van de
  Ven}}]{cappellari2006}
{Cappellari}, M., {Bacon}, R., {Bureau}, M., {et~al.} 2006, \mnras, 366, 1126,
  \dodoi{10.1111/j.1365-2966.2005.09981.x}

\bibitem[{{Cappellari} {et~al.}(2013){Cappellari}, {Scott}, {Alatalo}, {Blitz},
  {Bois}, {Bournaud}, {Bureau}, {Crocker}, {Davies}, {Davis}, {de Zeeuw},
  {Duc}, {Emsellem}, {Khochfar}, {Krajnovi{\'c}}, {Kuntschner}, {McDermid},
  {Morganti}, {Naab}, {Oosterloo}, {Sarzi}, {Serra}, {Weijmans}, \&
  {Young}}]{cappellari2013}
{Cappellari}, M., {Scott}, N., {Alatalo}, K., {et~al.} 2013, \mnras, 432, 1709,
  \dodoi{10.1093/mnras/stt562}

\bibitem[{Carpenter {et~al.}(2017)Carpenter, Gelman, Hoffman, Lee, Goodrich,
  Betancourt, Brubaker, Guo, Li, \& Riddell}]{carpenter2017}
Carpenter, B., Gelman, A., Hoffman, M.~D., {et~al.} 2017, Journal of
  Statistical Software, 76, 1–32, \dodoi{10.18637/jss.v076.i01}

\bibitem[{{Carrick} {et~al.}(2015){Carrick}, {Turnbull}, {Lavaux}, \&
  {Hudson}}]{carrick2015}
{Carrick}, J., {Turnbull}, S.~J., {Lavaux}, G., \& {Hudson}, M.~J. 2015,
  \mnras, 450, 317, \dodoi{10.1093/mnras/stv547}

\bibitem[{{Colless} {et~al.}(2001){Colless}, {Saglia}, {Burstein}, {Davies},
  {McMahan}, \& {Wegner}}]{colless2001}
{Colless}, M., {Saglia}, R.~P., {Burstein}, D., {et~al.} 2001, \mnras, 321,
  277, \dodoi{10.1046/j.1365-8711.2001.04044.x}

\bibitem[{{Cortese} {et~al.}(2014){Cortese}, {Fogarty}, {Ho}, {Bekki},
  {Bland-Hawthorn}, {Colless}, {Couch}, {Croom}, {Glazebrook}, {Mould},
  {Scott}, {Sharp}, {Tonini}, {Allen}, {Bloom}, {Bryant}, {Cluver}, {Davies},
  {Drinkwater}, {Goodwin}, {Green}, {Kewley}, {Kostantopoulos}, {Lawrence},
  {Mahajan}, {Medling}, {Owers}, {Richards}, {Sweet}, \& {Wong}}]{cortese2014}
{Cortese}, L., {Fogarty}, L.~M.~R., {Ho}, I.~T., {et~al.} 2014, \apjl, 795,
  L37, \dodoi{10.1088/2041-8205/795/2/L37}

\bibitem[{{Courteau} {et~al.}(2014){Courteau}, {Cappellari}, {de Jong},
  {Dutton}, {Emsellem}, {Hoekstra}, {Koopmans}, {Mamon}, {Maraston}, {Treu}, \&
  {Widrow}}]{courteau2014}
{Courteau}, S., {Cappellari}, M., {de Jong}, R.~S., {et~al.} 2014, Reviews of
  Modern Physics, 86, 47, \dodoi{10.1103/RevModPhys.86.47}

\bibitem[{{Dam}(2020)}]{dam2020}
{Dam}, L. 2020, \mnras, 497, 1301, \dodoi{10.1093/mnras/staa2040}

\bibitem[{{de Graaff} {et~al.}(2020){de Graaff}, {Bezanson}, {Franx}, {van der
  Wel}, {Bell}, {D'Eugenio}, {Holden}, {Maseda}, {Muzzin}, {Pacifici}, {van de
  Sande}, {Sobral}, {Straatman}, \& {Wu}}]{degraaff2020}
{de Graaff}, A., {Bezanson}, R., {Franx}, M., {et~al.} 2020, \apjl, 903, L30,
  \dodoi{10.3847/2041-8213/abc428}

\bibitem[{{de Graaff} {et~al.}(2021){de Graaff}, {Bezanson}, {Franx}, {van der
  Wel}, {Holden}, {van de Sande}, {Bell}, {D'Eugenio}, {Maseda}, {Muzzin},
  {Sobral}, {Straatman}, \& {Wu}}]{degraaff2021}
---. 2021, \apj, 913, 103, \dodoi{10.3847/1538-4357/abf1e7}

\bibitem[{{Djorgovski} \& {Davis}(1987)}]{djor1987}
{Djorgovski}, S., \& {Davis}, M. 1987, \apj, 313, 59, \dodoi{10.1086/164948}

\bibitem[{{Dogruel} {et~al.}(2023){Dogruel}, {Taylor}, {Cluver}, {D'Eugenio},
  {de Graaff}, {Colless}, \& {Sonnenfeld}}]{dogruel2023}
{Dogruel}, M.~B., {Taylor}, E.~N., {Cluver}, M., {et~al.} 2023, \apj, 953, 45,
  \dodoi{10.3847/1538-4357/acde56}

\bibitem[{{Dressler} {et~al.}(1987){Dressler}, {Lynden-Bell}, {Burstein},
  {Davies}, {Faber}, {Terlevich}, \& {Wegner}}]{dressler1987}
{Dressler}, A., {Lynden-Bell}, D., {Burstein}, D., {et~al.} 1987, \apj, 313,
  42, \dodoi{10.1086/164947}

\bibitem[{{Falc{\'o}n-Barroso} {et~al.}(2011){Falc{\'o}n-Barroso},
  {S{\'a}nchez-Bl{\'a}zquez}, {Vazdekis}, {Ricciardelli}, {Cardiel}, {Cenarro},
  {Gorgas}, \& {Peletier}}]{falcon-barroso2011}
{Falc{\'o}n-Barroso}, J., {S{\'a}nchez-Bl{\'a}zquez}, P., {Vazdekis}, A.,
  {et~al.} 2011, \aap, 532, A95, \dodoi{10.1051/0004-6361/201116842}

\bibitem[{{Graves} \& {Faber}(2010)}]{graves2010_3}
{Graves}, G.~J., \& {Faber}, S.~M. 2010, \apj, 717, 803,
  \dodoi{10.1088/0004-637X/717/2/803}

\bibitem[{{Graves} {et~al.}(2009){Graves}, {Faber}, \&
  {Schiavon}}]{graves2009_2}
{Graves}, G.~J., {Faber}, S.~M., \& {Schiavon}, R.~P. 2009, \apj, 698, 1590,
  \dodoi{10.1088/0004-637X/698/2/1590}

\bibitem[{{Graziani} {et~al.}(2019){Graziani}, {Courtois}, {Lavaux}, {Hoffman},
  {Tully}, {Copin}, \& {Pomar{\`e}de}}]{graziani2019}
{Graziani}, R., {Courtois}, H.~M., {Lavaux}, G., {et~al.} 2019, \mnras, 488,
  5438, \dodoi{10.1093/mnras/stz078}

\bibitem[{{Harrison}(1974)}]{harrison1974}
{Harrison}, E.~R. 1974, \apjl, 191, L51, \dodoi{10.1086/181545}

\bibitem[{{Hong} {et~al.}(2014){Hong}, {Springob}, {Staveley-Smith},
  {Scrimgeour}, {Masters}, {Macri}, {Koribalski}, {Jones}, \&
  {Jarrett}}]{hong2014}
{Hong}, T., {Springob}, C.~M., {Staveley-Smith}, L., {et~al.} 2014, \mnras,
  445, 402, \dodoi{10.1093/mnras/stu1774}

\bibitem[{{Hopkins} {et~al.}(2013){Hopkins}, {Driver}, {Brough}, {Owers},
  {Bauer}, {Gunawardhana}, {Cluver}, {Colless}, {Foster}, {Lara-L{\'o}pez},
  {Roseboom}, {Sharp}, {Steele}, {Thomas}, {Baldry}, {Brown}, {Liske},
  {Norberg}, {Robotham}, {Bamford}, {Bland-Hawthorn}, {Drinkwater}, {Loveday},
  {Meyer}, {Peacock}, {Tuffs}, {Agius}, {Alpaslan}, {Andrae}, {Cameron},
  {Cole}, {Ching}, {Christodoulou}, {Conselice}, {Croom}, {Cross}, {De
  Propris}, {Delhaize}, {Dunne}, {Eales}, {Ellis}, {Frenk}, {Graham},
  {Grootes}, {H{\"a}u{\ss}ler}, {Heymans}, {Hill}, {Hoyle}, {Hudson}, {Jarvis},
  {Johansson}, {Jones}, {van Kampen}, {Kelvin}, {Kuijken},
  {L{\'o}pez-S{\'a}nchez}, {Maddox}, {Madore}, {Maraston}, {McNaught-Roberts},
  {Nichol}, {Oliver}, {Parkinson}, {Penny}, {Phillipps}, {Pimbblet}, {Ponman},
  {Popescu}, {Prescott}, {Proctor}, {Sadler}, {Sansom}, {Seibert},
  {Staveley-Smith}, {Sutherland}, {Taylor}, {Van Waerbeke}, {V{\'a}zquez-Mata},
  {Warren}, {Wijesinghe}, {Wild}, \& {Wilkins}}]{hopkins2013}
{Hopkins}, A.~M., {Driver}, S.~P., {Brough}, S., {et~al.} 2013, \mnras, 430,
  2047, \dodoi{10.1093/mnras/stt030}

\bibitem[{{Howlett} {et~al.}(2022){Howlett}, {Said}, {Lucey}, {Colless}, {Qin},
  {Lai}, {Tully}, \& {Davis}}]{howlett2022}
{Howlett}, C., {Said}, K., {Lucey}, J.~R., {et~al.} 2022, \mnras, 515, 953,
  \dodoi{10.1093/mnras/stac1681}

\bibitem[{{Howlett} {et~al.}(2017){Howlett}, {Staveley-Smith}, {Elahi}, {Hong},
  {Jarrett}, {Jones}, {Koribalski}, {Macri}, {Masters}, \&
  {Springob}}]{howlett2017}
{Howlett}, C., {Staveley-Smith}, L., {Elahi}, P.~J., {et~al.} 2017, \mnras,
  471, 3135, \dodoi{10.1093/mnras/stx1521}

\bibitem[{{Hubble}(1929)}]{hubble1929}
{Hubble}, E. 1929, Proceedings of the National Academy of Science, 15, 168,
  \dodoi{10.1073/pnas.15.3.168}

\bibitem[{{Hyde} \& {Bernardi}(2009)}]{hyde2009}
{Hyde}, J.~B., \& {Bernardi}, M. 2009, \mnras, 396, 1171,
  \dodoi{10.1111/j.1365-2966.2009.14783.x}

\bibitem[{{Jones} {et~al.}(2009){Jones}, {Read}, {Saunders}, {Colless},
  {Jarrett}, {Parker}, {Fairall}, {Mauch}, {Sadler}, {Watson}, {Burton},
  {Campbell}, {Cass}, {Croom}, {Dawe}, {Fiegert}, {Frankcombe}, {Hartley},
  {Huchra}, {James}, {Kirby}, {Lahav}, {Lucey}, {Mamon}, {Moore}, {Peterson},
  {Prior}, {Proust}, {Russell}, {Safouris}, {Wakamatsu}, {Westra}, \&
  {Williams}}]{jones2009}
{Jones}, D.~H., {Read}, M.~A., {Saunders}, W., {et~al.} 2009, \mnras, 399, 683,
  \dodoi{10.1111/j.1365-2966.2009.15338.x}

\bibitem[{{Jorgensen} {et~al.}(1995){Jorgensen}, {Franx}, \&
  {Kjaergaard}}]{jorgensen1995}
{Jorgensen}, I., {Franx}, M., \& {Kjaergaard}, P. 1995, \mnras, 276, 1341,
  \dodoi{10.1093/mnras/276.4.1341}

\bibitem[{{Kaiser}(1987)}]{kaiser1987}
{Kaiser}, N. 1987, \mnras, 227, 1, \dodoi{10.1093/mnras/227.1.1}

\bibitem[{{Koda} {et~al.}(2014){Koda}, {Blake}, {Davis}, {Magoulas},
  {Springob}, {Scrimgeour}, {Johnson}, {Poole}, \& {Staveley-Smith}}]{koda2014}
{Koda}, J., {Blake}, C., {Davis}, T., {et~al.} 2014, \mnras, 445, 4267,
  \dodoi{10.1093/mnras/stu1610}

\bibitem[{{Kourkchi} {et~al.}(2020){Kourkchi}, {Tully}, {Eftekharzadeh},
  {Llop}, {Courtois}, {Guinet}, {Dupuy}, {Neill}, {Seibert}, {Andrews},
  {Chuang}, {Danesh}, {Gonzalez}, {Holthaus}, {Mokelke}, {Schoen}, \&
  {Urasaki}}]{kourkchi2020}
{Kourkchi}, E., {Tully}, R.~B., {Eftekharzadeh}, S., {et~al.} 2020, \apj, 902,
  145, \dodoi{10.3847/1538-4357/abb66b}

\bibitem[{{Lange} {et~al.}(2015){Lange}, {Driver}, {Robotham}, {Kelvin},
  {Graham}, {Alpaslan}, {Andrews}, {Baldry}, {Bamford}, {Bland-Hawthorn},
  {Brough}, {Cluver}, {Conselice}, {Davies}, {Haeussler}, {Konstantopoulos},
  {Loveday}, {Moffett}, {Norberg}, {Phillipps}, {Taylor},
  {L{\'o}pez-S{\'a}nchez}, \& {Wilkins}}]{lange2015}
{Lange}, R., {Driver}, S.~P., {Robotham}, A. S.~G., {et~al.} 2015, \mnras, 447,
  2603, \dodoi{10.1093/mnras/stu2467}

\bibitem[{{Leavitt} \& {Pickering}(1912)}]{leavitt1912}
{Leavitt}, H.~S., \& {Pickering}, E.~C. 1912, Harvard College Observatory
  Circular, 173, 1

\bibitem[{{Magoulas} {et~al.}(2012){Magoulas}, {Springob}, {Colless}, {Jones},
  {Campbell}, {Lucey}, {Mould}, {Jarrett}, {Merson}, \&
  {Brough}}]{magoulas2012}
{Magoulas}, C., {Springob}, C.~M., {Colless}, M., {et~al.} 2012, \mnras, 427,
  245, \dodoi{10.1111/j.1365-2966.2012.21421.x}

\bibitem[{Owen(1980)}]{owen1980}
Owen, D.~B. 1980, Communications in Statistics - Simulation and Computation, 9,
  389, \dodoi{10.1080/03610918008812164}

\bibitem[{{Phillips}(1993)}]{phillips1993}
{Phillips}, M.~M. 1993, \apjl, 413, L105, \dodoi{10.1086/186970}

\bibitem[{{Qin} {et~al.}(2021){Qin}, {Parkinson}, {Howlett}, \&
  {Said}}]{qin2021}
{Qin}, F., {Parkinson}, D., {Howlett}, C., \& {Said}, K. 2021, \apj, 922, 59,
  \dodoi{10.3847/1538-4357/ac249d}

\bibitem[{{Riess} {et~al.}(2021){Riess}, {Casertano}, {Yuan}, {Bowers},
  {Macri}, {Zinn}, \& {Scolnic}}]{riess2021}
{Riess}, A.~G., {Casertano}, S., {Yuan}, W., {et~al.} 2021, \apjl, 908, L6,
  \dodoi{10.3847/2041-8213/abdbaf}

\bibitem[{{Riess} {et~al.}(2016){Riess}, {Macri}, {Hoffmann}, {Scolnic},
  {Casertano}, {Filippenko}, {Tucker}, {Reid}, {Jones}, {Silverman},
  {Chornock}, {Challis}, {Yuan}, {Brown}, \& {Foley}}]{riess2016}
{Riess}, A.~G., {Macri}, L.~M., {Hoffmann}, S.~L., {et~al.} 2016, \apj, 826,
  56, \dodoi{10.3847/0004-637X/826/1/56}

\bibitem[{{Robotham} {et~al.}(2011){Robotham}, {Norberg}, {Driver}, {Baldry},
  {Bamford}, {Hopkins}, {Liske}, {Loveday}, {Merson}, {Peacock}, {Brough},
  {Cameron}, {Conselice}, {Croom}, {Frenk}, {Gunawardhana}, {Hill}, {Jones},
  {Kelvin}, {Kuijken}, {Nichol}, {Parkinson}, {Pimbblet}, {Phillipps},
  {Popescu}, {Prescott}, {Sharp}, {Sutherland}, {Taylor}, {Thomas}, {Tuffs},
  {van Kampen}, \& {Wijesinghe}}]{robotham2011}
{Robotham}, A.~S.~G., {Norberg}, P., {Driver}, S.~P., {et~al.} 2011, \mnras,
  416, 2640, \dodoi{10.1111/j.1365-2966.2011.19217.x}

\bibitem[{{Said} {et~al.}(2020){Said}, {Colless}, {Magoulas}, {Lucey}, \&
  {Hudson}}]{khaled2020}
{Said}, K., {Colless}, M., {Magoulas}, C., {Lucey}, J.~R., \& {Hudson}, M.~J.
  2020, \mnras, 497, 1275, \dodoi{10.1093/mnras/staa2032}

\bibitem[{{S{\'a}nchez-Bl{\'a}zquez} {et~al.}(2006){S{\'a}nchez-Bl{\'a}zquez},
  {Peletier}, {Jim{\'e}nez-Vicente}, {Cardiel}, {Cenarro},
  {Falc{\'o}n-Barroso}, {Gorgas}, {Selam}, \&
  {Vazdekis}}]{sanchez-blazquez2006}
{S{\'a}nchez-Bl{\'a}zquez}, P., {Peletier}, R.~F., {Jim{\'e}nez-Vicente}, J.,
  {et~al.} 2006, \mnras, 371, 703, \dodoi{10.1111/j.1365-2966.2006.10699.x}

\bibitem[{{Schmidt}(1968)}]{schmidt1968}
{Schmidt}, M. 1968, \apj, 151, 393, \dodoi{10.1086/149446}

\bibitem[{{Scolnic} {et~al.}(2018){Scolnic}, {Jones}, {Rest}, {Pan},
  {Chornock}, {Foley}, {Huber}, {Kessler}, {Narayan}, {Riess}, {Rodney},
  {Berger}, {Brout}, {Challis}, {Drout}, {Finkbeiner}, {Lunnan}, {Kirshner},
  {Sanders}, {Schlafly}, {Smartt}, {Stubbs}, {Tonry}, {Wood-Vasey}, {Foley},
  {Hand}, {Johnson}, {Burgett}, {Chambers}, {Draper}, {Hodapp}, {Kaiser},
  {Kudritzki}, {Magnier}, {Metcalfe}, {Bresolin}, {Gall}, {Kotak}, {McCrum}, \&
  {Smith}}]{scolnic2018}
{Scolnic}, D.~M., {Jones}, D.~O., {Rest}, A., {et~al.} 2018, \apj, 859, 101,
  \dodoi{10.3847/1538-4357/aab9bb}

\bibitem[{{Springob} {et~al.}(2012){Springob}, {Magoulas}, {Proctor},
  {Colless}, {Jones}, {Kobayashi}, {Campbell}, {Lucey}, \&
  {Mould}}]{springob2012}
{Springob}, C.~M., {Magoulas}, C., {Proctor}, R., {et~al.} 2012, \mnras, 420,
  2773, \dodoi{10.1111/j.1365-2966.2011.19900.x}

\bibitem[{{Springob} {et~al.}(2014){Springob}, {Magoulas}, {Colless}, {Mould},
  {Erdo{\u{g}}du}, {Jones}, {Lucey}, {Campbell}, \& {Fluke}}]{springob2014}
{Springob}, C.~M., {Magoulas}, C., {Colless}, M., {et~al.} 2014, \mnras, 445,
  2677, \dodoi{10.1093/mnras/stu1743}

\bibitem[{{Taylor} {et~al.}(2010){Taylor}, {Franx}, {Brinchmann}, {van der
  Wel}, \& {van Dokkum}}]{taylor2010}
{Taylor}, E.~N., {Franx}, M., {Brinchmann}, J., {van der Wel}, A., \& {van
  Dokkum}, P.~G. 2010, \apj, 722, 1, \dodoi{10.1088/0004-637X/722/1/1}

\bibitem[{{Taylor} {et~al.}(2011){Taylor}, {Hopkins}, {Baldry}, {Brown},
  {Driver}, {Kelvin}, {Hill}, {Robotham}, {Bland -Hawthorn}, {Jones}, {Sharp},
  {Thomas}, {Liske}, {Loveday}, {Norberg}, {Peacock}, {Bamford}, {Brough},
  {Colless}, {Cameron}, {Conselice}, {Croom}, {Frenk}, {Gunawardhana},
  {Kuijken}, {Nichol}, {Parkinson}, {Phillipps}, {Pimbblet}, {Popescu},
  {Prescott}, {Sutherland}, {Tuffs}, {van Kampen}, \&
  {Wijesinghe}}]{taylor2011}
{Taylor}, E.~N., {Hopkins}, A.~M., {Baldry}, I.~K., {et~al.} 2011, \mnras, 418,
  1587, \dodoi{10.1111/j.1365-2966.2011.19536.x}

\bibitem[{{Taylor} {et~al.}(2015){Taylor}, {Hopkins}, {Baldry},
  {Bland-Hawthorn}, {Brown}, {Colless}, {Driver}, {Norberg}, {Robotham},
  {Alpaslan}, {Brough}, {Cluver}, {Gunawardhana}, {Kelvin}, {Liske},
  {Conselice}, {Croom}, {Foster}, {Jarrett}, {Lara-Lopez}, \&
  {Loveday}}]{taylor2015}
---. 2015, \mnras, 446, 2144, \dodoi{10.1093/mnras/stu1900}

\bibitem[{{Taylor} {et~al.}(2023){Taylor}, {Cluver}, {Bell}, {Brinchmann},
  {Colless}, {Courtois}, {Hoekstra}, {Kannappan}, {Lagos}, {Liske}, {Tempel},
  {Howlett}, {McGee}, {Said}, {Skelton}, {Gunawardhana}, {Bellstedt}, {Hunt},
  {Jarrett}, {Lidman}, {Lucey}, {Alam}, {Bilicki}, {de Graaff}, {Hellwing},
  {Leslie}, {Loubser}, {Marchetti}, {Maseda}, {Mogotsi}, {Norberg},
  {Sonnenfeld}, {Sorce}, \& {4HS Team}}]{taylor2023}
{Taylor}, E.~N., {Cluver}, M., {Bell}, E., {et~al.} 2023, The Messenger, 190,
  46, \dodoi{10.18727/0722-6691/5312}

\bibitem[{{Tonry} {et~al.}(2000){Tonry}, {Blakeslee}, {Ajhar}, \&
  {Dressler}}]{tonry2000}
{Tonry}, J.~L., {Blakeslee}, J.~P., {Ajhar}, E.~A., \& {Dressler}, A. 2000,
  \apj, 530, 625, \dodoi{10.1086/308409}

\bibitem[{{Tully} {et~al.}(2014){Tully}, {Courtois}, {Hoffman}, \&
  {Pomar{\`e}de}}]{tully2014}
{Tully}, R.~B., {Courtois}, H., {Hoffman}, Y., \& {Pomar{\`e}de}, D. 2014,
  \nat, 513, 71, \dodoi{10.1038/nature13674}

\bibitem[{{Tully} {et~al.}(2016){Tully}, {Courtois}, \& {Sorce}}]{tully2016}
{Tully}, R.~B., {Courtois}, H.~M., \& {Sorce}, J.~G. 2016, \aj, 152, 50,
  \dodoi{10.3847/0004-6256/152/2/50}

\bibitem[{{Tully} \& {Fisher}(1977)}]{tullyfisher1977}
{Tully}, R.~B., \& {Fisher}, J.~R. 1977, \aap, 500, 105

\bibitem[{{Turnbull} {et~al.}(2012){Turnbull}, {Hudson}, {Feldman}, {Hicken},
  {Kirshner}, \& {Watkins}}]{turnbull2012}
{Turnbull}, S.~J., {Hudson}, M.~J., {Feldman}, H.~A., {et~al.} 2012, \mnras,
  420, 447, \dodoi{10.1111/j.1365-2966.2011.20050.x}

\bibitem[{{van der Wel} {et~al.}(2022){van der Wel}, {van Houdt}, {Bezanson},
  {Franx}, {D'Eugenio}, {Straatman}, {Bell}, {Muzzin}, {Sobral}, {Maseda}, {de
  Graaff}, \& {Holden}}]{wel2022}
{van der Wel}, A., {van Houdt}, J., {Bezanson}, R., {et~al.} 2022, \apj, 936,
  9, \dodoi{10.3847/1538-4357/ac83c5}

\bibitem[{{Watkins} \& {Feldman}(2015)}]{watkins2015}
{Watkins}, R., \& {Feldman}, H.~A. 2015, \mnras, 450, 1868,
  \dodoi{10.1093/mnras/stv651}

\bibitem[{{Weiner} {et~al.}(2006){Weiner}, {Willmer}, {Faber}, {Melbourne},
  {Kassin}, {Phillips}, {Harker}, {Metevier}, {Vogt}, \& {Koo}}]{weiner2006}
{Weiner}, B.~J., {Willmer}, C. N.~A., {Faber}, S.~M., {et~al.} 2006, \apj, 653,
  1027, \dodoi{10.1086/508921}

\bibitem[{{Zaritsky} {et~al.}(2008){Zaritsky}, {Zabludoff}, \&
  {Gonzalez}}]{zaritsky2008}
{Zaritsky}, D., {Zabludoff}, A.~I., \& {Gonzalez}, A.~H. 2008, \apj, 682, 68,
  \dodoi{10.1086/529577}

\bibitem[{{Zibetti} {et~al.}(2009){Zibetti}, {Charlot}, \& {Rix}}]{zibetti2009}
{Zibetti}, S., {Charlot}, S., \& {Rix}, H.-W. 2009, \mnras, 400, 1181,
  \dodoi{10.1111/j.1365-2966.2009.15528.x}

\end{thebibliography}
\bibliographystyle{aasjournal}

\appendix

\section{Consistent velocity dispersion measurements across heterogeneous spectroscopic data}

Many targets in the GAMA sample have been previously observed by other spectroscopic surveys; most notably SDSS, 2dFGRS, and 6dFGS.
The spectra from these surveys have been retrieved and incorporated into the GAMA database, including for the purpose of measuring redshifts, emission lines, etc.  
Some of these targets have been re-observed by GAMA, mostly as filler and/or low priority targets.
This means that, while the GAMA redshift sample is well-defined and virtually complete, the spectroscopic data quality is inhomogeneous, and there are potential systematic effects across the selection boundaries of the different surveys.
This Appendix describes the process by which we have measured velocity dispersions from this heterogeneous spectral dataset, and how we have ensured consistency across the measurements obtained from different surveys' spectra.

Our velocity dispersion measurements are derived using \textsc{pPXF} \citep{cappellari2017}, which forward-models the observed data as a linear combination of template stellar spectra broadened with a Gaussian convolution kernel.
We use the MILES stellar library \citep{sanchez-blazquez2006} as our template set, adopting 2.51~\AA\ as a consensus value for the effective spectral resolution \citep{falcon-barroso2011}.
One important decision is the inclusion additive and multiplicative polynomials in the fits: we use an 10th order additive Legendre polynomial and a 5th order multiplicative Legendre polynomial.
This step is strictly necessary for the 2dFGRS and 6dFGS spectra, which are not flux calibrated.
As well as sidestepping potential issues in background subtraction and/or flux calibration in the spectra, the inclusion of these polynomials mean that the results are constrained by the shapes of broad absorption features in the stellar continuum rather than overall continuum shape/color or the absolute/relative strengths of different spectral features.

The spectral resolution for the data is a critical input to the process, since this cannot be empirically separated from the true intrinsic broadening in the spectrum.
Our assumed spectral resolutions for each survey are given in Table 1, including an independent characterization of the spectral resolution for 6dF, which is based on sky line measurements.
Compared to the estimates from \citet{jones2009}, this reduces the inferred 6dF velocities dispersions by $\approx 5$ \%, and is necessary to bring the 6dF-derived measurements into good agreement with other surveys.
The template spectra are smoothed to match this resolution and then both templates and data are rebinned to a common log-wavelength grid before the \textsc{pPXF} fits.

We use a two-stage process to measure and subtract strong emission lines and isolate the stellar continuum in the observed spectra.
In the first stage, spectra are fit as a combination of stellar templates plus kinematically distinct sets of both Balmer and also forbidden emission line templates.
Any emission lines that are detected at $>5 \sigma$ in this first stage are marked and retained; all other lines are discounted.
This initial stage protects against over-fitting in the second and final stage.

\begin{figure}
    \centering
    \includegraphics[height=0.315\textwidth]{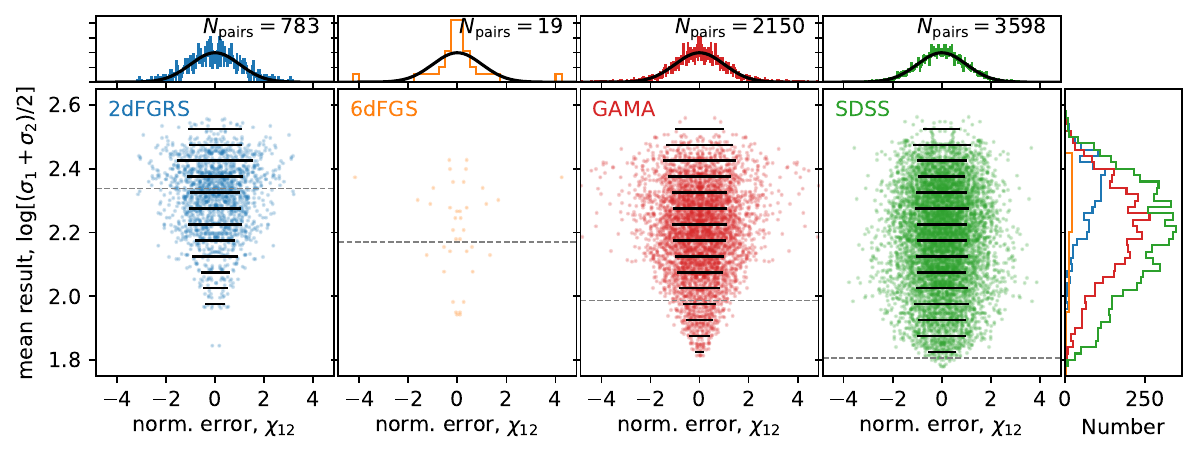}
    \caption{Intra-survey comparisons of velocity dispersion measurements from repeat observations.  From left to right, we show pair-wise comparisons of velocity dispersion measurements from repeat observations by 2dFGRS, 6dFGS, GAMA, and SDSS, each plotted as a function of the pair-wise mean value, and normalised by the reported errors, added in quadrature.  These results can be used to rescale/calibrate the reported uncertainties on the measurements from each data source.}
    \label{fig:random_errors}
\end{figure}

\begin{figure}
    \centering
    \includegraphics[height=0.315\textwidth]{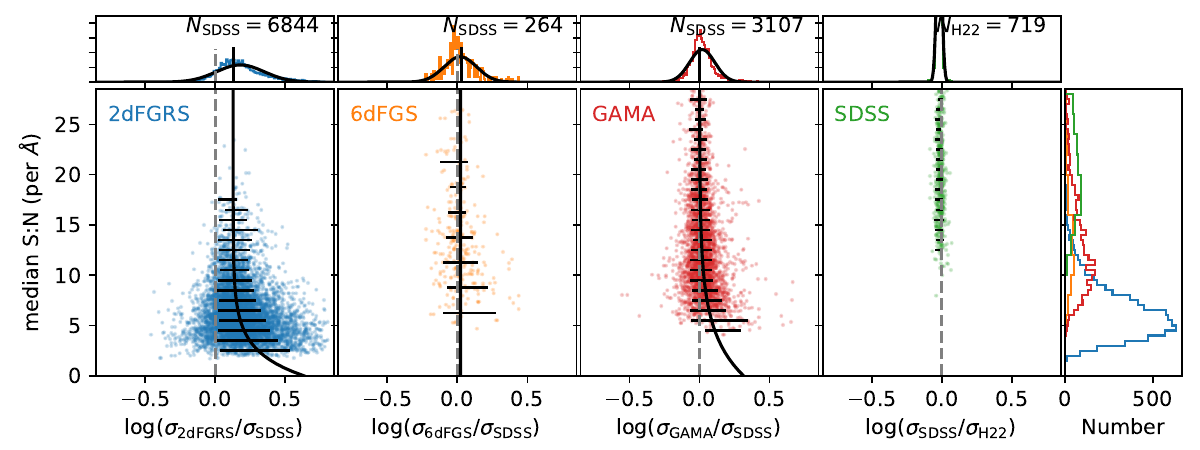}
    \caption{({\em left:}) Cross-survey comparisons of velocity dispersion measurements for common targets between SDSS and other surveys.  In the first three panels we show, From left to right, pair-wise comparisons of velocity dispersion measurements for common targets between SDSS and each of 2dFGRS, 6dFGS, and GAMA, each plotted as a function of median S/N in that survey. These results can be used to quantify/recalibrate bias in the measurements at low S/N, as shown by the black curves. In the final panel we compare our measurements based on SDSS spectra to the measurements from \citet{howlett2022}, which are based on a similar (but not identical) process applied to the same spectra.  This provides a means to calibrate both random and systematic errors associated with the choice of algorithm. }
    \label{fig:systematic_errors}
\end{figure}

The error propagation from counts on the detector to velocity dispersion measurements is not linear, and there are significant unmodelled sources of error in the measurements (including through seeing variations).  
Thus is some re-scaling of the estimated uncertainties on the inferred velocity dispersions is justified.
We use repeat observations of the same targets within each survey to calibrate the quoted errors/uncertainties, by finding the scalar value, $u$, required to bring the NMAD spread between repeat observations (of the same target by the same survey) to have a nominal $\chi^2$ equal to 1.
Note that for 6dFGS, we have performed this exercise for the full 6dFGS spectral database, not only the 19 6dFGS galaxies that were re-observed by GAMA.
The results,  shown in Figure {\ref{fig:random_errors}}, do not show significant variation as a function of measured velocity dispersion, except where the inferred values are comparable to the resolution limit of the spectra.
(We have also checked this explicitly by computing the scaling factor in decile bins of the pairwise mean observed velocity dispersion and signal-to-noise, and see no effect.)

Next we have compared cross-survey repeats of the same targets to test for potential systematic errors across the heterogeneous dataset, as shown in Figure {\ref{fig:systematic_errors}}.
Given that observed velocity dispersion peaks in the center and decreases approximately monotonically with projected radius, some systematic differences are to be expected across surveys, due to differing fiber sizes, different median seeing, etc.
To bring measurements from the different surveys to a common standard, we re-calibrate each survey to match the values derived from SDSS spectra, since this is the survey with the highest spectral resolution and widest area.
We fit a function of the form $f(S; a, b, c) = b \cdot \exp[-a \, (S - 5)] + c$ to the cross-survey data comparisons, where $S$ is the median S/N across the available spectrum in units of \AA$^{-1}$, and the values of the coefficients $a$, $b$, and $c$ are survey specific.
For 6dFGS, we find the value of the coefficient $b$ to be statistically indistinguishable from zero, so we drop this value for the fit.
Further, for GAMA, we find the values of both $b$ and $c$ to be consistent with 0; that is, we find no clear statistical evidence of the need for a correction, and so no correction is applied.
These cross-calibrated values, which are reported in the GAMA \texttt{VelocityDispersions} DMU as \texttt{SIG\_STARCORR} are what we use in this paper, and what are recommended for use where homogeneous velocity dispersion measurements are required.

In the final panel of Figure \ref{fig:systematic_errors}, we show a spectrum-by-spectrum match between 719 of the velocity dispersion measurements described here and those presented by \citet{howlett2022}, which are derived following a very similar process.
Noting that this comparison is restricted to a high S/N subset of the spectra we consider here, the mean and RMS difference between our measurements and those from \citet{howlett2022} are -0.018 and 0.015 dex respectively.  
The systematic difference is small, but non-negligible, and left uncorrected would induce a systematic bias in the inferred distances/velocities between our two catalogues.
Similarly, the RMS differences are small, but significant, considering that the two values are derived from the same high signal-to-noise measurements; they are in fact comparable in size to the median statistical uncertainty on each measurement, which is $\approx 0.02$ dex.
This shows how, for S/N $\gtrsim 15$, both the precision and the accuracy of these values are limited by the analytic methods used to derive the velocity dispersion measurements, at least as much if not more so than statistical measurement errors in the spectra themselves.

\begin{table}
    \centering
    \begin{tabular}{lcccccccccc}
    \hline
       Survey & $\lambda$ (\AA) & $\lambda R(\lambda)$ & $\sigma_\mathrm{lim}$ (km/s) & $N_\mathrm{targ}$ &
       $N_\mathrm{meas}$ & 
            $u$ & $a$ & $b$  & $c$    \\
        (1)   & (2)             & (3)                   &  (4)                  &  
            (5) & (6) & (7)  & (8)  & (9) & (10) \\
    \hline
       2dFGRS & 3546--6849      &  9.0                  & 218                   &  14720 & 13782 & 
            1.01 & 0.479  & 0.215 & 0.107  \\
       6dFGS  & 3772--7220      & 1$3.764 (\lambda / 10^4 \text{\AA} + 1.1295)$ & 148                   &
           974 & 952 & ---  & 0.738  & 0  & 0.0035  \\
       GAMA   & 3772--7220      & 4.8 (4.5) @ 4800 (7250) \AA &  97                    &  
           88504 & 85687 & 1.59 & 0  & 0  & 0    \\
       SDSS   & 3541--7407      & from \texttt{WDISP} FITS ext. &  64                    &  
            26818 & 23122 & 0.93 & --- & --- & ---   \\
    \hline
    \end{tabular}
    \caption{Summary of spectral data sources and re-calibration parameters.  Columns: (1)~name of survey; (2)~median useful wavelength range (constrained by both the data and the MILES template library); (3)~adopted spectral resolutions, which for 2dFGRS, GAMA, and SDSS are as described by \citet{colless2001}, \citet{hopkins2013}, and \citet{abazajian2009}, respectively; (4)~effective velocity dispersion measurement limit as determined by the median spectral resolution; (5)~number of velocity dispersion measurements from each survey in the catalog; (6)~number of unique galaxies with velocity dispersion measurements from each survey in the catalog; (7)~error rescaling factor for each survey, derived from intra-survey repeat observations as shown in Fig.\ \ref{fig:random_errors}; (8--10)~bias correction parameters for each survey, derived from cross-survey comparisons with SDSS as shown in Fig.\ \ref{fig:systematic_errors}. The final catalog contains a total of 132326 independent measurements for 111831 unique targets.
    \label{tab:my_label} }
\end{table}

\section{Direct vs. orthogonal coefficients in distance determination from the FP}\label{sec:direct_vs_orth_planes}

The importance of obtaining the direct coefficients in distance determinations, instead of the orthogonal ones that define the underlying distribution, can be better understood when we algebraically find the coefficients that maximize $p(r | s, i)$ given our best-fitting model. In the 3D Gaussian case, we have $\bm{x}-\bm{\bar{x}} = (r-\bar{r}, s-\bar{s}, i-\bar{i})$ and the best-fitting model that can be easily obtained from our parent model is $\bm{x}\sim \mathcal{N}(\bm{\bar{x}}, \bm{\Sigma_\text{fp}})$, which makes the log-likelihood,
\begin{equation}
    \ln p(r | s, i, \bm{\bar{x}}, \bm{\Sigma_\text{fp}} ) = -\frac{1}{2} (\bm{x}-\bm{\bar{x}})\bm{\Sigma_\text{fp}}^{-1}(\bm{x}-\bm{\bar{x}})^\intercal + \text{constant}.
    \label{eq:loglike}
\end{equation}
After expanding the matrix multiplication and considering that $s$ and $i$ are fixed at their observed values, this equation can be reduced to
\begin{equation}
    \ln p(r | s, i, \bm{\bar{x}}, \bm{\Sigma_\text{fp}} ) = \gamma_0 + \gamma_1 (r-\bar{r}) + \gamma_2 (r-\bar{r})^2,
\end{equation}
where,
\begin{align}
    \gamma_1 &= 2 [(s-\bar{s})\Lambda_{12} + (i-\bar{i})\Lambda_{13}] ,\nonumber \\
    \gamma_2 &= \Lambda_{13},~ \Lambda = \bm{\Sigma_\text{fp}}^{-1},
\end{align}
and $\gamma_0$ encapsulates all the constants arising during the process. We are looking for the values, $r_*$, that will maximize equation~(\ref{eq:loglike}), in other words, $\partial \ln p/\partial r = 0$. Therefore, $2\gamma_2 (r_* - \bar{r}) + \gamma_1 = 0$, which leads to,
\begin{equation}
    r_* = -\frac{\Lambda_{12}}{\Lambda_{11}}s - \frac{\Lambda_{13}}{\Lambda_{11}}i + \bar{r} + \frac{\Lambda_{12}}{\Lambda_{11}}\bar{s} + \frac{\Lambda_{13}}{\Lambda_{11}}\bar{i}.
    \label{eq:fplambda}
\end{equation}
This is just $r_* = as + bi + c$ with $a=-\Lambda_{12}/\Lambda_{11}$, $b=-\Lambda_{13}/\Lambda_{11}$ and $c=\bar{r}-a\bar{s}-b\bar{i}$. Now, expanding $\Lambda=\bm{\Sigma_\text{fp}}^{-1}$,
\begin{align}
    \Lambda_{11} &= \frac{\sigma_s^2 \sigma_i^2 - \sigma_{si}^2}{|\bm{\Sigma_\text{fp}}|},\quad 
    \Lambda_{12} = \frac{\sigma_{ri} \sigma_{si} - \sigma_{rs} \sigma_i^2}{|\bm{\Sigma_\text{fp}}|},\nonumber\\
    \Lambda_{13} &= \frac{\sigma_{rs} \sigma_{si} - \sigma_i^2 \sigma_{ri}}{|\bm{\Sigma_\text{fp}}|}
\end{align}
which finally gives,
\begin{equation}
    a = \frac{\sigma_{ri}\sigma_{si} - \sigma_{rs}\sigma_i^2}{\sigma_{si}^2 - \sigma_s^2\sigma_i^2},\quad b=\frac{\sigma_{rs}\sigma_{si} - \sigma_{ri}\sigma_s^2}{\sigma_{si}^2 - \sigma_s^2\sigma_i^2}.
    \label{eq:fpdirect}
\end{equation}
These are the direct coefficients given by \cite{bernardi2003c} and they minimize the residuals in $r-$direction (i.e., ordinary least squares -- OLS), which is what really matters for distance measurements. 

\section{Mock Galaxy Catalogues}\label{sec:mocks}

We generate mock samples starting from the algorithm of \cite{magoulas2012} to create FP parameters $(r, s, i)\equiv(\log R_e, \log\sigma_e, \log\langle I_e\rangle)$, then, we calculate other galaxy parameters using relevant scaling relations. These samples will be used in validation of the fitting method (i.e., determination of the accuracy and precision) we describe in section \ref{sec:method} and also in estimating the errors in the best-fit parameters, as in \cite{magoulas2012, khaled2020}.

The $v-$space comprises of three orthonormal vectors $(\bm{\hat{v}_1}, \bm{\hat{v}_2}, \bm{\hat{v}_3})$ that define the axes of the 3D Gaussian of the FP and they are expressed in terms of the FP slopes $a$ and $b$ \citep[][]{colless2001, magoulas2012}. The distribution of galaxies in the FP space will therefore be oriented around these axes with variances $\sigma_1, \sigma_2, \sigma_3$ and the center of the 3D Gaussian will be at the mean values $(\bar{r}, \bar{s}, \bar{i})$ of the FP observables.

To generate other galaxy parameters, we use four scaling relations: stellar-mass vs color \citep[][]{taylor2015}, stellar-mass-to-light vs color \citep[][]{taylor2011}, stellar-mass vs dynamical mass \citep[][]{taylor2010} and size vs S\'ersic index \citep[][]{caon1993} each having slope $(A_{XY})$, intercept $(B_{XY})$ and intrinsic scatter $(\sigma_{XY})$. These relations have the form $Y=A_{XY} X + B_{XY} + \epsilon_{XY}$ where $\epsilon_{XY}=\mathcal{N}(0,\sigma_{XY})$ represents Gaussian scatter around the mean linear relation. As such, we use 20 parameters in total to generate an extensive mock galaxy catalogue. The input values for these parameters are obtained from our best-fitting model to the GAMA sample defined in section \ref{sec:data_method}.

\begin{figure*}
    \centering
    \includegraphics[width=0.88\textwidth]{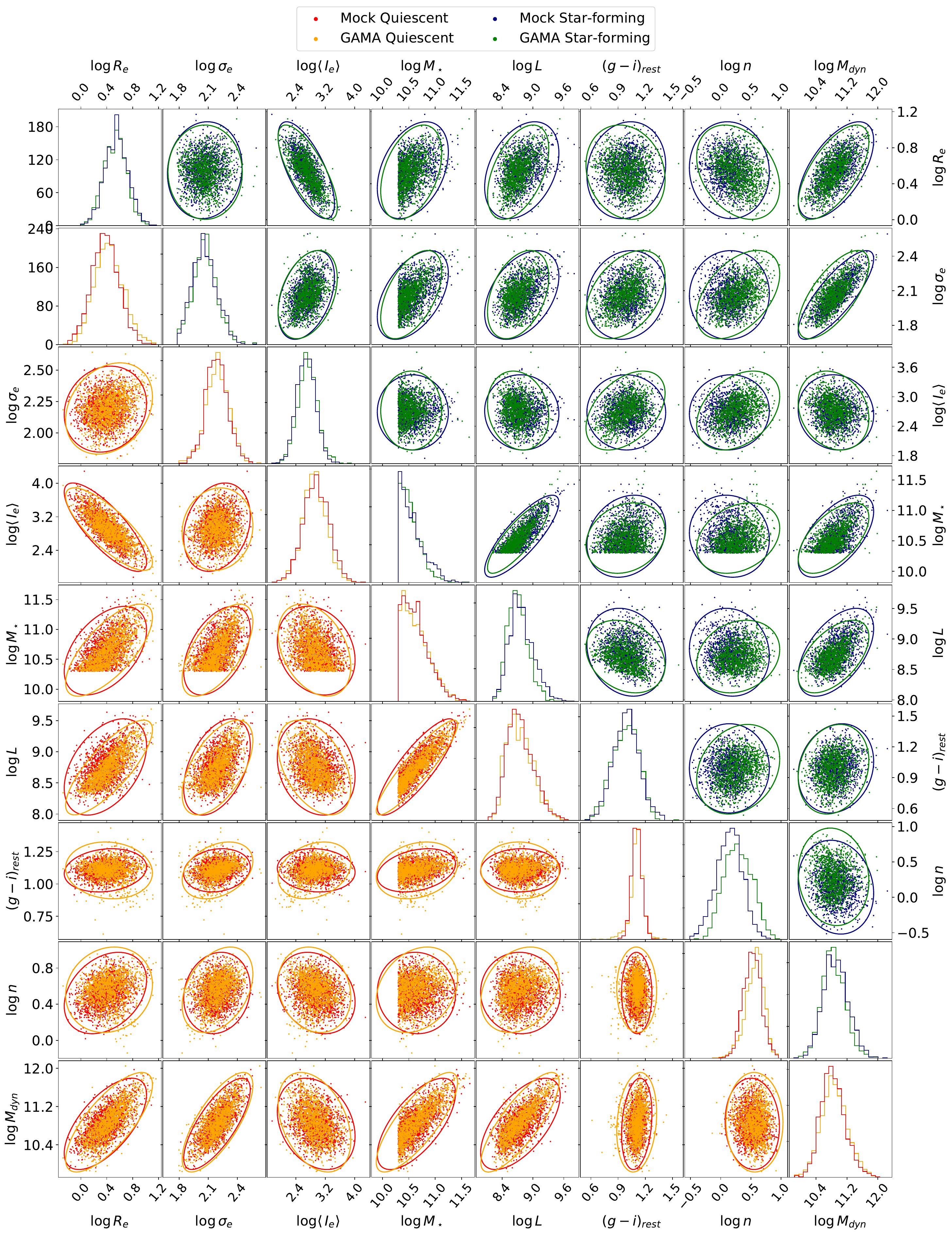}
    \caption{Illustration of mock samples for both early (lower left panels) and late type galaxies (upper right panels) in comparison to GAMA samples overlaid with 99\% confidence ellipses showing the underlying Gaussian distribution, corresponding to each sample.}
    \label{fig:mocks}
\end{figure*}

The algorithm for generating the mock samples is outlined below.
\begin{enumerate} \setlength{\itemsep}{0pt}
    \item Calculate $v-$axes using the FP slopes $a$ and $b$
    \item Calculate the covariance matrix by $\bm{\Sigma}_\text{FP}=\bm{V}\bm{\Lambda}\bm{V}^\intercal$ where $\bm{V}$ is the matrix containing $\bm{\hat{v}_1}, \bm{\hat{v}_2}, \bm{\hat{v}_3}$ as columns and $\bm{\Lambda}=\text{diag}(\sigma_1^2, \sigma_2^2, \sigma_3^2)$.
    \item Draw $\{r, s, i \}$ values randomly from a 3D Gaussian with mean $(\bar{r}, \bar{s}, \bar{i})$ and covariance matrix $\bm{\Sigma}_\text{FP}$ \label{item:rsi_true}
    \item \textit{Distances and redshifts}: Draw values randomly from a uniform distribution for comoving volume $V_\text{com}$, derive comoving distances $D_\text{com}$ from $V_\text{com}$, then derive redshifts from $D_\text{com}$. Finally, convert $D_\text{com}$ to angular diameter distance $D_A$.\label{item:zDist}
    \item \textit{Magnitudes}: Derive apparent $(m_\lambda)$ and absolute magnitudes $(M_\lambda)$ using surface brightness and effective radius in step \ref{item:rsi_true} and distances in step \ref{item:zDist}. \label{item:mags}
    \item \textit{Error estimates}: Generate 2\% uncertainties for $i$ and $s$, calculate uncertainty in $r$ with $\varepsilon_r=0.5\varepsilon_i$, then generate correlated errors from these uncertainties using a correlation coefficient -0.92 between $\varepsilon_r$ and $\varepsilon_i$ \label{item:error_rsi}
    \item Add measurement errors in step \ref{item:error_rsi} to $\{r, s, i \}$ in step \ref{item:rsi_true} to obtain observed FP parameters.
    \item Calculate \textit{selection probability} from equation (10) of \cite{magoulas2012} using the $r-$band magnitude limit of GAMA: $r_\text{lim}=19.8$ mag (AB units).
    \item Generate S\'ersic indices with $\log n = A_{\nu r}\log R_e + B_{\nu r} + \epsilon_{\nu r}$ \label{item:sersic}
    \item Estimate uncertainty $\epsilon\log n\equiv\epsilon_\nu$ using $\epsilon_r$ and the relevant scaling relation, then generate correlated errors using these uncertainties and correlation coefficient 0.76. \label{item:n_errors}
    \item Add the errors to $\log n$ from step \ref{item:sersic} to obtain observed S\'ersic indices.
    \item Calculate $M_\text{dyn}$ using $k(n)$ from \cite{bertin2002} with $r$ and $s$ from step \ref{item:rsi_true} and S\'ersic index from step \ref{item:sersic}
    \item Generate stellar-masses with $\log M_\star = A_\text{msd} \log M_\text{dyn} + B_\text{msd} + \epsilon_\text{msd}$
    \item Generate color from stellar-mass with $\text{color} = A_\text{mc}\log M_\star + B_\text{mc} + \epsilon_\text{mc}$\label{item:masscolor}
    \item Obtain observed values using correlated errors as in step \ref{item:n_errors} with correlation coefficient 0.5 \label{item:mc_obs}
    \item Generate stellar-mass-to-light ratio from rest-frame color with $\log M_\star/L = A_\text{mlc} \text{color} + B_\text{mlc} + \epsilon_\text{mlc}$, then obtain observed values as in steps \ref{item:n_errors} and \ref{item:mc_obs} \label{item:mlc}
    \item Luminosity can be calculated using either absolute magnitudes in step \ref{item:mags} or using $M_\star/L$ and $M_\star$ from steps \ref{item:masscolor}-\ref{item:mlc}.
    \item Finally, apply selection limits discussed in section \ref{sec:gama_sample}: $0.01<z<0.12\text{ and } 10.3\leqslant\log M_\star$
\end{enumerate}
Same algorithm can be applied to generate samples for SFs, using the input parameters derived from the corresponding best-fitting model to the GAMA SF sample. One of the mock samples for both Qs and SFs together with the actual GAMA Q/SF samples are shown in Figure \ref{fig:mocks} where the close agreement between mocks and actual data can be seen.

\begin{figure*} \centering
    \includegraphics[width=0.8\textwidth]{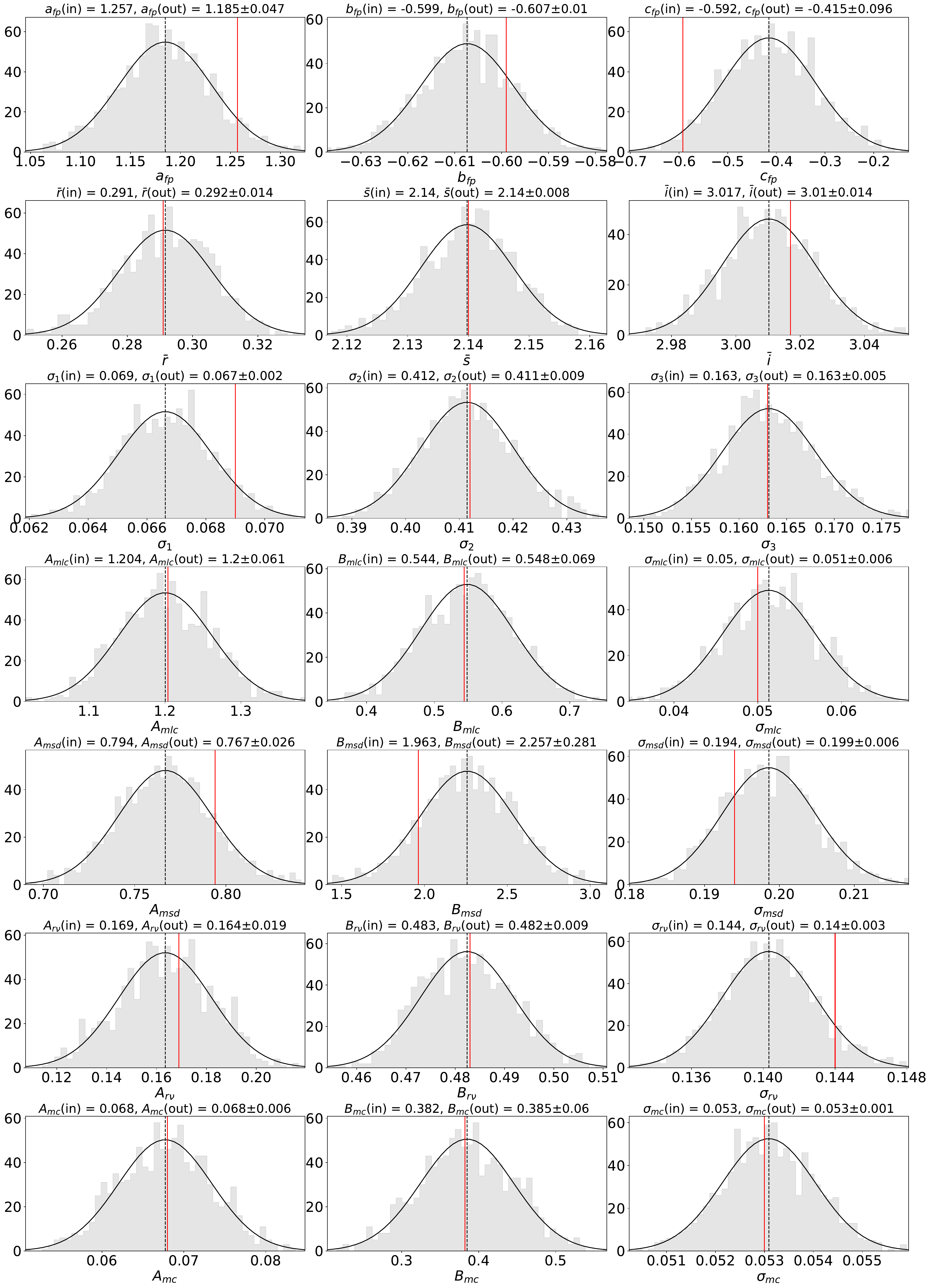}
    \caption{Distributions of the fitted parameters obtained from 1000 simulations of the best-fitting model for Qs. The first three rows are the FP parameters, the remaining four rows show the slopes, intercepts and intrinsic scatters of the scaling relations mass-to-light vs $g-i$ color (mlc), stellar vs dynamical mass (msd), S\'ersic index vs size $(r\nu)$ and $g-i$ color vs stellar-mass (mc) respectively. Best-fitting Gaussian to the distribution for each parameter is shown with black curve. Vertical dashed line shows the mean fitted value and vertical red line shows the input value. Each panel is centered on the mean and the $x-$axis spans $\pm3\sigma$.}
    \label{fig:1000mocks}
\end{figure*}

\section{Derivation of the normalization Factor}\label{sec:fj_derivation}

In this appendix, we show how to derive an expression for the normalization factor $f_j$ in equation~(\ref{eq:fj_w}), in terms of CDF and PDF of Gaussians, that not only makes efficient sampling possible in \textit{PySTAN}, but also makes the calculations faster using the corresponding optimized functions in \textit{SciPy} instead of numerical integration which is much slower.

Considering the integration limits given in equation \ref{eq:ucut}, the normalization integral in equation~(\ref{eq:fj_w}) becomes,
\begin{equation}
    f_j = \int\limits_{u_{\text{cut}, j}}^\infty du_j \int\limits_{s_{\text{low}}}^{s_{\text{up}}} ds_j \int\limits_{-\infty}^\infty  di_j d\nu_j dc_j~p(u_j, s_j, i_j, \nu_j, c_j | \bm{\bar{w}}, \mathbf{C}^w_j)~.
    \label{eq:fj_usi}
\end{equation}
Since no cut-offs are applied to $\bm{i}, \bm{\nu}$ and $\bm{c}$, this integral is just the marginalization of the 5D PDF over these parameters,
\begin{equation}
    \int\limits_{-\infty}^\infty p(\bm{w_j} | \bm{\bar{w}}, \mathbf{C}^w_j) di_j d\nu_j dc_j = p(u_j, s_j | \bm{\bar{w}'}, \bm{C'}^w_j)~,
\end{equation}
which results in the bivariate Gaussian distribution of $\bm{w'_j}=(u_j, s_j)^\intercal$ where the mean vector is $\bm{\bar{w}'} = (\bar{u}, \bar{s})^\intercal =  (\bar{i} + 2\bar{r}, \bar{s})^\intercal$ while the covariance matrix $\bm{C'}^w_j$ can be obtained by just striking out the rows and columns of $\mathbf{C}^w_j$ corresponding to $\bm{i}, \bm{\nu}$ and $\bm{c}$. Equation~(\ref{eq:fj_usi}) now becomes,
\begin{equation}
    f_j = \int\limits_{u_{\text{cut}, j}}^\infty \int\limits_{s_{\text{low}}}^{s_{\text{up}}} \frac{ \exp{ \left[ -\frac{1}{2} (\bm{w'_j} - \bm{\bar{w}'})^\intercal (\bm{C'}^w_j)^{-1} (\bm{w'_j} - \bm{\bar{w}'})\right]} }{2\pi \sqrt{|\bm{C'}^w_j}|} du_j ds_j ~.
    \label{eq:fj_bvn}
\end{equation}
Using the Cholesky decomposition of $\bm{C'}^w_j = L_j L_j^\intercal$ in equation~(\ref{eq:fj_bvn}), the exponent can be written as,
\begin{align}
    \chi_j^2  &= (\bm{w'_j} - \bm{\bar{w}'})^\intercal (L_j L_j^\intercal)^{-1} (\bm{w'_j} - \bm{\bar{w}'}) \nonumber \\
    &= (\bm{w'_j} - \bm{\bar{w}'})^\intercal (L_j^{-1})^\intercal L_j^{-1} (\bm{w'_j} - \bm{\bar{w}'}) \nonumber \\
    &=\left[ L_j^{-1} (\bm{w'_j} - \bm{\bar{w}'}) \right]^\intercal \left[ L_j^{-1} (\bm{w'_j} - \bm{\bar{w}'}) \right].
    \label{eq:chisq}
\end{align}
Here, the lower triangular Cholesky factor $L_j$ and its inverse $L_j^{-1}$ are,
\begin{equation}
    L_j = \begin{pmatrix} L_{j,11} & 0 \\ L_{j,21} & L_{j,22} \end{pmatrix}~,~    L_j^{-1} = \begin{pmatrix} \frac{1}{L_{j,11}} & 0 \\ \frac{-L_{j,21}}{L_{j,11}L_{j,22}} & \frac{1}{L_{j,22}} \end{pmatrix} \equiv \begin{pmatrix} H_{11} & 0 \\ H_{21} & H_{22} \end{pmatrix}.
    \label{eq:chol_inv}
\end{equation}
Then,
\begin{align}
    Q_j \equiv L_j^{-1} (\bm{w'_j} - \bm{\bar{w}'}) = \begin{pmatrix} H_{11} & 0 \\ H_{21} & H_{22} \end{pmatrix} \begin{pmatrix} u_j - \bar{u} \\ s_j - \bar{s} \end{pmatrix}     = \begin{pmatrix} H_{11}(u_j - \bar{u}) \\ H_{21}(u_j - \bar{u}) + H_{22}(s_j - \bar{s}) \end{pmatrix} = \begin{pmatrix} Q_{j,1} \\ Q_{j,2} \end{pmatrix}~.
    \label{eq:qvector}
\end{align}
Inserting equation~(\ref{eq:qvector}) into (\ref{eq:chisq}), then renaming the constants result in;
\begin{align}
    \chi_j^2 &= Q_j^\intercal Q_j = (Q_{j,1}, Q_{j,2})\begin{pmatrix} Q_{j,1}\\Q_{j,2} \end{pmatrix} = Q_{j,1}^2 + Q_{j,2}^2 \nonumber\\
    &= \underbrace{(H_{11}^2 + H_{21}^2)}_{2A_1^2} (u_j - \bar{u})^2 + 2\underbrace{H_{21}H_{22}}_{A_2} (u_j - \bar{u})(s_j - \bar{s})+ \underbrace{H_{22}^2}_{2A_3^2} (s_j - \bar{s})~.
    \label{eq:chisqfinal}
\end{align}
Substituting (\ref{eq:chisqfinal}) along with $|\mathbf{C}'^w_j| = |L_j L_j^\intercal| = |L_j|^2 = (L_{j,11}L_{j,22})^2$ in the integral (\ref{eq:fj_bvn}) gives, 
\begin{align}
    f_j &= \int\limits_{u_{\text{cut}, j}}^\infty \int\limits_{s_{\text{cut}}}^\infty \frac{ e^{-\left[ A_1^2 (u_j - \bar{u})^2 + A_2 (u_j - \bar{u}) (s_j - \bar{s}) + A_3^2 (s_j - \bar{s})^2 \right]} }{2\pi L_{j,11}L_{j,22}} ds_j du_j \nonumber\\
    &= \frac{1}{2\pi L_{j,11}L_{j,22}} \int\limits_{u_{\text{cut}, j}}^\infty e^{-A_1^2 (u_j - \bar{u})^2} \left[\,\, \int\limits_{s_\text{low}}^{s_\text{up}} e^{-A_3^2 (s_j - \bar{s})^2 - A_2 (u_j - \bar{u}) (s_j - \bar{s}) } ds_j \right] du_j~.
    \label{eq:fj_decomp}
\end{align}
The integral over $s$ is in the form 
\begin{equation}
    \int e^{-ax^2 - bx} dx = \frac{1}{2} \sqrt{\frac{\pi}{a}} e^{b^2/4a} \text{erf}\left[\sqrt{a}\left(x + \frac{b}{2a}\right) \right]\text{ for } a>0
\end{equation}
where erf is the error function. Therefore, for $a=A_3^2>0$ and $b=A_2 (u_j - \bar{u})$, the inner integral over $s$ becomes,
\begin{align}
    I_s = \frac{1}{2} \sqrt{\frac{\pi}{a}} e^{b^2/4a} \text{erf}\left[\sqrt{a}\left((s_j - \bar{s}) + \frac{b}{2a}\right) \right]_{s_\text{low}}^{s_\text{up}} 
\end{align}
Error function can be expressed as $\text{erf}(x) = 2\Phi(x\sqrt{2}) - 1$, where 
\begin{equation}
    \Phi(x) = \frac{1}{\sqrt{2\pi}}\int_{-\infty}^x e^{-t^2/2} dt = \int_{-\infty}^x \phi(t) dt \text{ and } \phi(x) = \frac{d\Phi(x)}{dx}
    \label{eq:int_s_erf}
\end{equation}
Then, equation~(\ref{eq:int_s_erf}) becomes:
\begin{equation}
    I_s = \sqrt{\frac{\pi}{a}} e^{b^2/4a} \Phi \left[ \sqrt{2a}(s_j - \bar{s}) + \frac{b}{\sqrt{2a}}\right]_{s_\text{low}}^{s_\text{up}}.
    \label{eq:int_s_phi}
\end{equation}
Plugging equation~(\ref{eq:int_s_phi}) into (\ref{eq:fj_decomp}) along with the values of $a$ and $b$ constants:
\begin{equation}
    f_j = \frac{\sqrt{\pi}}{2\pi L_{j,11}L_{j,22} A_3 } \int\limits_{u_{\text{cut}, j}}^\infty e^{-\left[ A_1^2 - \frac{A_2^2}{4A_3^2}\right](u_j - \bar{u})^2 } \Phi \left[ \sqrt{2}A_3(s_j - \bar{s}) + \frac{A_2 (u_j - \bar{u})}{\sqrt{2}A_3}\right]_{s_\text{low}}^{s_\text{up}} du_j.
    \label{eq:fj_uint}
\end{equation}
The definitions of the constants $A_1, A_2, A_3$ from equations~(\ref{eq:chisqfinal}) and (\ref{eq:chol_inv}):
\begin{align}
    & A_1^2 = \frac{H_{11}^2 + H_{21}^2}{2} = \frac{1}{2L_{j,11}^2} + \left( \frac{-L_{j,21}}{2L_{j,11}L_{j,22}} \right)^2~, \nonumber\\
    & A_2 = H_{21} H_{22} = \frac{-L_{j,21}}{L_{j,11}L_{j,22}^2}~,\nonumber\\
    & A_3^2 = \frac{H_{22}^2}{2} = \frac{1}{2L_{j,22}^2} \Rightarrow \sqrt{2}A_3 = \frac{1}{L_{j,22}}~, \nonumber\\
    & A_1^2 - \frac{A_2^2}{4A_3^2} = \frac{H_{11}^2 + H_{21}^2}{2} - \frac{H_{21}^2 H_{22}^2}{2 H_{22}^2} = \frac{H_{11}^2}{2} = \frac{1}{2L_{j,11}^2}~.
    \label{eq:int_const}
\end{align}
Finally, plugging equation~(\ref{eq:int_const}) into (\ref{eq:fj_uint}),
\begin{equation}
    f_j = \frac{1}{\sqrt{2\pi} L_{j,11}} \int\limits_{u_{\text{cut}, j}}^\infty e^{-\frac{(u_j - \bar{u})^2}{2L_{j,11}^2} } \Phi \left( \frac{s_j - \bar{s}}{L_{j,22}} - \frac{L_{j,21}}{L_{j,22}} \frac{u_j - \bar{u}}{L_{j,11}}\right)\Bigg|_{s_\text{low}}^{s_\text{up}} du_j
\end{equation}
and using the PDF of normal distribution, we can write
\begin{equation}
    f_j = \frac{1}{L_{j,11}} \int\limits_{u_{\text{cut}, j}}^\infty \phi \left( \frac{u_j - \bar{u}}{L_{j,11}} \right) \Phi \left( \frac{s_j - \bar{s}}{L_{j,22}} - \frac{L_{j,21}}{L_{j,22}} \frac{u_j - \bar{u}}{L_{j,11}}\right)\Bigg|_{s_\text{low}}^{s_\text{up}} du_j.
    \label{eq:fj_1mPhi}
\end{equation}
We can make even further simplifications by making the following definitions,
\begin{equation}
    \alpha_{j, \text{up}} \equiv \frac{s_\text{up} - \bar{s}}{L_{j,22}}, \quad \alpha_{j, \text{low}} \equiv \frac{s_\text{low} - \bar{s}}{L_{j,22}}, \quad \beta_j \equiv \frac{-L_{j,21}}{L_{j,22}}
    \label{eq:alpha_beta_def}
\end{equation}
with change of variables to $t_j = \frac{u_j - \bar{u}}{L_{j,11}} \Rightarrow dt_j = \frac{du_j}{L_{j,11}},\, t_\text{cut,j} = \frac{u_{\text{cut}, j} - \bar{u}}{L_{j,11}}$,
\begin{align}
    f_j = \int\limits_{t_{\text{cut,j}}}^\infty \phi(t_j) \left[ \Phi ( \alpha_{j, \text{up}} + \beta_j t_j ) - \Phi ( \alpha_{j, \text{low}} + \beta_j t_j )\right] dt_j~.
    \label{eq:fj_t}
\end{align}
This integral cannot be expressed in a closed form, however, it can be broken down into
\begin{equation}
    \int\limits_{t_{\text{cut,j}}}^\infty \phi(t_j)\Phi ( \alpha + \beta t_j )dt_j = \int\limits_{-\infty}^\infty \phi(t_j)\Phi ( \alpha + \beta t_j )dt_j - \int\limits^{t_{\text{cut,j}}}_{-\infty} \phi(t_j)\Phi ( \alpha + \beta t_j )dt_j~.
    \label{eq:int_pdfcdf}
\end{equation}
Using the tables provided by \cite{owen1980}, the first integral on the right hand side of equation~(\ref{eq:int_pdfcdf}) can be expressed in a closed form and the second integral can be expressed in terms of the CDF of the bivariate normal distribution as follows:
\begin{equation}
    \int\limits_{t_{\text{cut,j}}}^\infty \phi(t_j)\Phi ( \alpha + \beta t_j )dt_j = \Phi\left( \frac{\alpha}{\sqrt{1 + \beta^2}} \right) - \text{BvN}\left( \frac{\alpha}{\sqrt{1 + \beta^2}}, t_{\text{cut,j}}, \rho=\frac{-\beta}{\sqrt{1 + \beta^2}} \right).
    \label{eq:int_pdfcdf2}
\end{equation}
where BvN is the CDF of the (standard) bivariate normal distribution with correlation $\rho$ given by,
\begin{equation}
    \text{BvN}(h, k, \rho) = \frac{1}{2\pi\sqrt{1 - \rho^2}} \int\limits_{-\infty}^k \int\limits_{-\infty}^h \text{exp}\left[ -\frac{1}{2}\left( \frac{x^2 - 2\rho xy + y^2}{1 - \rho^2}\right) \right]dx dy.
\end{equation}
and it can be numerically evaluated using the readily available algorithms such as the one of \cite{richardj1989}. Finally, after rearranging the definitions in equation~(\ref{eq:alpha_beta_def}) as,
\begin{align}
    h_{j, \text{up/low}} &\equiv \frac{\alpha_{j, \text{up/low}}}{\sqrt{1 + \beta_j^2}} = \frac{s_\text{up/low}-\bar{s}}{\sqrt{L_{j,22}^2 + L_{j,11}^2}}\nonumber\\
    k_j &\equiv t_\text{cut, j} = \frac{u_{\text{cut}, j} - \bar{u}}{L_{j,11}} \nonumber \\
    \rho_j & \equiv \frac{-\beta_j}{\sqrt{1 + \beta_j^2}} = \frac{L_{j,21}}{\sqrt{L_{j,22}^2 + L_{j,11}^2}}
\end{align}
we can insert the results from equations~(\ref{eq:int_pdfcdf}) and~(\ref{eq:int_pdfcdf2}) into equation~(\ref{eq:fj_t}), which gives us the final expression for $f_j$ as follows:
\begin{equation}
    f_j = \Phi\left( h_{j, \text{up}} \right)- \text{BvN}\left( h_{j, \text{up}}, k_j, \rho_j \right) - \Phi\left( h_{j, \text{low}} \right) + \text{BvN}\left( h_{j, \text{low}}, k_j, \rho_j \right)~.
    \label{eq_app:fj_final}
\end{equation}
Notice that in case of $s_\text{up}\rightarrow \infty$ this equation reduces to 
\begin{equation}
    f_j = 1 - \Phi\left( k_j \right) - \Phi\left( h_{j, \text{low}} \right) + \text{BvN}\left( h_{j, \text{low}}, k_j, \rho_j \right)~.
    \label{eq_app:fj_s_low_final}
\end{equation}
Given the upper limit that we adopted for velocity dispersion is $\sigma_e<450$ km/s, equations~(\ref{eq_app:fj_final}) and~(\ref{eq_app:fj_s_low_final}) differ by $\lesssim 10^{-4}$ for all $j$.

\begin{figure}
    \centering
    \includegraphics[width=0.47\textwidth]{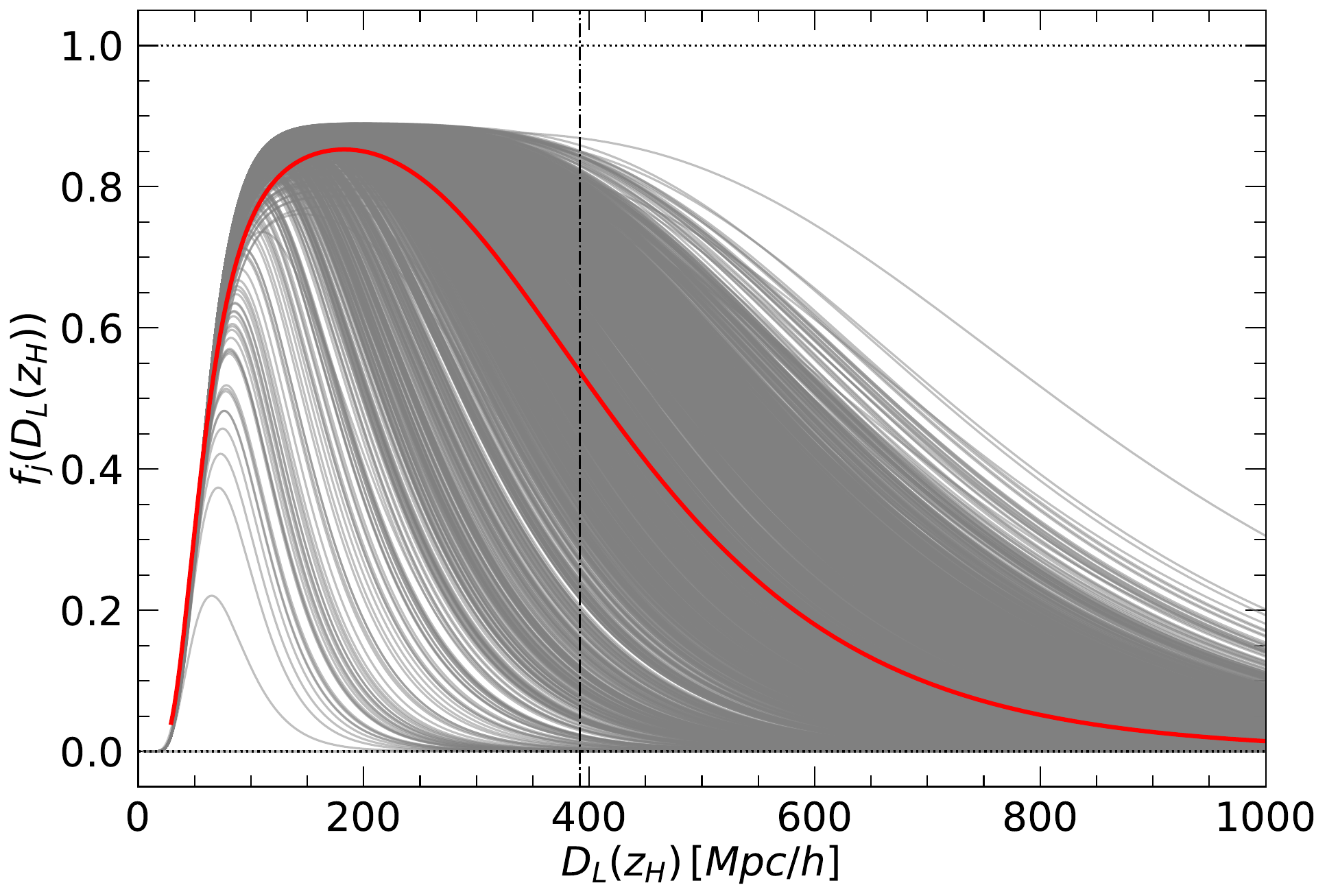}
    \caption{Normalization factor $f_j$ as a function of luminosity distance $(D_L)$ at proposed cosmological redshift $(z_H)$ for each quiescent galaxy in our sample. The thick red curve shows the median $f_j$. The vertical dash-dotted line shows the luminosity distance corresponding to the redshift limit of our sample, $z=0.12$.}
    \label{fig:fj_distance}
\end{figure}

We present this exact solution for each galaxy in Figure \ref{fig:fj_distance}, which shows the variation of the normalization factor $f_j$ as a function of proposed distance. 

\section{Peculiar Velocities From Separate and Independent Modeling of the Samples}\label{sec:pvs_separate}

\begin{figure*}[htbp]
    \centering
\begin{minipage}[c]{0.485\linewidth}    
\includegraphics[width=\textwidth]{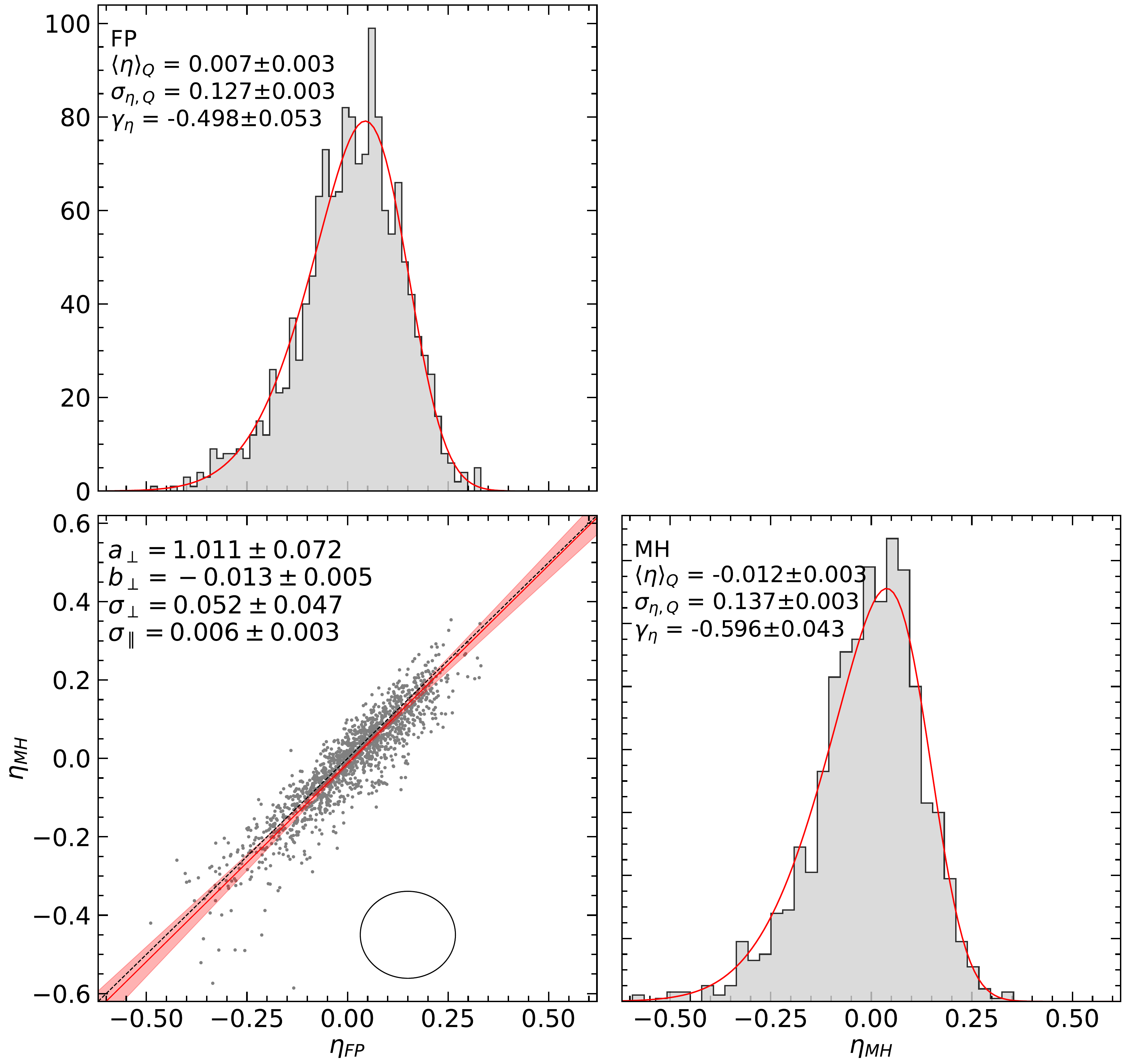}
    \caption{Same as Figure \ref{fig:fp_mh_comparison_q} but for Qs, when the Q and SF populations are modeled separately and independently.}
    \label{fig:fp_mh_comparison_q_separate}
\end{minipage}
\hfill
\begin{minipage}[c]{0.485\linewidth}
    \centering
    \includegraphics[width=\textwidth]{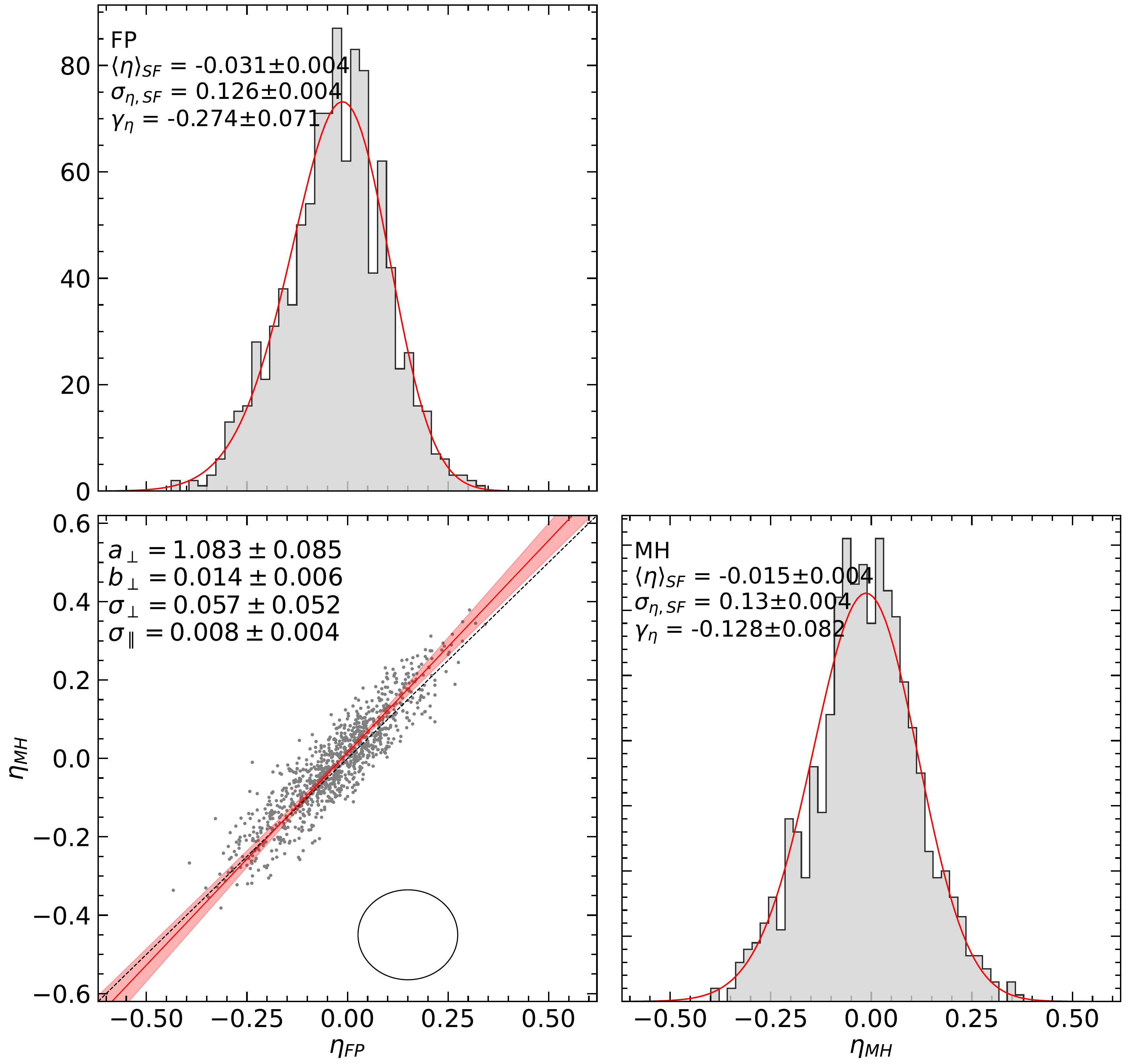}
    \caption{Same as Figure \ref{fig:fp_mh_comparison_q_separate} but for the SFs.}
    \label{fig:fp_mh_comparison_sf_separate}
\end{minipage}
\end{figure*}

In this appendix, we make comparisons between the PVs derived from the FP and the MP, as we have done in section \ref{sec:mp_vs_fp}, but for when the quiescent and star-forming galaxy samples are treated separately and independently. We present this comparison in Figures \ref{fig:fp_mh_comparison_q_separate} and \ref{fig:fp_mh_comparison_sf_separate}, which show that in this case, the FP and MH work approximately as well for both galaxy populations, with similar RMS scatter and skewness for all measurements, and small but not necessarily negligible mean offsets between the alternative analysis.  The skewness in these distributions, like those in Figures \ref{fig:fp_mh_comparison_q} and \ref{fig:fp_mh_comparison_sf}, is driven by a few outliers at very low $\eta$.  Noting that in our FP/MH fitting such outliers are objectively identified and down-weighted according the good/bad mixture modeling, these points can be seen in, e.g., Figures \ref{fig:lfpnc_trends}, \ref{fig:sps_residuals_fp}, and \ref{fig:q_z_residuals}, as massive, red, and very high velocity dispersion galaxies towards the upper end of our redshift window.

\end{document}